\documentclass[twocolumn]{aastex62}
\pdfoutput=1 
\usepackage{amsmath,amstext}
\usepackage[T1]{fontenc}
\usepackage[figure,figure*]{hypcap}
\usepackage{mwe}
\usepackage{lipsum}
\usepackage{graphicx} 
\usepackage{multirow}
\usepackage{xcolor,colortbl}
\usepackage{xspace}
\usepackage{csvsimple}
\usepackage{afterpage}

\newcommand{\twelveCO}{$^{12}$CO}
\newcommand{\thirtCO}{$^{13}$CO}

\newcommand{\nh}{$n_{\text{H}_2}$\xspace}
\newcommand{\NCO}{$N_\text{CO}$\xspace}
\newcommand{\tkin}{$T_\text{kin}$\xspace}
\newcommand{\XCO}{$X_\text{CO}$\xspace}
\newcommand{\XthCO}{$X_\text{13CO}$\xspace}
\newcommand{\radex}{\texttt{RADEX}\xspace}

\shorttitle{Non-LTE Fitting in Molecular Ridge}
\shortauthors{Finn et al.}

\begin{document}

\title{Physical Conditions in the LMC's Quiescent Molecular Ridge: Fitting Non-LTE Models to CO Emission}

\author{Molly K. Finn} %
\affiliation{Department of Astronomy, University of Virginia, Charlottesville, VA 22904, USA}

\author{Remy Indebetouw} %
\affiliation{Department of Astronomy, University of Virginia, Charlottesville, VA 22904, USA}
\affiliation{National Radio Astronomy Observatory, 520 Edgemont Road, Charlottesville, VA 22903, USA}

\author{Kelsey E. Johnson} %
\affiliation{Department of Astronomy, University of Virginia, Charlottesville, VA 22904, USA}

\author{Allison H. Costa} %
\affiliation{Department of Astronomy, University of Virginia, Charlottesville, VA 22904, USA}

\author{C.-H. Rosie Chen} %
\affiliation{Max-Planck-Institut f\"{u}r Radioastronomie, Auf dem H\"{u}gel 69, D-53121 Bonn, Germany}

\author{Akiko Kawamura} %
\affiliation{National Astronomical Observatory of Japan, National Institutes of Natural Sciences, 2-21-1 Osawa, Mitaka, Tokyo 181-8588, Japan}

\author{Toshikazu Onishi} %
\affiliation{Department of Physical Science, Graduate School of Science, Osaka Prefecture University, 1-1 Gakuen-cho, Naka-ku, Sakai, Osaka 599-8531, Japan}

\author{J\"{u}rgen Ott} %
\affiliation{National Radio Astronomy Observatory, P.O. Box O, Socorro, NM 87801, USA}
\affiliation{Physics Department, New Mexico Institute of Mining and Technology, 801 Leroy Pl., Socorro, NM 87801, USA}

\author{Kazuki Tokuda} %
\affiliation{National Astronomical Observatory of Japan, National Institutes of Natural Sciences, 2-21-1 Osawa, Mitaka, Tokyo 181-8588, Japan}
\affiliation{Department of Physical Science, Graduate School of Science, Osaka Prefecture University, 1-1 Gakuen-cho, Naka-ku, Sakai, Osaka 599-8531, Japan}

\author{Tony Wong} %
\affiliation{Astronomy Department, University of Illinois, 1002 W. Green Street, Urbana, IL 61801, USA}

\author{Sarolta Zahorecz} %
\affiliation{National Astronomical Observatory of Japan, National Institutes of Natural Sciences, 2-21-1 Osawa, Mitaka, Tokyo 181-8588, Japan}
\affiliation{Department of Physical Science, Graduate School of Science, Osaka Prefecture University, 1-1 Gakuen-cho, Naka-ku, Sakai, Osaka 599-8531, Japan}

\begin{abstract}

The Molecular Ridge in the LMC extends several kiloparsecs south from 30 Doradus, and it contains $\sim30$\% of the molecular gas in the entire galaxy. However, the southern end of the Molecular Ridge is quiescent---it contains almost no massive star formation, which is a dramatic decrease from the very active massive star-forming regions 30 Doradus, N159, and N160.
We present new ALMA and APEX observations of the Molecular Ridge at a resolution as high as $\sim$16\arcsec\ ($\sim$3.9 pc) with molecular lines \twelveCO(1-0), \thirtCO(1-0), \twelveCO(2-1), \thirtCO(2-1), and CS(2-1). We analyze these emission lines with our new multi-line non-LTE fitting tool to produce maps of \tkin, \nh, and \NCO across the region based on models from \radex. Using simulated data for a range of parameter space for each of these variables, we evaluate how well our fitting method can recover these physical parameters for the given set of molecular lines. We then compare the results of this fitting with LTE and \XCO methods of obtaining mass estimates and how line ratios correspond with physical conditions. We find that this fitting tool allows us to more directly probe the physical conditions of the gas and estimate values of \tkin, \nh, and \NCO that are less subject to the effects of optical depth and line-of-sight projection than previous methods. The fitted \nh values show a strong correlation with the presence of YSOs, and with the total and average mass of the associated YSOs. Typical star formation diagnostics, such as mean density, dense gas fraction, and virial parameter do not show a strong correlation with YSO properties.

\end{abstract}


\section{Introduction} \label{sec:intro} 

Our understanding of star formation is heavily dependent on our understanding of molecular clouds and the physics that governs them. It is difficult, and in many cases impossible, to determine those physical conditions without relying on assumptions or scaling relations \citep[e.g., ][]{Kennicutt98}. These assumptions appear to be sufficient in many cases, but there are examples of clouds and regions of galaxies that are not forming stars as we would expect based on these scaling relations. For example, ``Maddalena's Cloud'' G126-2.5 is a giant molecular cloud in the Milky Way that has unusually low star formation \citep{Maddalena85}, and the star formation rate in the Central Molecular Zone in the Galactic Center is an order of magnitude lower than would be predicted by Galactic trends \citep{Longmore13}. To understand why, we must determine physical conditions without making assumptions that molecular clouds are behaving in the "typical" way.

One of the most common mass estimates for molecular clouds comes from the CO-to-H$_2$ conversion factor, \XCO, which is discussed in detail in \citet{Bolatto13}. It is often calibrated with the virial mass or dust mass and is used to convert the integrated intensity of \twelveCO(1-0) emission to a column density of H$_2$. The \XCO factor method is most valid when determining masses on large size scales where many molecular clouds are smoothed together, averaging over the varying physical conditions. On the scale of individual star-forming clouds or individual lines-of-sight, the conversion of CO flux to H$_2$ column density with an adopted \XCO factor cannot be expected to be constant \citep[][and references therein]{Bolatto13}.

Another measure of mass can be made by assuming local thermal equilibrium (LTE) to get excitation temperature, optical depth, and column density \citep{MangumShirley15}. This method is based on the assumption that the gas is sufficiently dense for the molecular excitation levels to have a Boltzmann distribution corresponding to an excitation temperature, $T_\text{ex}$, and that the excitation temperatures of \twelveCO\ and \thirtCO\ are equal. This method also often assumes that \twelveCO\ is optically thick, allowing for an easy estimate of the excitation temperature ($T_{ex}$) from the brightness temperature ($T_B$), while \thirtCO\ is optically thin, which makes it possible to determine the optical depth with an assumption of the relative abundance of \twelveCO\ and \thirtCO\ \citep{KoeppenKegel80}.

However, these calculations break down if \twelveCO\ becomes optically thin, or if either line's level population is not well described by a Boltzmann distribution. Studies have shown regimes in which the LTE calculations overestimate the column density by up to a factor of two in bright ($T_B$ > 40 K) clouds \citep{Indebetouw20}, and underestimate the mass by up to a factor of 7 when the \thirtCO\ becomes sub-thermally excited \citep{Castets90,Padoan00,Heyer15}. Assuming LTE also requires that the density is sufficiently high such that the excitation is entirely governed by temperature, meaning that any dependence on density drops out of the equations and so cannot be solved for.

Ratios of isotopologues (e.g. \thirtCO/\twelveCO) can trace volume density in the case where one line is optically thick and the other line is sub-thermally excited \citep{Nishimura15}. Ratios of upper to lower excitation levels of CO (e.g. \twelveCO(2-1)/\twelveCO(1-0)) scale with excitation temperature and density when both lines are optically thin, and the ratio approaches unity as the lines get increasingly optically thick \citep{Sakamoto94,Nishimura15,Penaloza17}. These ratios are also dependent on optical depth and local abundance ratios and so can only provide rough diagnostics of the density and temperature \citep{Penaloza17}.

In this study, we fit molecular line observations to the results of non-LTE escape probability models from \radex \citep{radex}. This avoids many of the assumptions required for other methods, such as those listed above, and so allows us to better characterize the actual physical conditions of the gas. With this method, we obtain estimates of not just the temperature and column density, but also the volume density. Our only assumptions in this case are that the different molecular lines are tracing the same gas with a constant abundance ratio throughout the cloud, and that the gas in each voxel is homogeneous---we fit only one set of physical conditions for each pixel and velocity channel despite the fact that temperature and density almost certainly vary along the line of sight and within the beam. Through this model-fitting study, we determine the physical conditions of molecular clouds in the Large Magellanic Cloud (LMC) and compare those results to other common methods: adopting an \XCO factor, assuming LTE, or using line diagnostics.

\begin{figure*}
    \centering
    \includegraphics[width=0.56\textwidth]{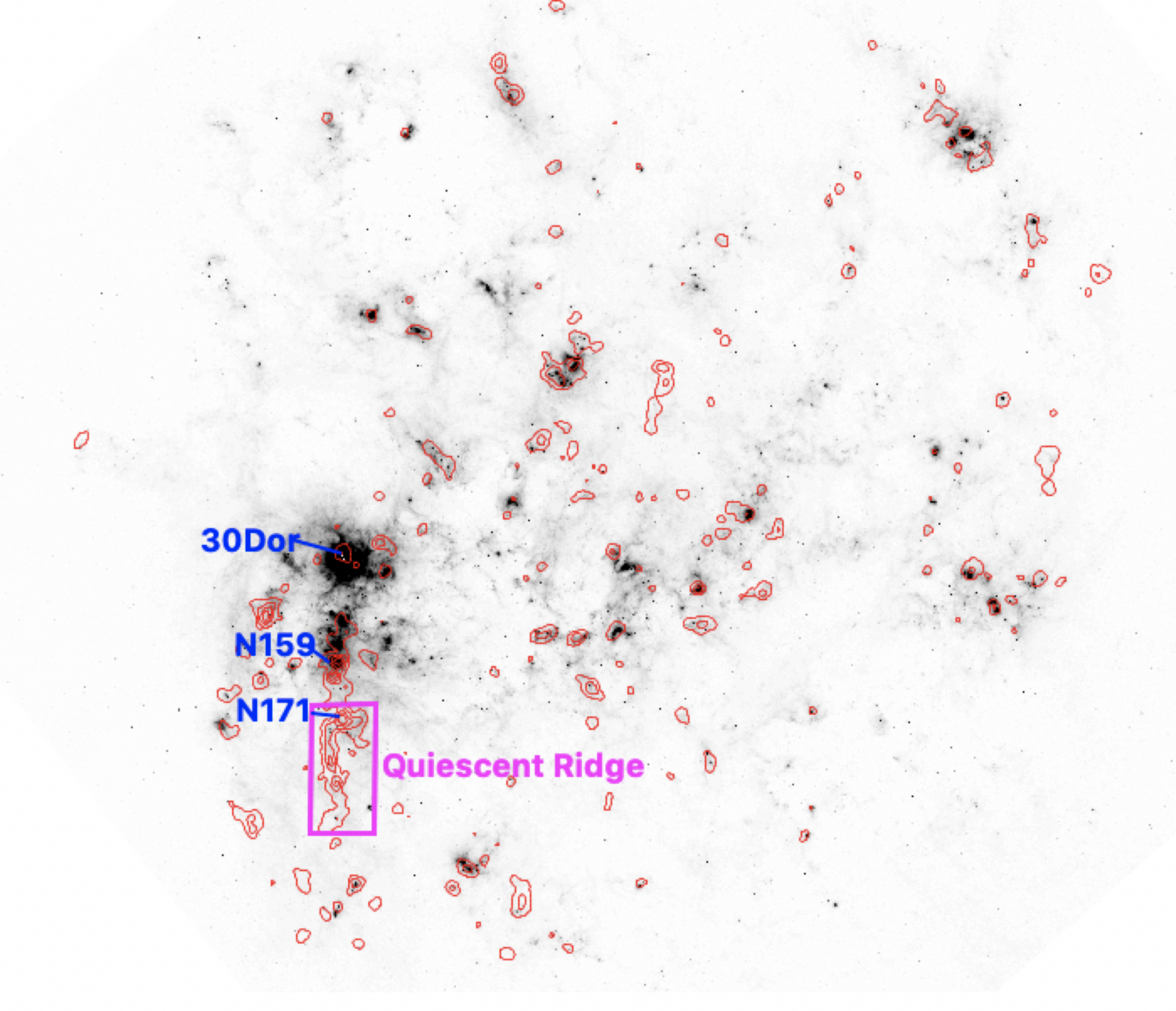}
    \includegraphics[width=0.43\textwidth]{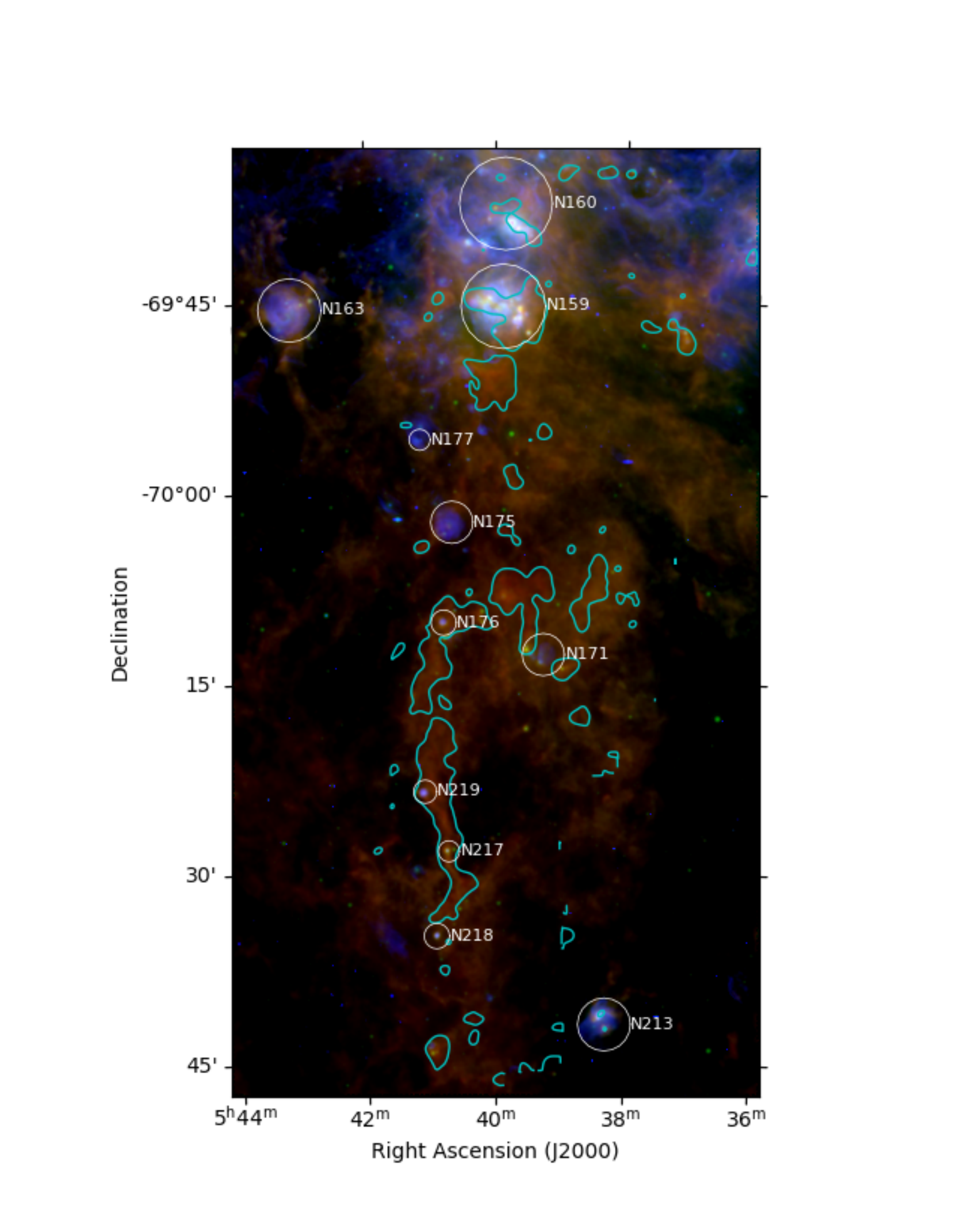}
    \caption{\emph{Left: } The LMC with the Molecular Ridge highlighted in pink and active star-forming regions 30 Doradus (30 Dor), N159, and N171 in blue. The grayscale is MIPS 24 $\mu$m from the SAGE survey \citep{Meixner06}, and the red contours are \twelveCO(1-0) from the NANTEN survey \citep{NANTEN}. We can see from this that the Ridge is a unique feature, showing up prominently in the red \twelveCO(1-0) contours, but lacking any strong emission in the 24 $\mu$m grayscale, which is a common star formation tracer. \emph{Right: } RGB image zoomed in on the Ridge. The red is PACS 250 $\mu$m from the HERITAGE survey, green is the same 24 $\mu$m as the grayscale on the left, and blue is H$\alpha$ from MCELS survey \citep{SmithMCELS98}. The cyan contours are \twelveCO(1-0) from the MAGMA survey, and HII regions identified by \citet{Henize56} are shown as white circles. }
    \label{fig:LMC zoom in}
\end{figure*}

We use as a case study the quiescent southern end of the Molecular Ridge in the LMC, extending 1-2 kpcs south from 30 Doradus (Figure\,\ref{fig:LMC zoom in}). We assume a distance to the LMC of 50 kpc \citep{Schaefer08}. \citet{Cohen88} first noted the Ridge as a striking feature in low resolution maps of \twelveCO(1-0), and further observations of \twelveCO\ by the NANTEN survey \citep{NANTEN} revealed that it contained $\sim$30\% of all CO-bright molecular gas mass in the LMC \citep{Mizuno01}. Despite the large reservoir of molecular gas, the Ridge is surprisingly quiescent, showing little sign of star formation based on the presence of young optical clusters or H$\alpha$ emission \citep{Davies76,Bica96,Yamaguchi01}. \citet{Indebetouw08} find the H$\alpha$ emission \citep{Calzetti07} would predict a star formation rate of $2.6\times10^{-4}$ M$_\odot$ yr$^{-1}$, while the star formation rate predicted by the molecular gas surface density and the Schmidt-Kennicutt law \citep{Kennicutt98} would be $8\times10^{-3}$ M$_\odot$ yr$^{-1}$, over a factor of 30 larger.
There are only five HII regions in the Ridge that were identified and named by \citet{Henize56}, most notably N171, as well as some fainter HII regions (see Figure\,\ref{fig:LMC zoom in}).  

By looking for embedded stellar objects in the Ridge from the \textit{Spitzer} SAGE survey \citep{Meixner06}, \citet{Indebetouw08} showed that the lack of young, blue clusters and low H$\alpha$ emission is likely due to the Ridge preferentially forming relatively low mass star clusters rather than having deeply-embedded high mass objects or simply not forming stars at all (the star formation measured by modeling the YSO population was a factor of two lower than that predicted from the extragalactic Schmitt-Kennicutt law \citep{Kennicutt98}, but agreed within the uncertainties). This is a stark contrast to 30 Doradus and the active massive star formation regions N159 and N160 directly to the north of the Ridge that are forming massive stars prodigiously. This makes the Molecular Ridge a particularly interesting region for studies of the molecular gas properties. 

The quiescence of the Ridge could be due to atypical gas conditions, so a robust, assumption-minimizing approach is needed to analyze its physical conditions. To do this, we use four molecular lines - \twelveCO(1-0), \thirtCO(1-0), \twelveCO(2-1), and \thirtCO(2-1), the observations of which are described in \S\ref{sec:obs}. We fit non-LTE \radex models to those observed lines as described in \S\ref{sec: Radex Fitting}, and so avoid assumptions about stability, local excitation, or optical depth. We evaluate the performance of this fitting and details of methodology choices in the Appendices. 

In \S\ref{sec: clumps}, we segment the emission into clumps and determine the physical properties of these clumps. We then discuss YSOs detected in the Ridge and match them to those CO clumps in \S\ref{sec:ysos}.
We evaluate how the derived properties of the clumps compare with other common methods of determining physical conditions in \S\ref{sec:Comparisons}, and how the derived properties correlate with star formation as traced by the presence of associated YSOs in \S\ref{sec: SF trends}. Our major results are summarized in  \S\ref{sec:conclusions}.

\section{Observations} \label{sec:obs}

\begin{figure*}
    \centering
    \includegraphics[width=0.34\textwidth]{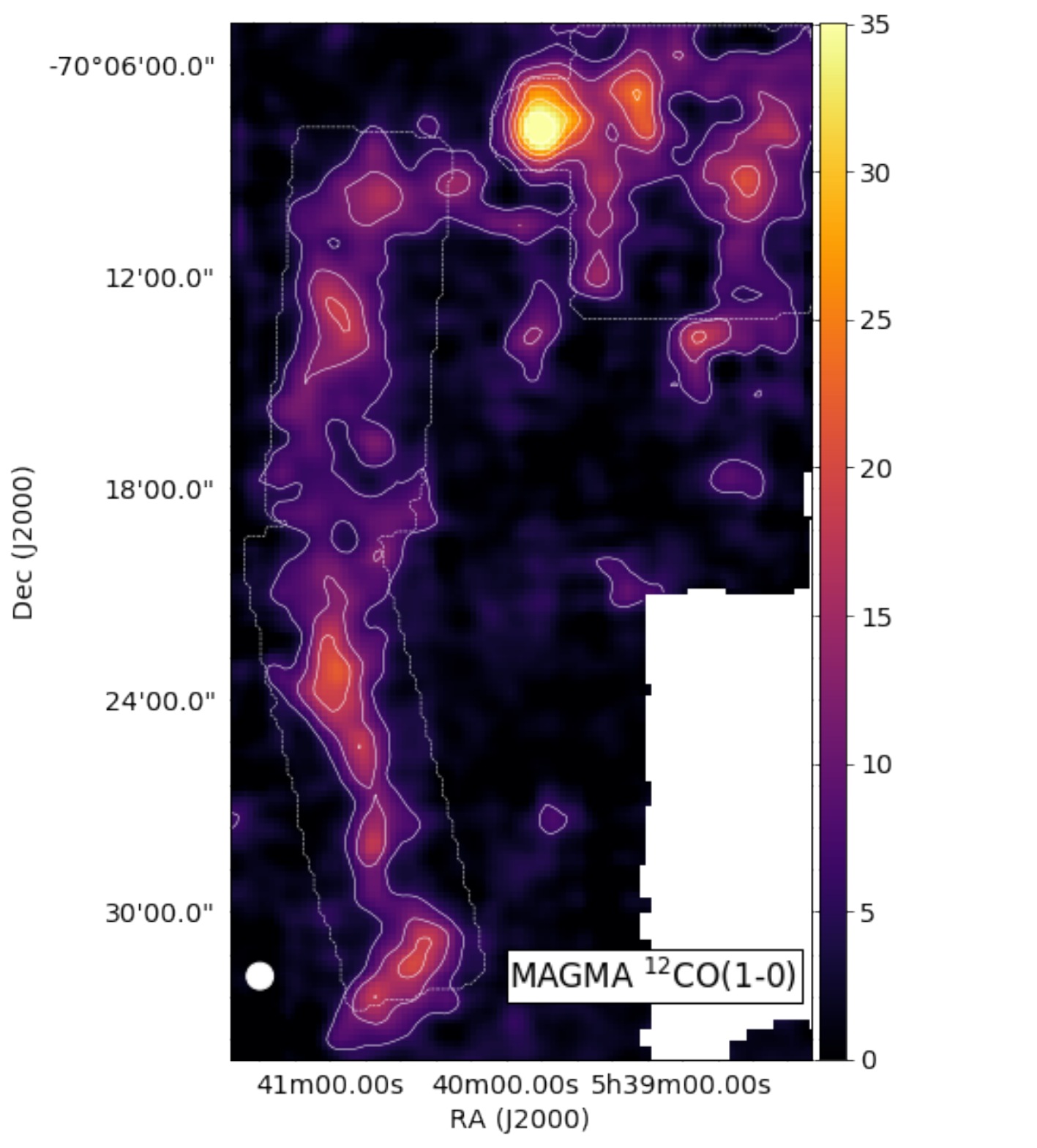}
    \includegraphics[width=0.32\textwidth]{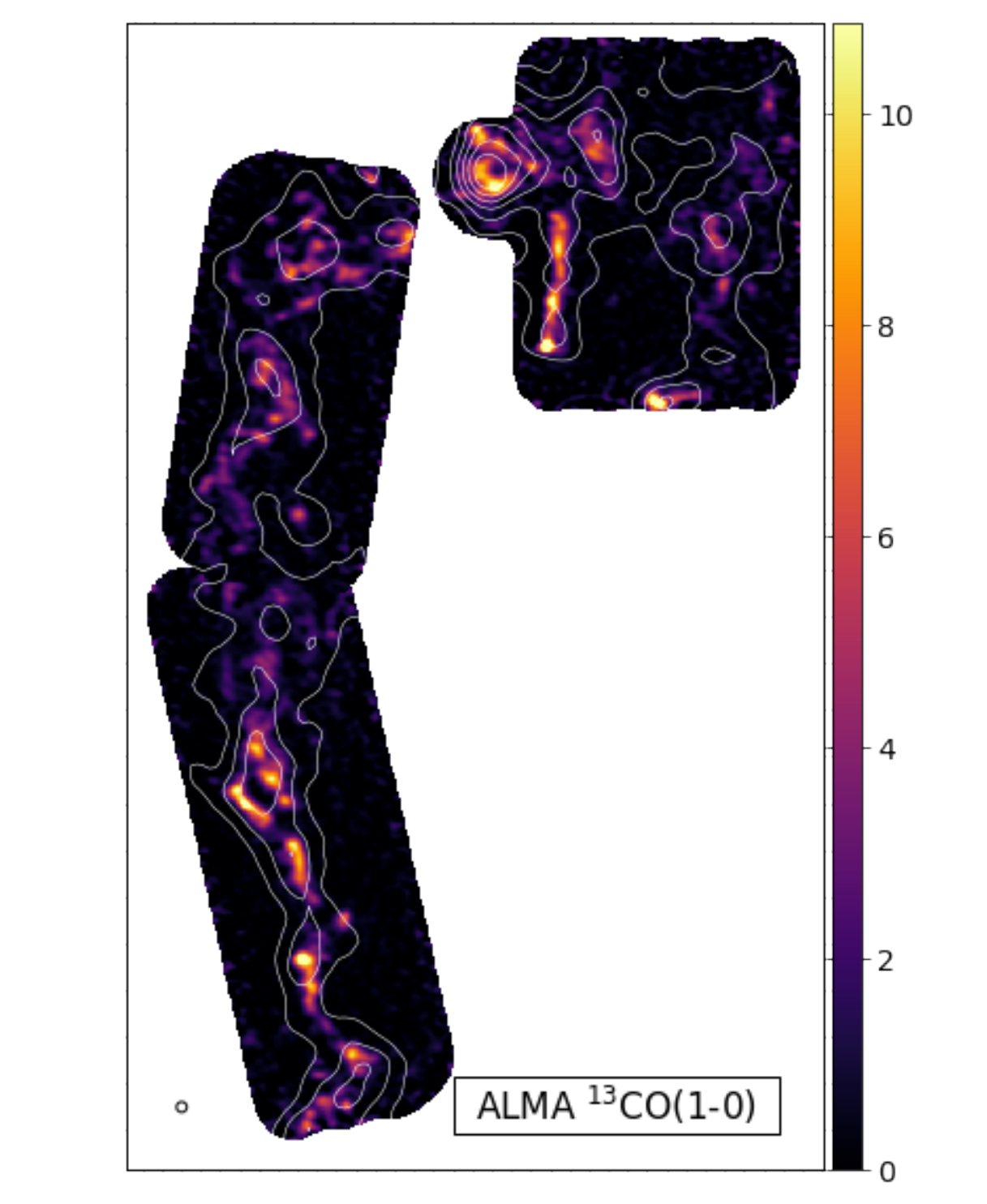}\\
    \includegraphics[width=0.32\textwidth]{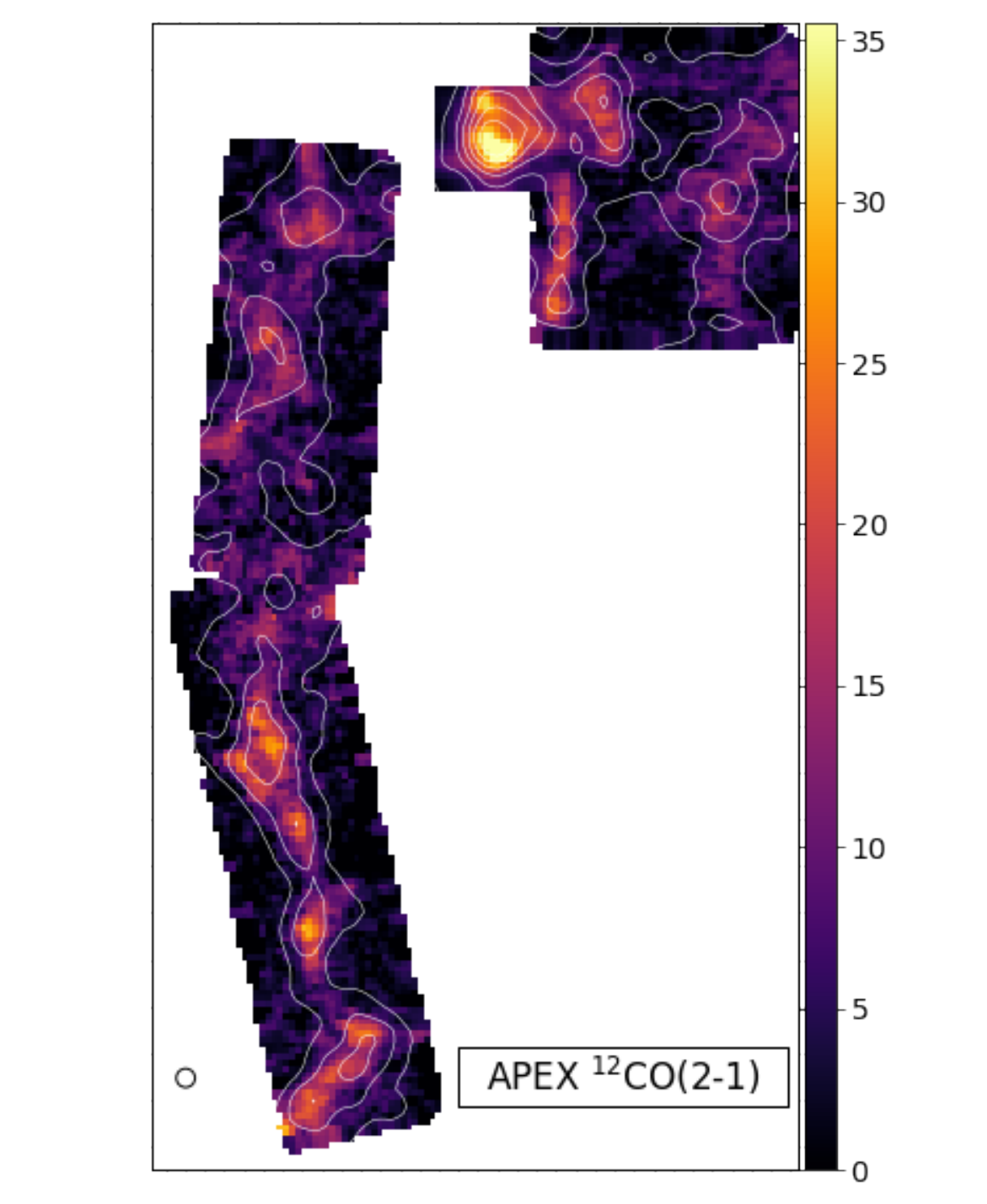}
    \includegraphics[width=0.32\textwidth]{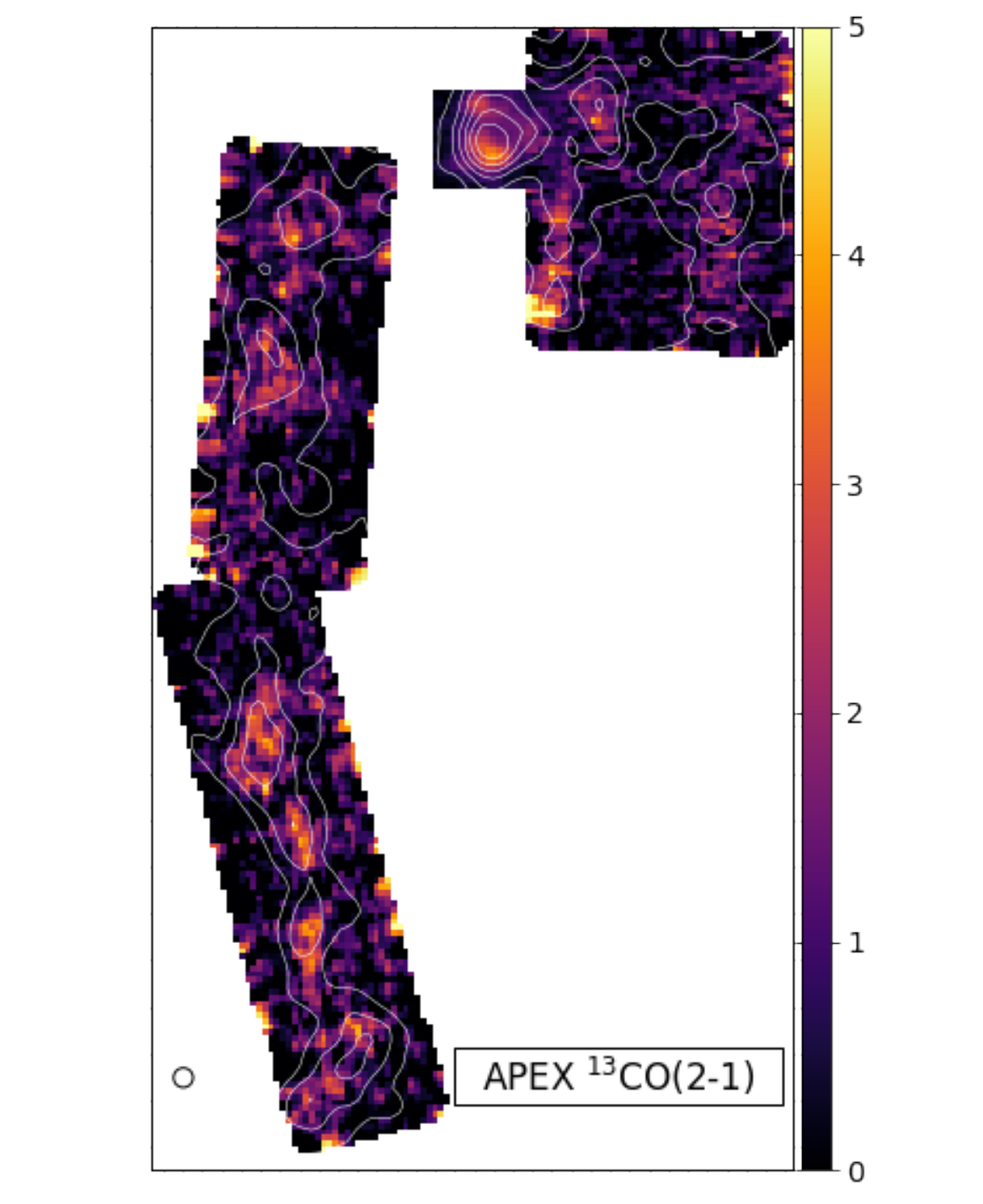}
    \includegraphics[width=0.32\textwidth]{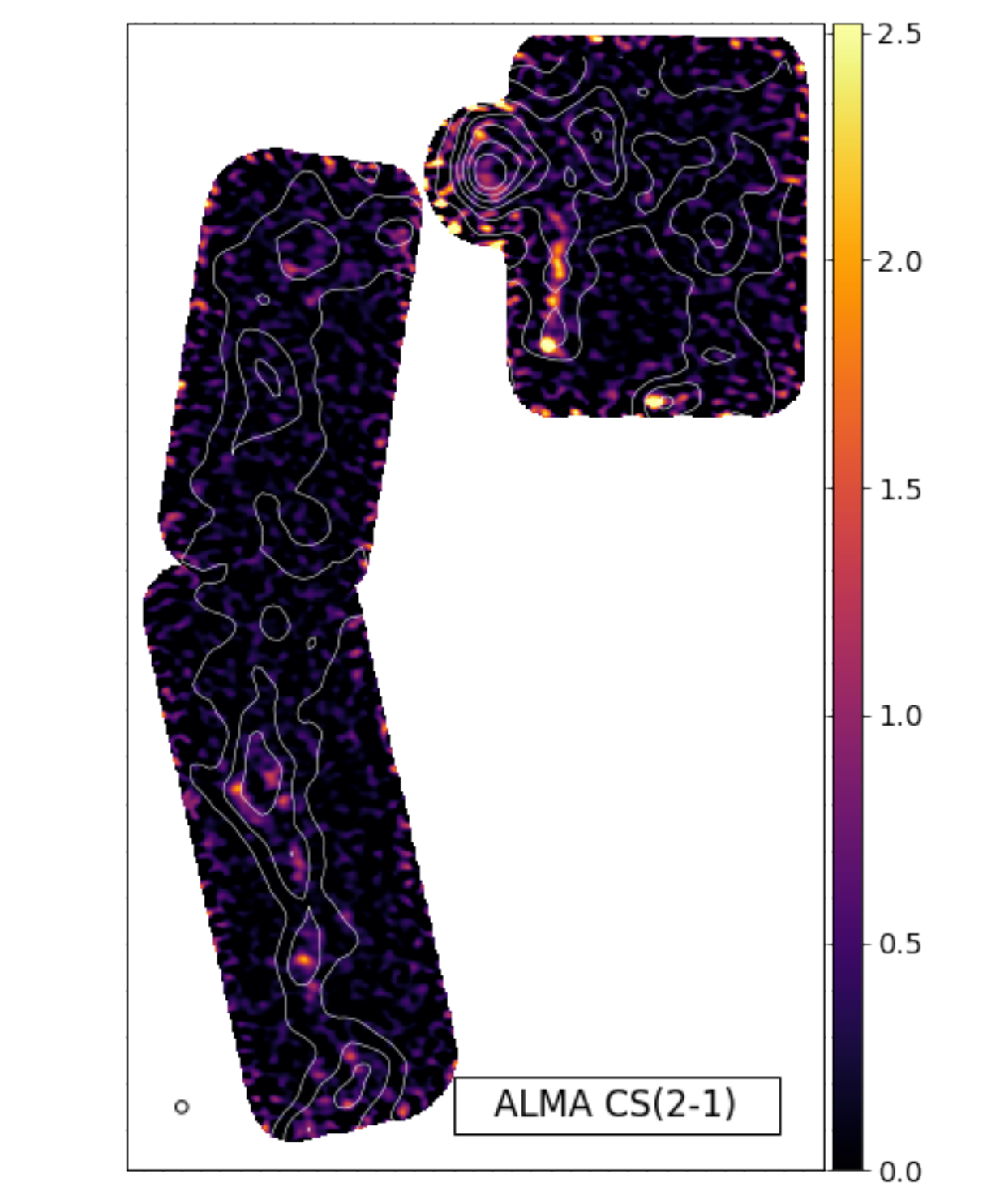}
    \caption{Integrated intensity maps of the observations used in this analysis. The contours are the integrated intensity of \twelveCO(1-0) at intervals of 6, 12, 18, 24, and 30 K km s$^{-1}$. The dotted contour in the MAGMA \twelveCO(1-0) map shows the common observational footprint of all the maps. All maps are in units of  K km s$^{-1}$, and the beams are shown in the lower left corners.}
    \label{fig:moments}
\end{figure*}

In this analysis, we make use of new \thirtCO(1-0) and CS(2-1) data from the Atacama Large Millimeter/submillimeter Array (ALMA) 7m Atacama Compact Array (ACA), described in \S\ref{subsec:alma}. We also use \twelveCO(1-0) data from the Mopra Telescope taken as part of the Magellanic Mopra Assessment (MAGMA) survey \citep{MAGMA}, and new observations of \twelveCO(2-1) and \thirtCO(2-1) from the Atacama Pathfinder Experiment (APEX), described in \S\ref{subsec:apex}. These observations are summarized in Table\,\ref{tab:observations}, and the integrated intensity maps are shown in Figure\,\ref{fig:moments}. The errors reported in Table\,\ref{tab:observations} are the rms noise in line-free regions of the data cubes in a single channel of 1 km s$^{-1}$. Though Table\,\ref{tab:observations} and Figure\,\ref{fig:moments} show the resolutions obtained for each set of observations, the majority of the analysis presented in this paper is performed with all data sets convolved to 45\arcsec\ and 1.0 km s$^{-1}$ velocity resolution to compare among the data sets. The final data cubes used in the analysis are available as supplementary material\footnote{\url{https://doi.org/10.5281/zenodo.4838414}}.

\begin{table}
    \centering
    \caption{Observations used in this analysis}
    \begin{tabular}{c|c|c|c|c}
        \hline
        \hline
         Source & Line & Beam & RMS & Velocity\\
          & & (\arcsec) & (K) & Channel \\
         \hline
         ALMA ACA & \thirtCO(1-0) & 16 & 0.033  & 0.5 km/s \\
         ALMA ACA & CS(2-1) & 18 & 0.025 & 0.5 km/s \\
         ALMA TP & \thirtCO(1-0) & 63 & 0.0078 & 0.19 km/s \\
         ALMA TP & CS(2-1) & 70 & 0.0062 & 0.19 km/s \\
         MAGMA & \twelveCO(1-0) & 45 & 0.11 & 0.5 km/s \\
         APEX & \twelveCO(2-1) & 29 & 0.23 & 1.0 km/s \\
         APEX & \thirtCO(2-1) & 30 & 0.065 & 1.0 km/s \\
    \end{tabular}
    \label{tab:observations}
\end{table}

\subsection{ALMA data} \label{subsec:alma}

Interferometric data were obtained in three maps with the ALMA 7m ACA for project 2017.1.00271.S.
The data contain 3 spectral windows centered on \thirtCO(1-0), C$^{18}$O(1-0), and CS(2-1), each with 2048 61.035~kHz channels (125~MHz bandwidth).  An additional 2~GHz wide spectral window with coarse channels (0.98~MHz) was observed centered on H40$\alpha$ at 99~GHz.  
The north-west 96-pointing map was observed 9 times between 2017-11-07 and 2017-11-15 for a total of 438 minutes on source.  J0522-3627 (5-5.5~Jy) and J0529-7245 (600-700~mJy) were used for bandpass and amplitude, and for phase calibration, respectively.  The central 76-pointing map was observed 9 times between 2017-10-15 and 2017-11-06 for a total of 346 minutes on source, and the southern 106-pointing map 11 times between 2017-10-15 and 2017-11-06 for a total of 470 minutes on source.  Those maps used the same bandpass and amplitude calibrator as the northern.  In a given execution, either  J0635-7516 (1.25~Jy) or J0529-7245 was used for phase calibration. 

The data were calibrated with the ALMA data pipeline Pipeline-CASA51-P2-B, v.40896\footnote{\url{https://almascience.nrao.edu/processing/}} \citep{Davis21}, packaged with CASA 5.1.1-5\footnote{\url{casa.nrao.edu}} \citep{McMullin07}.
The standard pipeline recipe and default parameters were used as described in the ALMA pipeline User's Guide.
Visibilities are calibrated at full spectral resolution. Time-varying gains are solved on the phase calibrator using the 2~GHz wide spectral window, and transferred to the narrow spectral windows using a constant spw-spw phase offset during each 1.5-hour execution block. Gains are transferred to the science target on the scan timescale with linear interpolation in time.  Weights are set correctly by the ALMA correlator and propagated through the calibration process, so no statwt is required. Continuum and line spectral channels are found in each spectral windows by the pipeline task findCont described in the manual.  A linear per-visibility fit is performed and subtracted in the uv domain.  The pipeline images data at full spectral resolution, but we re-imaged the calibrated continuum-subtracted visibilities as described below. 

By design, project 2017.1.00271.S did not cover the $\sim2$ arcminute HII region at 5:39:50 -70:08:00 because it was already observed with ALMA ACA by projects 2012.1.00603.S and 2015.1.00196.S. 
These projects also have three narrow spectral windows centered on \thirtCO(1-0), C$^{18}$O(1-0), and CS(2-1), and a 2~GHz wide spectral window.  The narrow windows have 30.518~kHz and 122.07~kHz channels in 2012.1.00603.S and 2015.1.00196.S, respectively, but all have 125MHz bandwidth like the primary dataset.  The wide window is centered at 96.8~GHz in project 2015.1.00196.S, but that spectral window is not analyzed here. 
Project 2012.1.00603.S was observed 8 times between 2013-12-17 and 2015-04-28, using Ganymede, J0519-4546 (1.3Jy), Uranus, Callisto or Mars as the amplitude calibrator, J0538-4405, J0519-4546, J0635-7516, or J1037-2934 as the bandpass calibrator, and J0635-7516 or J0601-7036 as the phase calibrator. Those data were calibrated manually by ALMA staff, using a script accessible in the ALMA archive. That script solves for time-varying gains on each spectral window individually, and transfers the gains from phase calibrator to science target, but not between spectral windows.

Project 2015.1.00196.S was observed 8 times between 2016-05-01 and 2016-06-12, using J0538-4405 (2.6~Jy) or J1107-4449 (1.3~Jy) for bandpass calibration, J0538-4405 or Uranus for amplitude calibration, and J0529-7245 (700-850~mJy) for phase calibration. The data were calibrated with Pipeline-CASA56-P1-B v.42866 packaged with CASA 5.6.1-8, following the same procedure as the primary dataset 2017.1.00271.S.

There were no detections in C$^{18}$O(1-0) above a 3$\sigma$ upper limit of 200~mJy. 
The \thirtCO(1-0) visibility data from 2012.1.00603.S 
and the CS(2-1) data from 2012.1.00603.S and 2015.1.00196.S were added to the 2017.1.00271.S NW tile data before imaging. 

For all projects, total power ALMA data was obtained for rectangular regions corresponding to the interferometric maps, extended by one primary beam in both dimensions. Project 2012.1.00603.S was observed 4 times between 2013-12-16 and 2014-12-14, and processed with Pipeline-Cycle2-R1-B v.31667 in CASA 4.2.2.  Project 2015.1.00196.S was observed 13 times between 2016-03-23 and 2016-04-08 and processed with Pipeline-Cycle3-R4-B v.36660 in CASA 4.5.3.  Project 2017.1.00271.S NW, central, and southern maps were observed 23 times between 2018-03-30 and 2018-04-23, 25 times between 2018-01-09 and 2018-04-05, and 22 times between 2018-01-24 and 2018-03-21, respectively. All 2017.1.00271.S data were calibrated and imaged with Pipeline-CASA51-P2-B v.40896 packaged with CASA 5.1.1-5. The ALMA single dish pipeline is also described in the User's Manual, and the standard procedure was used: application of system temperature amplitude calibration, subtraction of an OFF position, line detection by clustering analysis, fitting and removal of a polynomial baseline, and a second iteration of line detection and baseline removal. The spectra are then gridded to produce image cubes at native spectral resolution, with beams and rms as noted in Table$\;$\ref{tab:observations}.

The interferometric data were imaged and combined with the total power data using CASA 5.6.1-8.  For the line cubes, total power images were Hanning smoothed and used as a starting model for interferometric deconvolution with the {\tt tclean} task. 
The images have a cell size of 2.1$\arcsec\times$2.1$\arcsec\times$0.5~km$\,$s$^{-1}$, were cleaned to a 1 $\sigma$ threshold (0.6 Jy/bm for \thirtCO(1-0) and 0.3 Jy/bm for CS(2-1)) using the mosaic gridder, hogbom deconvolver, briggs weighting with robust$=0.5$, and auto-multithresh masking using the pipeline default automasking parameters. 
Use of the total power starting model increases the signal-to-noise on the ``overlap'' spatial scales to which the interferometric and total power are both sensitive, but can overestimate the total flux density after nonlinear deconvolution. To correct the flux, the final deconvolved image is combined with the total power image (multiplied by the interferometric sensitivity map) using the {\tt feather} task, which adds the two images in the Fourier domain and ensures the correct total flux density on all spatial scales. The interferometer recovered 40\% of the total flux across the region, with individual clouds recovering between 15 and 99\%. The final combined image is then divided by the interferometric sensitivity map to obtain the correct flux scale as a function of position.  
We then convolved the three regions to a common circular beam of 16.0\arcsec\ and 18.2\arcsec\ for \thirtCO(1-0) and CS(2-1), respectively, and mosaiced them into a single map, linearly weighted by each tile's sensitivity image.

\subsection{APEX data} \label{subsec:apex}

\twelveCO(2-1) and \thirtCO(2-1) were observed with the Atacama Pathfinder Experiment 12m telescope, APEX, between August 14 and 24, 2017 under project number 0100.F-9313(A).  The observations were taken with the APEX-1 receiver, resulting in a beam size of 27.8\arcsec--29.0\arcsec.  Three maps were obtained, corresponding to the three ALMA maps, using on-the-fly (OTF) mapping. Standard calibration was performed using R-Dor, Venus, RAFGL1235, and 07454-7112.  Data reduction was carried out using GILDAS/CLASS; to increase the signal-to-noise ratios in individual channels, contiguous channels were smoothed to a velocity resolution of 1.0 km~s$^{-1}$ and then baseline subtracted, resulting rms $\sim$0.24 and 0.09~K for \twelveCO\ and \thirtCO, respectively. The APEX data cubes were gridded to 9\arcsec$\times$9\arcsec\ ($\sim$3 pixels per beam) to facilitate comparisons with the ALMA data cubes. As with the interferometric ALMA data, the 2 arcmin region at 5:39:50 -70:08:00 was not observed with APEX, but instead, the archival ALMA total power data for \twelveCO(2-1) and \thirtCO(2-1) from projects 2012.1.00603.S and 2015.1.00196.S were added to our APEX mosaic.  
The APEX and ALMA images were combined after convolving the images to a common beam size and gridding as an average weighted by each image's sensitivity map.

\section{Radex Fitting} \label{sec: Radex Fitting}

\subsection{Fitting Method} \label{subsec: fitting method}

\begin{figure}
    \centering
    \includegraphics[width=0.5\textwidth]{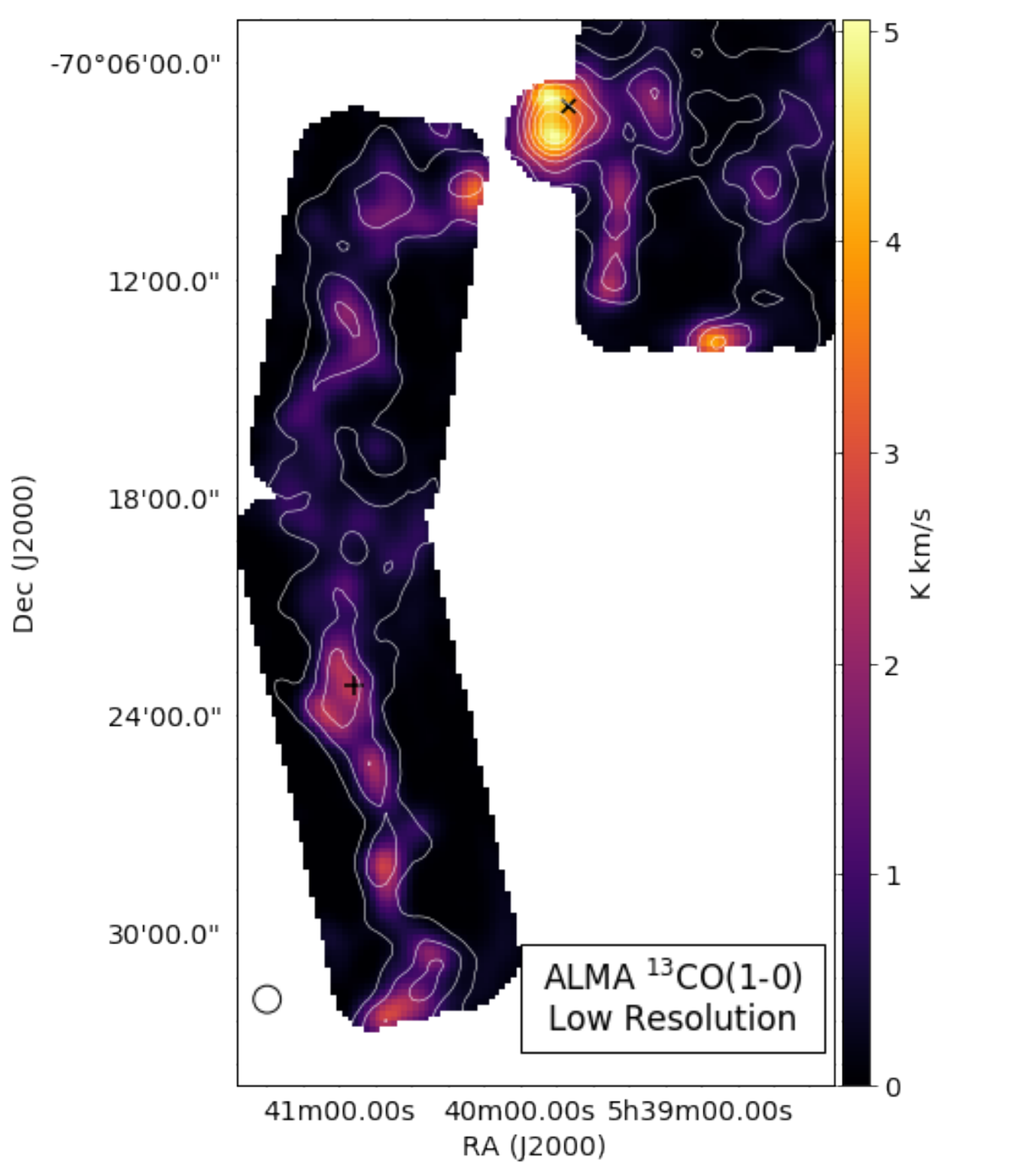}
    \caption{Integrated intensity map of \thirtCO(1-0) convolved to a resolution of 45\arcsec\ (beam shown in lower left corner) to match the limiting resolution of the \twelveCO(1-0) map. The contours are the integrated intensity of \twelveCO(1-0) as shown in Figure\,\ref{fig:moments}, and locations of the example spectra shown in Figure\,\ref{fig:example spectra} are marked with a cross for panel (a) and an X for panel (b). The majority of the analysis in this paper is performed at this lowered resolution, including the \radex fitting.}
    \label{fig:lowres 13CO}
\end{figure}

\begin{figure}
    \centering
    \includegraphics[width=0.5\textwidth]{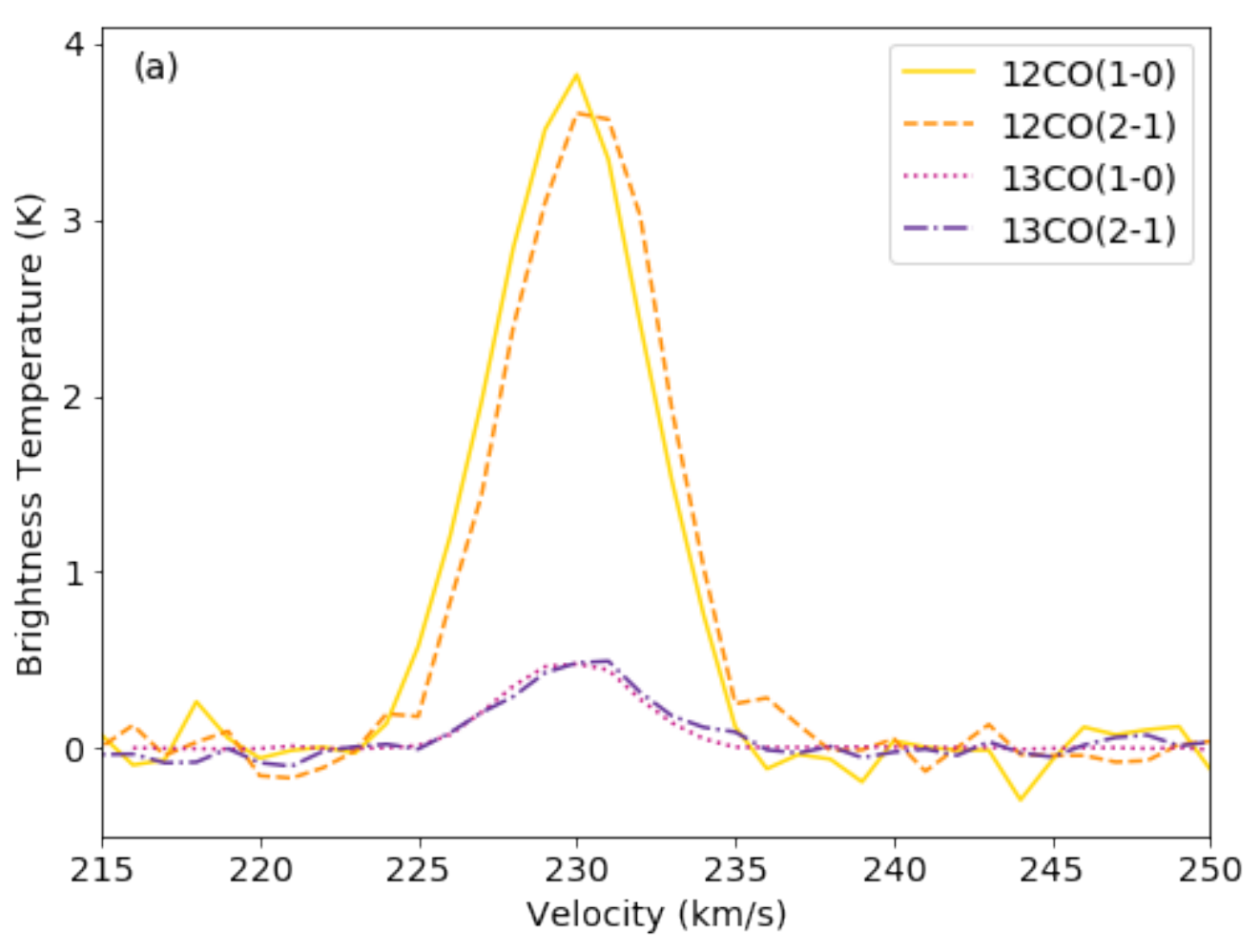}
    \includegraphics[width=0.5\textwidth]{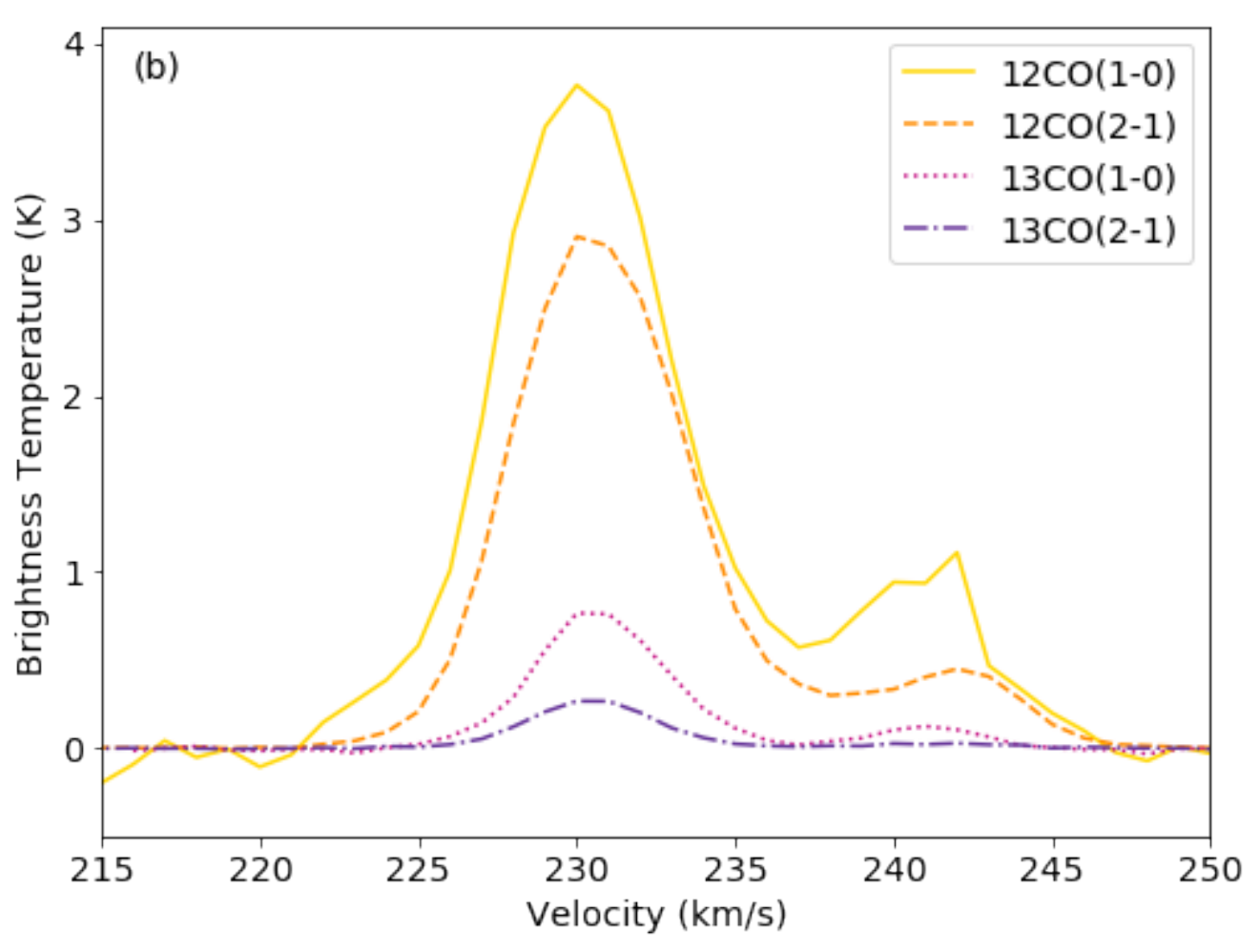}
    \caption{Example spectra of \twelveCO(1-0), \twelveCO(2-1), \thirtCO(1-0), and \thirtCO(2-1) from two different peaks. These are examples of spectra that are used in the \radex fitting, and so are taken from maps that have all been convolved to a beam size of 45\arcsec. The top panel shows a spectra that is typical throughout the region, while the bottom panel shows an example with more velocity structure from the northern region of the map. The locations of these two spectra are shown in the \thirtCO(1-0) map in Figure\,\ref{fig:lowres 13CO}.}
    \label{fig:example spectra}
\end{figure}

To determine physical parameters from the observed \twelveCO\ and \thirtCO\ emission lines, we compared the line intensities at each pixel and velocity to model intensities for a range of physical parameters from the non-LTE escape probability code \radex \citep{radex}. This was done by computing a three-dimensional grid of \radex models for a range of kinetic temperatures (\tkin), H$_2$ volume densities (\nh), and \twelveCO\ column densities (\NCO). The four emission cubes--\twelveCO(1-0), \thirtCO(1-0), \twelveCO(2-1), and \thirtCO(2-1)--were all convolved to 45\arcsec\ and 1 km s$^{-1}$ to match the lowest common resolutions among the data sets. The errors used in calculating probabilities are the rms errors in these newly convolved maps, measured in emission-free slices of the cubes. The lower-resolution errors for \twelveCO(1-0), \thirtCO(1-0), \twelveCO(2-1), and \thirtCO(2-1) are 0.11 K, 0.017 K, 0.1 K, and 0.035 K, respectively. We show the the \thirtCO(1-0) map at this lowered resolution in Figure\,\ref{fig:lowres 13CO} and example input spectra for the lines  in Figure\,\ref{fig:example spectra}. 

When computing \radex models, we used a homogeneous spherical escape probability geometry with a line width of 1 km s$^{-1}$ to match our observations' velocity channels, and a background temperature of 2.73 K. We also assume that the ratio of \twelveCO\ to \thirtCO\ ($R_{13}$) is in the range 50--100  \citep{Nikolic07} and that $N_\text{\twelveCO}$/$N_\text{\thirtCO} = R_{13}$.

We limited the ranges of the parameters to be \tkin between 2 and 200 K, \nh between $10^{1.5}$ and $10^7$ cm$^{-3}$, and \NCO between $10^{14}$ and $10^{18}$ cm$^{-2}$. The ranges of \nh and \NCO are evenly spaced in log space, \tkin is spaced linearly. When computing the \radex grid, we also excluded regions of the parameter space where \radex predictions are less reliable, such as where the optical depth gets very large ($\tau>300$), where \twelveCO\ becomes overpopulated and the excitation gets inflated \citep[{$T_{ex} > 2.5\times$\tkin,}][]{KoeppenKegel80}, where any output values become unphysically negative, and where \radex took more than 999,000 iterations to solve.

For each combination of the three parameters, $\Vec{p}$ = (\tkin, \nh, \NCO), the resultant model brightness temperatures from \radex, $R(\Vec{p})$, and the beam filling factor, $f$, were used to compute a probability given the observed brightness temperature for some voxel, $I$, and its error, $\delta$, for each observed line, $j$, using the equation

\begin{equation}
    P(\Vec{p}|I) = \prod_j \frac{-1}{\delta_j} \exp{\left[ \frac{1}{2} \left( \frac{(I_j - R(\Vec{p})_j\times f)}{\delta_j}\right)^2\right]}
\end{equation}

We find the combination of parameter values that yields the greatest probability, $\Vec{p}_\text{max}$. We then find the odds ratio for all other parameter combinations in the grid:

\begin{equation}
    O = \frac{P(\Vec{p}_\text{max}|I)}{P(\Vec{p}|I)}
\end{equation}
 
In the case of uniform priors (so $P(\Vec{p}_\text{max}) = P(\Vec{p})$, which we assume here for all parameters), this reduces to the Bayes Factor:

\begin{equation}
    B = \frac{P(I|\Vec{p}_\text{max})}{P(I|\Vec{p})}
\end{equation}

To compare $\Vec{p}_\text{max}$ with all other combinations of parameters, we use the ``Jeffreys'' scale \citep{Trotta08} to determine if $p_\text{max}$ is ``inconclusively'', ``weakly'', ``moderately'', or ``strongly'' preferred to the other parameter combinations. \cite{Trotta08} defines this empirically derived scale as follows: a value of $|\ln{B}| < 1$ corresponds to inconclusive evidence, $1 \leq |\ln{B}| < 2.5$ is weak evidence,  $2.5 \leq |\ln{B}| < 5$ is moderate evidence, and $\mathbf{|\ln{B}| \geq 5}$ is strong evidence. The value of $|\ln{B}|$ is zero for $\Vec{p}_\text{max}$, and increases for parameter combinations that have lower probabilities of matching the observed intensities. 

After excluding the regions of parameter space for which $|\ln{B}| \geq 5.0$ ($\Vec{p}_\text{max}$ is strongly preferred over all the excluded parameter combinations) we determine the ranges of the remaining parameter space for each parameter to obtain what we call here ``Bayesian intervals''. We do the same to get intervals excluding parameter combinations with $|\ln{B}| \geq 2.5$ and $|\ln{B}| \geq 1.0$ to get intervals outside of which $\Vec{p}_\text{max}$  is ``moderately'' and ``weakly'' preferred, respectively. In this case, the ``strong'' 5.0 Bayesian interval is the largest and least constrained of the three since parameter combinations that $\Vec{p}_\text{max}$ is only moderately or weakly preferred over are included within the interval. The ``weak'' 1.0 Bayesian interval is the narrowest and most constrained, since $\Vec{p}_\text{max}$ only needs to be weakly preferred over a parameter combination for it to be excluded.

The Bayesian intervals and $\Vec{p}_\text{max}$ all depend entirely on the 3-dimensional probability density function (PDF). In addition to these intervals, we also consider each individual parameter's probability density profile, integrated over the other two parameters. From these profiles, we determine one-sigma and two-sigma, 67\% and 95\%, confidence intervals, defined as the smallest ranges of the parameters for which the sum under their normalized probability profiles is 0.67 and 0.95, respectively. The confidence intervals depend only on the integrated 1-dimensional profiles instead of the 3-dimensional PDF and therefore depend on the spacing of intervals used in the parameter ranges. Since the \nh and \NCO ranges span several orders of magnitude, the confidence intervals are determined in log space so as not to overly weight the higher values. The confidence interval of \tkin is calculated with linear spacing.

\begin{figure*}
    \centering
    \includegraphics[width=0.75\textwidth]{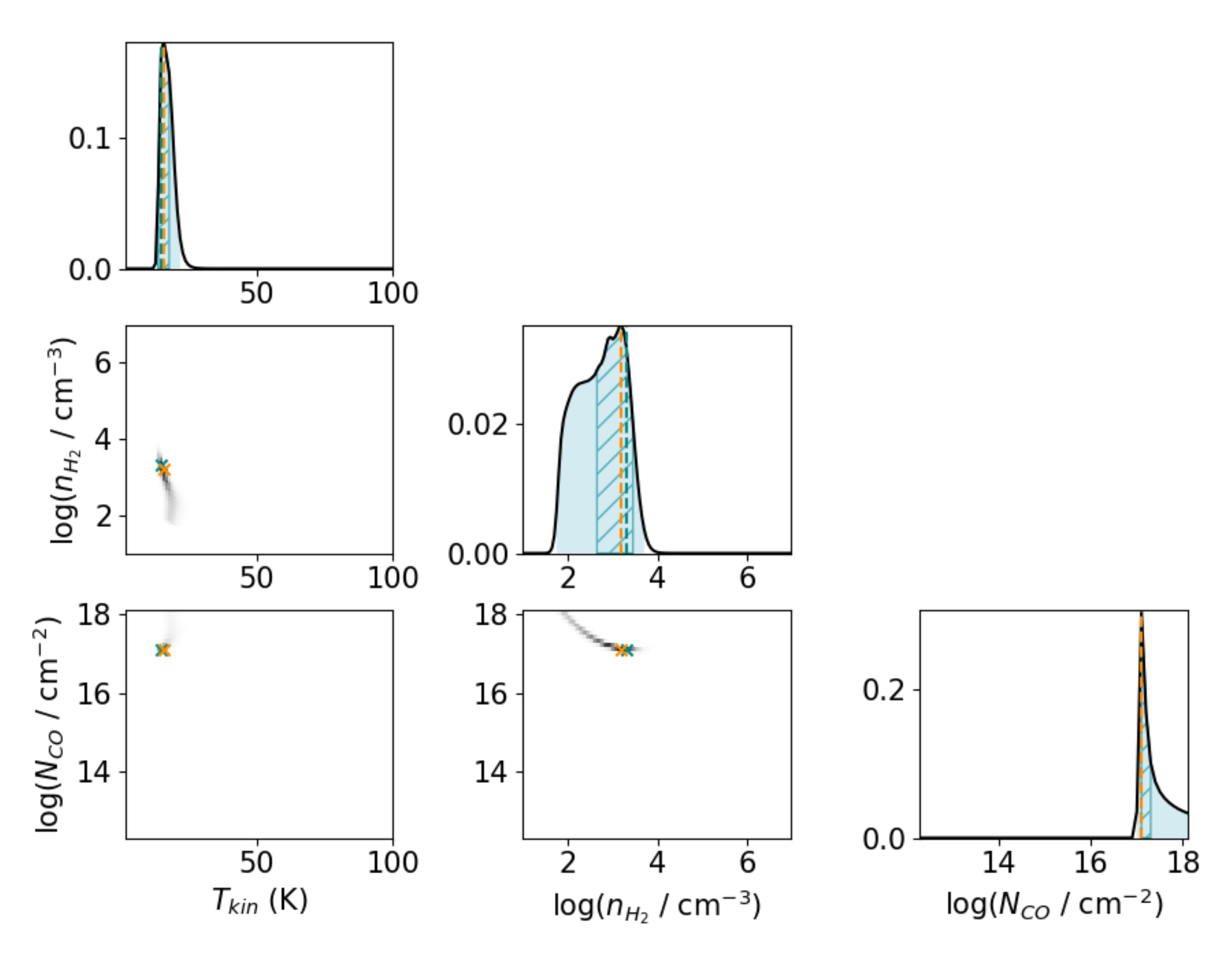}
    \caption{Example of a probability distribution from representative data with line intensities of 1.7 K, 0.2 K, 1.2 K, and 0.14 K for \twelveCO(1-0), \thirtCO(1-0), \twelveCO(2-1), and \thirtCO(2-1), respectively, using a 15\% beam filling factor, and $R_{13}=100$. The ranges on the axes show the full tested parameter space. The blue shading shows the 2.5 Bayesian intervals (``moderate'' evidence), and the blue hatching shows the 1.0 Bayesian intervals (``weak'' evidence). The vertical blue dashed lines and blue crosses indicate $p_\text{max}$, and the orange dashed lines and orange crosses indicate the maximum value of the probability profiles. This plot indicates that this pixel most likely has \tkin= 15 K, \nh= 10$^{3.3}$ cm$^{-3}$, and \NCO= 10$^{17.1}$ cm$^{-2}$.} 
    \label{fig:examplecorner}
\end{figure*}

An example corner plot showing a resultant distribution for one pixel of data is shown in Figure\,\ref{fig:examplecorner}. The profiles along the diagonal show some of the metrics described above: $\Vec{p}_\text{max}$, the collapsed 1-dimensional probability profiles, and the two smallest Bayesian intervals.

We attempted to independently constrain $R_{13}$ in the same way as the other parameters, adding it as a fourth dimension to the tested parameter space. However, $R_{13}$ was rarely constrained and including it significantly increased the computational requirements. We therefore assumed a range of $R_{13}$ = 50--100 instead \citep{Nikolic07} and performed the fitting once with $R_{13}$ = 50 and once with $R_{13}$ = 100. With additional observations of higher J \twelveCO\ and \thirtCO\ lines, we might be able to constrain $R_{13}$ on a pixel-by-pixel basis, making it worth the additional computational requirements. 

We similarly attempted to include fitting the beam filling factor as a fourth parameter with minimal success. Appendix\,\ref{append: bff} goes into detail about the various attempts at fitting and measuring the beam filling factor. After examining the results of these attempts, we used a range of filling factors from 10\%--20\%. The lower limit of this range comes from unphysical fitting solutions (primarily defined as excessively large line-of-sight path lengths) and the upper limit comes from measured upper limits when comparing the high resolution \thirtCO(1-0) observations at 13\arcsec\ to the low resolution maps at 45\arcsec. The exception to this is the $\sim2$ arcminute region at 5:39:50 -70:08:00, where we found a lower limit on the filling factor to be 15\% rather than 10\%. A filling factor of 10\% results in line-of-sight path lengths that are $>100$ pc while the radius of the clouds in the region are measured to be $\sim 20$ pc. We do not take into account any potential difference in beam filling factors between different lines, despite them likely having different spatial distributions.

Using simulated data from the full range of the parameter space based on expected emission from \radex and the measured rms error in the observed maps, we evaluated how well our data can be fit by the process described here. We also evaluated which of the Bayesian and confidence intervals best recovered true parameter values while still constraining their values. The full evaluation process is described in Appendix\,\ref{append: simulated data} and additional related plots are available as supplementary material\footnote{\url{https://doi.org/10.5281/zenodo.4646288}}. 
The true parameter values were almost always recovered at \NCO $> 10^{15}$ cm$^{-2}$ and for the full range of \tkin and \nh. The intervals that were determined to best characterize the true parameter values were a combination of the 95\% confidence interval and the 1.0 Bayesian interval. They showed similarly high recovery rates of the true parameter values and had the tightest constraints on those values. In some regions the 95\% confidence interval was better constrained than the 1.0 Bayesian interval, and vice versa, hence the combination of the two.

We also consider how the fitting process depends on including all four observed lines--\twelveCO(1-0), \thirtCO(1-0), \twelveCO(2-1), \thirtCO(2-1)--and how it would change if we included only the three lines with the best angular resolution and dropped \twelveCO(1-0), which has a resolution of 45\arcsec. This would allow us to do the entire fitting process at higher resolution since we would instead be limited by the \thirtCO(2-1) at 30\arcsec. We consider this case in Appendix\,\ref{append: 3 line fitting} using the fitting evaluation methods described in Appendix\,\ref{append: simulated data}. We find that dropping the \twelveCO(1-0) results in a loss of sensitivity to moderate values of \NCO. While the resulting fitted intervals still include the correct value almost all the time for \NCO $> 10^{15}$ cm$^{-2}$, they are only well-constrained for \NCO $> 10^{16}$ cm$^{-2}$. We decided that the improvement in resolution is not worth the loss in sensitivity to this range of column densities.

\subsection{Generating Maps of Physical Parameters} \label{subsec:getting maps}

\begin{figure*}
    \centering
    \includegraphics[width=0.33\textwidth]{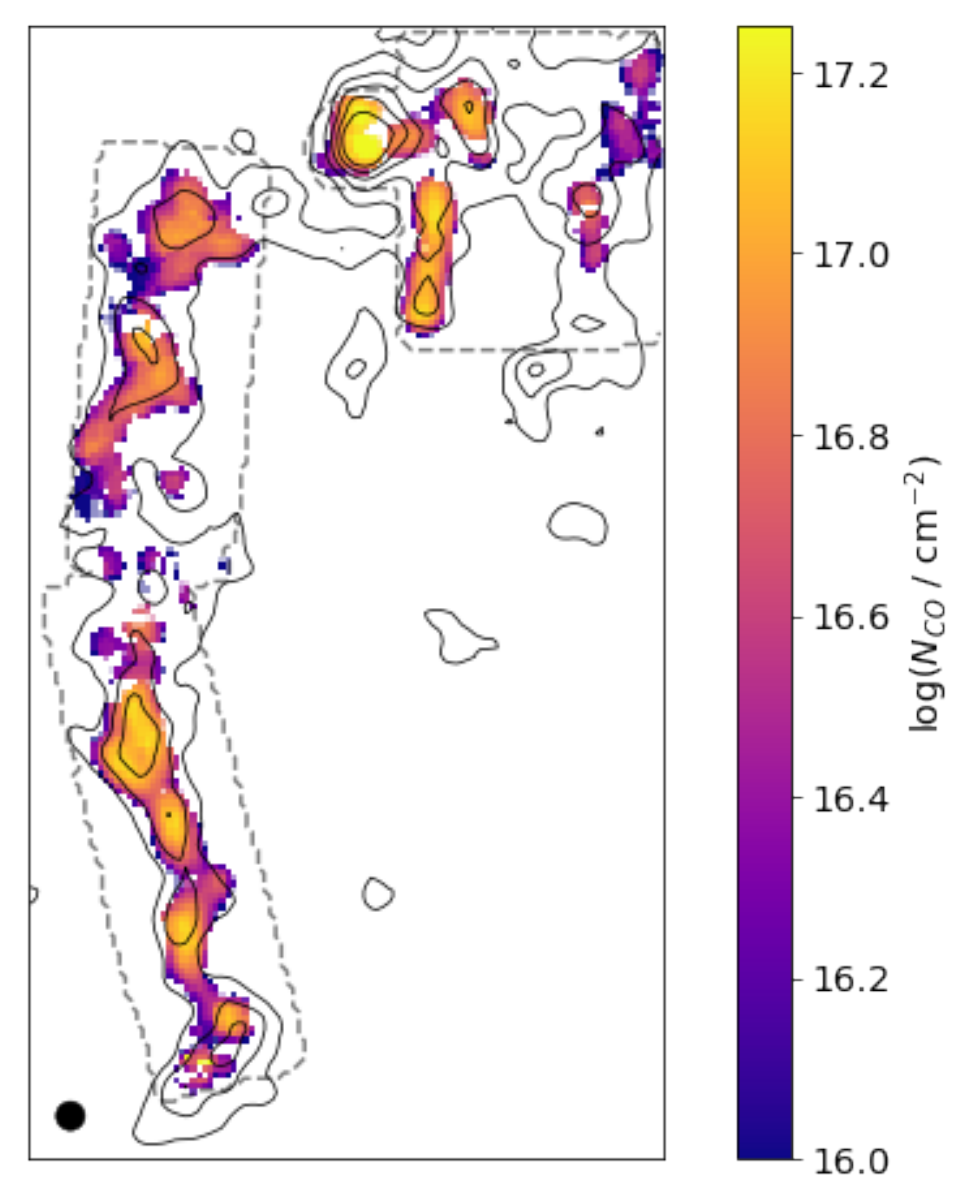}
    \includegraphics[width=0.32\textwidth]{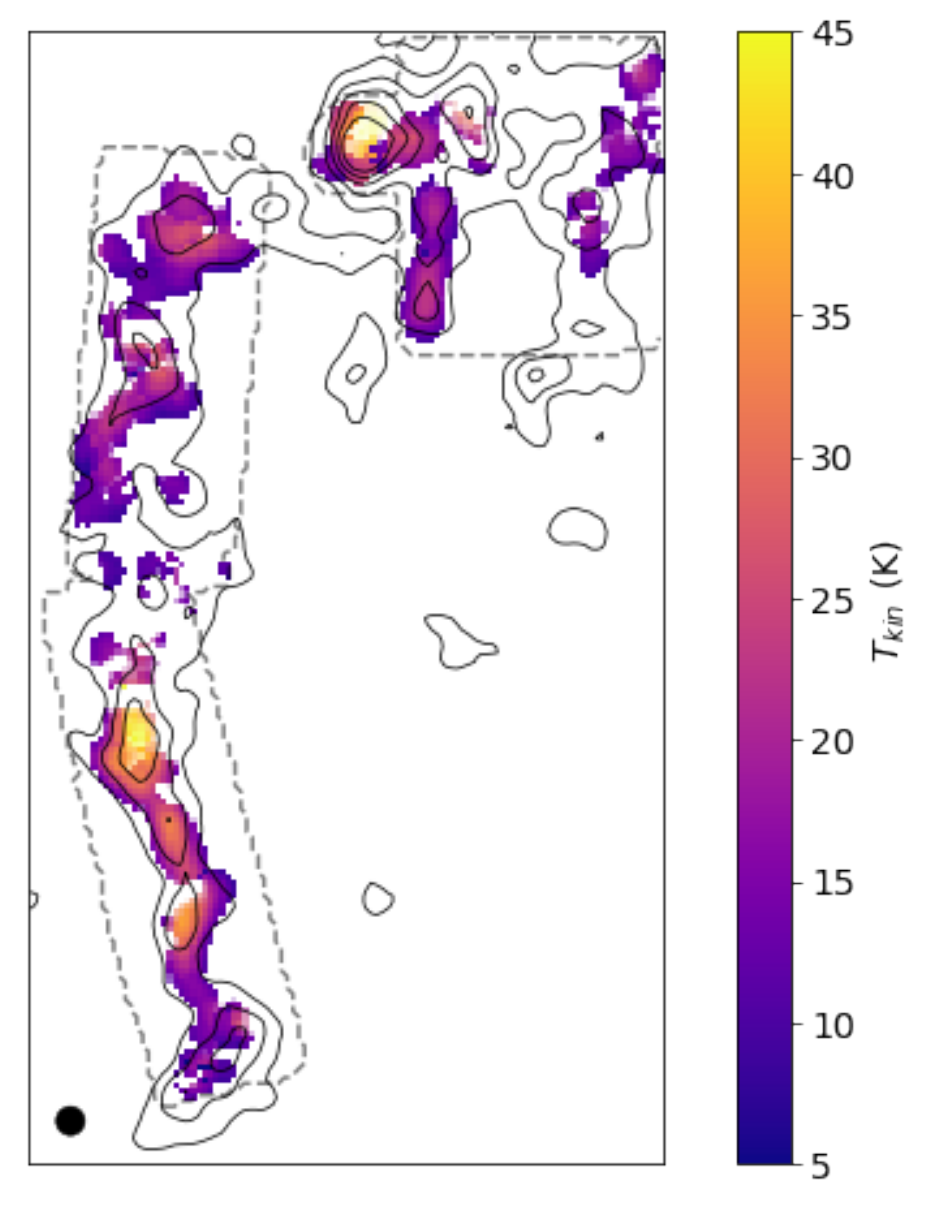}
    \includegraphics[width=0.33\textwidth]{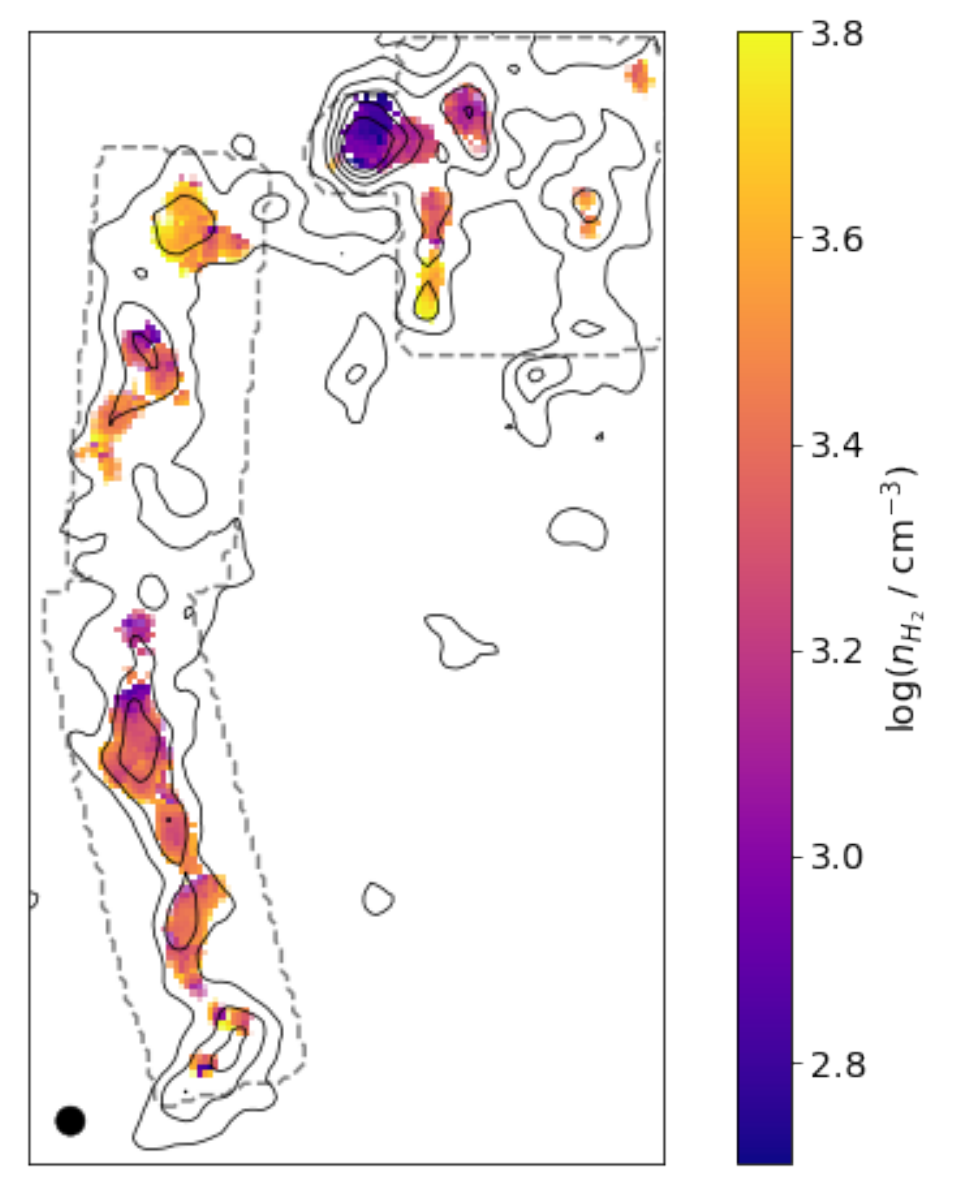}
    \includegraphics[width=0.327\textwidth]{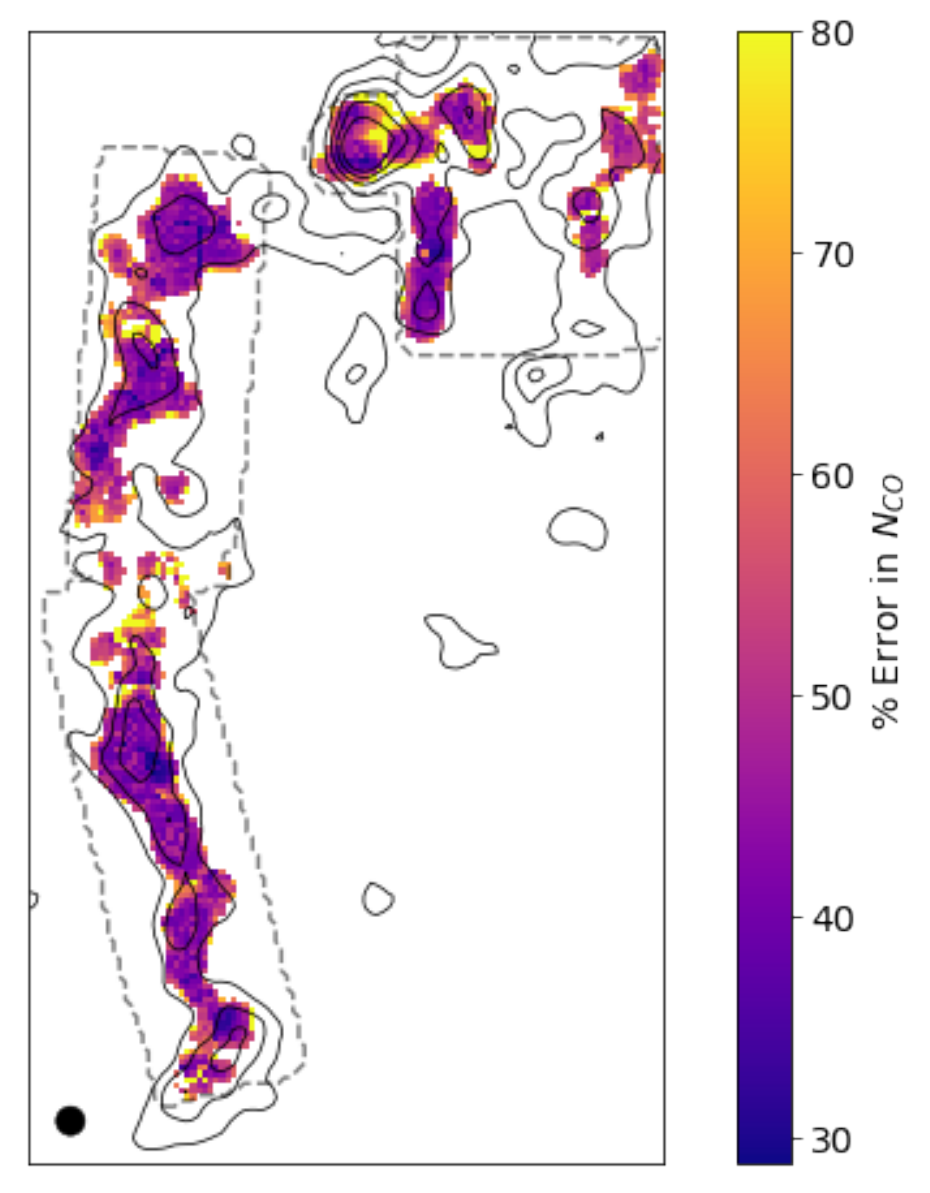}
    \includegraphics[width=0.327\textwidth]{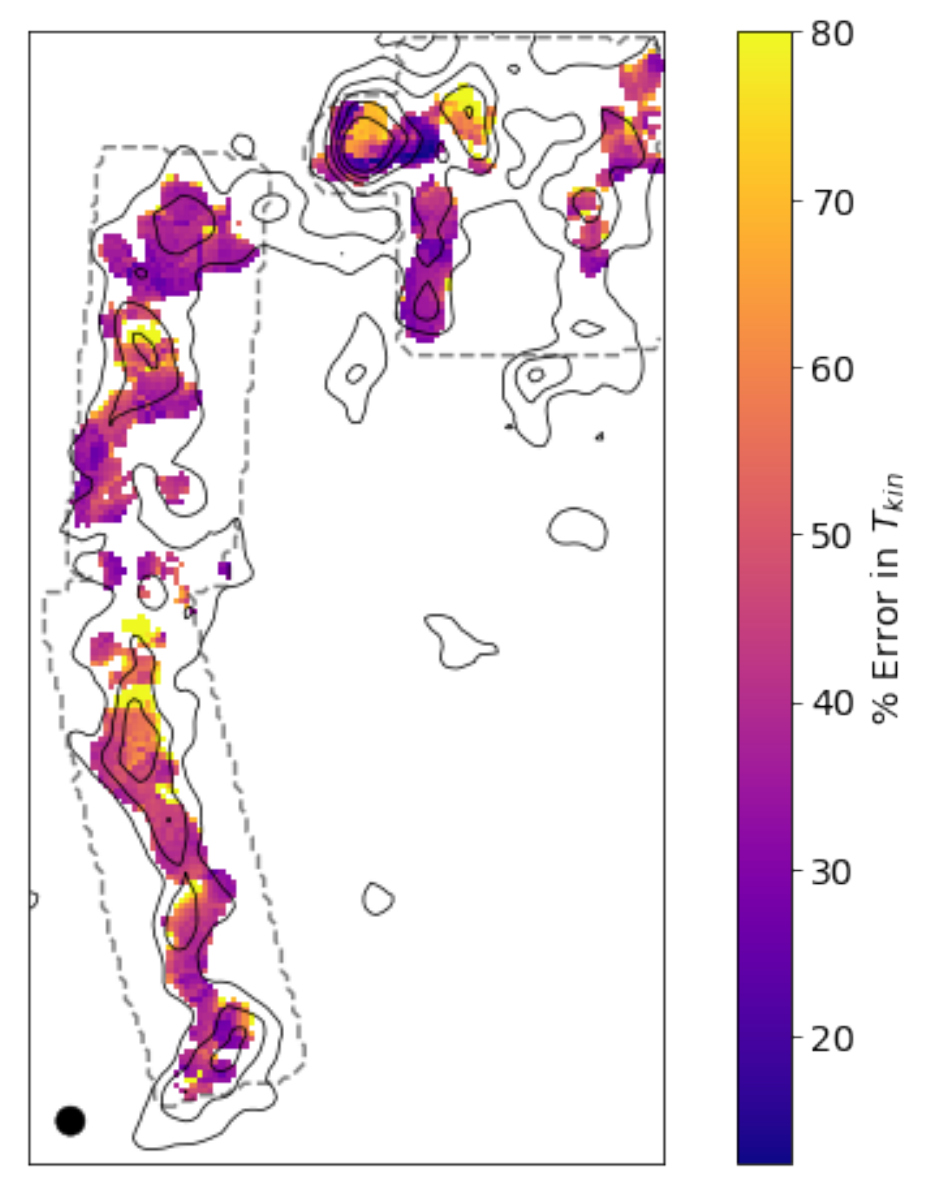}
    \includegraphics[width=0.327\textwidth]{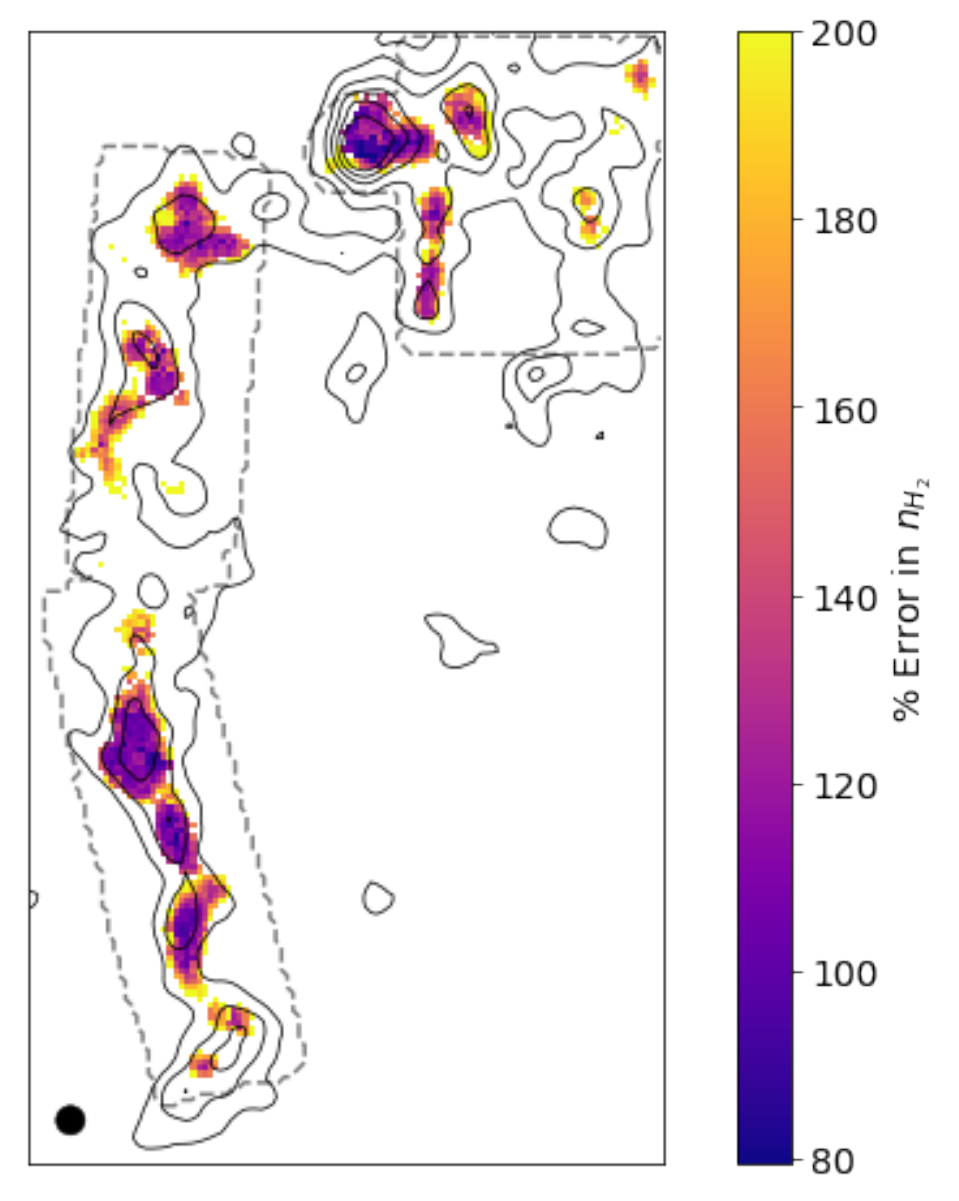}
    \caption{\radex-fitted maps of \NCO, \tkin, and \nh (top row, left to right) and the percent error in \NCO, \tkin, and \nh (bottom row, left to right). \NCO was masked where the error was more than 80\%, \tkin where the error was more than 50\%, and \nh where the error was more than 200\%.The values shown have been averaged between the values fit with $R_{13}=50$ and $R_{13}=100$ and filling factors of 10\% and 20\%, which can result in errors larger than the described cutoffs. The pixel transparency in the maps of the fitted parameters (top row) is proportional to the error in those parameters (bottom row). All \NCO values are corrected for the assumed beam filling factor, but the \nh values are those fitted to the structures at the filling factor scale (4.5\arcsec\ or 9\arcsec\ for filling factors of 10\% and 20\%, respectively). The solid contours are the integrated intensity of MAGMA \twelveCO(1-0), as shown in Figure\,\ref{fig:moments}, and the dashed contours show the common observational footprint. The 45\arcsec\ beam is shown in the bottom left corner.}
    \label{fig:Ridge full maps}
\end{figure*}

We perform the \radex fitting on each of the pixels in the map. For each velocity component within each pixel, we find the peak of the \thirtCO(1-0) line profile for a single velocity component and perform a full fit of the three physical parameters---\tkin, \nh, \NCO---for the peak value. We then assume that \tkin and \nh are constant for that line of sight and fit the rest of the line only considering the \tkin and \nh that were in the fitted interval of the line peak to get \NCO across the line. We include only velocities for which at least two lines have detections above 5$\sigma$. \NCO is then summed over the line to get a total value for the pixel. Upper and lower errors come from the upper and lower bounds of the fitted interval for each parameter, and for \NCO, the upper errors and lower errors are propagated separately to get upper and lower errors on the total \NCO for the line. 

We also considered two other methods for this fitting process: fitting all three parameters for each voxel along the line of sight or holding \tkin and \nh fixed for the entire cloud after segmenting the emission into clumps. These alternative methods are compared in Appendix\,\ref{append:comparing radex}, where we find that the method described here of holding \tkin and \nh fixed for the line of sight yielded the most reliable realistic results. 

This process was done once with $R_{13}$=50 and once with $R_{13}$=100, and also once each with a filling factor of 10\% and 20\% for a total of four runs, with the exception of the region around 5:39:50 -70:08:00, for which we used a lower limit on the filling factor of 15\% rather than 10\% as described in Appendix\,\ref{append: bff}. To correct our final results for this filling factor, we multiplied the fitted \NCO by the assumed filling factor for the clump. We did not correct the \nh values for the filling factor, so the values reported are those of clump structures on the scale of the assumed filling factor (4.5\arcsec\, 6.75\arcsec\, or 9\arcsec\ for filling factor of 10\%, 15\%, or 20\%, respectively).

When approaching a boundary between two velocity components, both spatially and in velocity space, we drew a hard barrier rather than doing any partial pixel assignments to account for overlapping line wings or spatial overlap. To check if this affected the fitting results, we did the \radex fitting for three overlapping velocity components, but this time fitting Gaussian line profiles to each pixel to appropriately assign partial emission to the overlapping clumps. The \radex fitting code used this partial emission assignment and continued the fitting as before. This did not result in any significant change in any of the derived quantities, and so fortunately the detailed accounting of multiple velocity components does not need to be added in general to this kind of analysis. This is likely because the line wing that was cut from Component A and assigned to Component B is well accounted for by the line wing of Component B that was assigned to Component A, so the amount of emission is not significantly changed. This result might change if there is a large temperature difference between overlapping components, but that seems unlikely to occur in most scenarios.

The results for each velocity component were combined into maps of the whole Ridge shown in Figure\,\ref{fig:Ridge full maps} by adding \NCO along each line of sight and using a mass-weighted average for \tkin and \nh. We masked fits that were not well-constrained since our results from Appendix\,\ref{append: simulated data} showed that poorly constrained fits often were also not accurate. How well each parameter could be expected to be constrained varied largely, as shown in Appendix\,\ref{append: simulated data}. \tkin and \NCO were both usually tightly constrained, while the fitted \nh is not as well constrained. This appears to be a reflection of how well the data at hand can inform the physical parameters rather than a reflection of how reliable the fitting process is. For both Figure\,\ref{fig:Ridge full maps} and deriving quantities in \S\ref{subsec:derived props}, we masked values of \NCO where the error was more than 80\%, \tkin where the error was more than 50\%, and \nh where the error was more than 200\%. We also masked values of \tkin that were less than 3 K, and values of \nh that rose sharply at the edges of clumps. In the case of the Molecular Ridge, we accomplished this by masking where \nh was greater than 10$^4$ cm$^{-3}$ since values larger than that only occurred in edge pixels, but this would change if the range of fitted \nh had been higher. 

After cutting pixels that had poorly constrained or unphysical fits, we combined fitted values from the runs with different $R_{13}$ and filling factors. The reported values for \tkin, \nh, and \NCO are the mean of the best fit values from the  $R_{13}$=50 and $R_{13}$=100 results and the 10\% and 20\% filling factor results. The upper and lower errors are from the highest and lowest values included in any of the fitted intervals (i.e. if \tkin is $26\pm3$~K for $R_{13}$=50 and $30\pm5$~K for $R_{13}$=100, the reported \tkin is $28_{-5}^{+7}$~K). When reporting a single error, we use the geometric mean of the upper and lower error. This results in the maps shown in Figure\,\ref{fig:Ridge full maps} sometimes having larger errors than the cutoffs described here.

\section{Clump Definitions and Properties} \label{sec: clumps}

\subsection{Clump Definitions} \label{subsec: clump defs}

\begin{figure}
    \centering
    \includegraphics[width=0.5\textwidth]{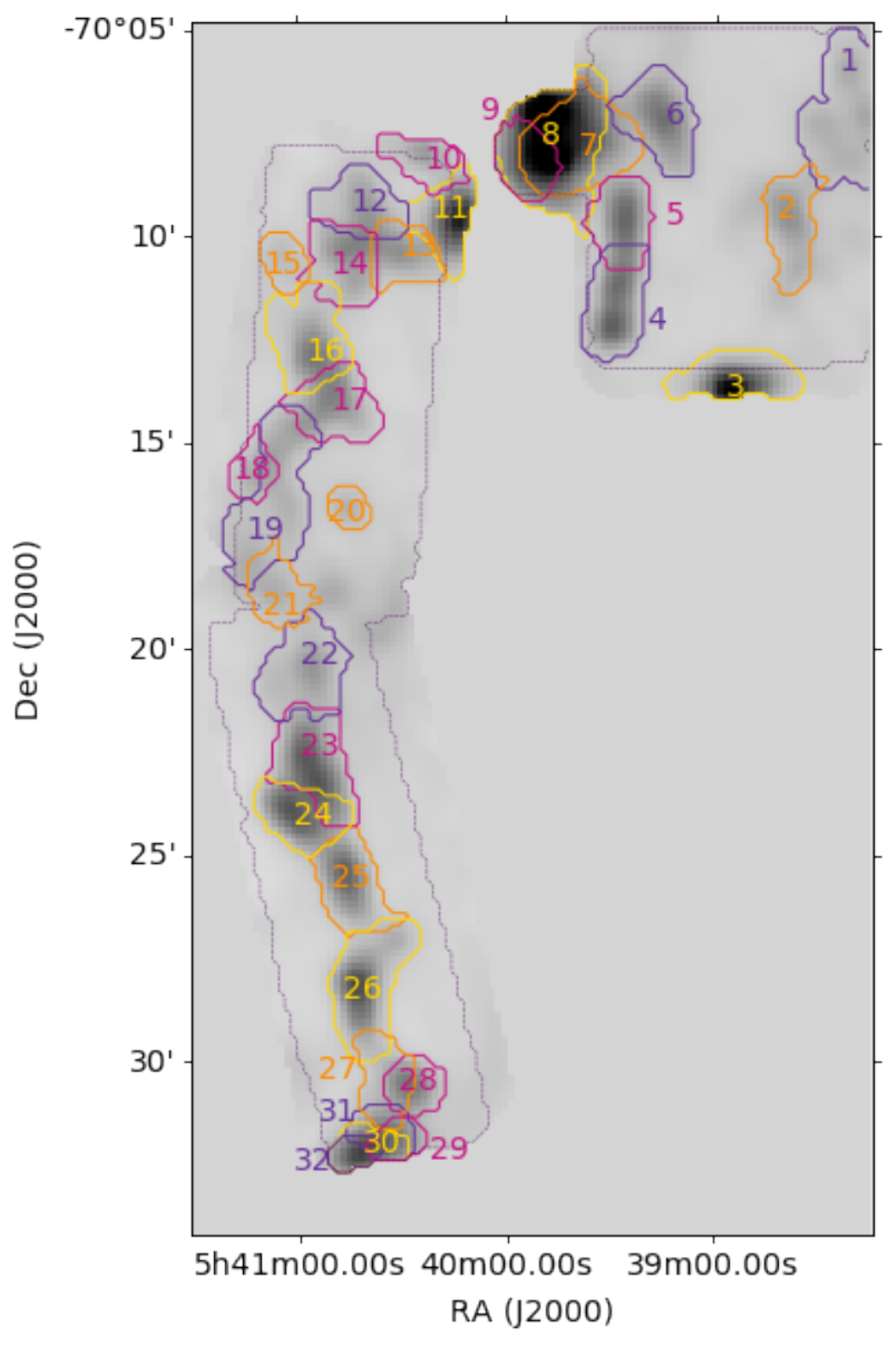}
    \caption{Projection of the 32 clumps identified by \texttt{quickclump} with identifying numbers. The grayscale is the integrated intensity of \thirtCO(1-0) (same map as Figure\,\ref{fig:lowres 13CO}). Overlapping clump borders indicate that the clumps overlap along the line of sight and are differentiated by their velocity structure. The dotted contour shows the common observational footprint of the four observed emission lines.}
    \label{fig:clump IDS}
\end{figure}

We used \texttt{quickclump}\footnote{\url{https://github.com/vojtech-sidorin/quickclump/}}\citep{Sidorin17}, which is a Python clump-finding algorithm that is similar in methodology to \texttt{clumpfind} \citep{clumpfind} and \texttt{DENDROFIND} \citep{Wunsch12}. These clumps are based on the ALMA \thirtCO(1-0) cube convolved to 45\arcsec\ to match the lowest resolution observation (the \twelveCO(1-0) from MAGMA), and the input parameters used were \texttt{Nlevels}=1000, \texttt{Tcutoff}=4$\sigma$=1.4 K, \texttt{dTleaf}=4$\sigma$=1.4 K, and \texttt{Npixmin}=5. A map of the 32 clumps identified by \texttt{quickclump} is shown in Figure\,\ref{fig:clump IDS} where they are also given identifying numbers.
Each voxel in the data cube was assigned to at most a single clump, so overlapping clump borders in Figure\,\ref{fig:clump IDS} indicate that the clumps overlap along the line of sight and are differentiated by their velocity structure. The integrated line fluxes for each of these clumps and each observed line are given in Table\,\ref{tab:line fluxes}. Clumps 3, 10, and 11 do not have corresponding APEX data and so were not included in the \radex fitting.

If any clump had more than 75\% of the pixels masked in any parameter, we removed the entire clump from the following analysis; this was only the case for \nh in clumps 1, 20, 29, 30, and 32. We also discarded the fits of clumps 9, 15, and 21 because they had very few pixels with successful fits, the fitted parameters had large variations from pixel-to-pixel, and they were major outliers in later trends. 
 
\begin{table*}[]
    \centering
    \caption{Integrated Line Fluxes of Clumps}
    \begin{filecontents*}{LineStrengthTable.csv}
ClumpNames, W12CO10, W13CO10, W12CO21, W13CO21, WCS21, W12CO10err, W13CO10err, W12CO21err, W13CO21err, WCS21err,test
1,70.9,5.4,62.7,6.45,0.411,0.4,0.04,0.3,0.08,0.04,1
2,46.3,4.84,42.7,4.18,0.229,0.3,0.04,0.3,0.07,0.03,1
3,42.4,6.97,0,0,1.15,0.3,0.05,0,0,0.03,0
4,75.7,10.1,81.7,14.4,3.07,0.4,0.06,0.4,0.1,0.04,1
5,58.1,8.67,57.9,7.8,1.83,0.3,0.05,0.3,0.08,0.03,1
6,100,9.57,83.1,8.1,0.699,0.4,0.05,0.4,0.09,0.04,1
7,119,10.5,98.1,8.78,1.12,0.5,0.06,0.4,0.1,0.04,1
8,277,20.1,202,14.6,2.68,0.8,0.08,0.6,0.1,0.06,1
9,31.6,2.47,28.4,2.13,0.371,0.3,0.03,0.2,0.06,0.03,1
10,12.4,1.4,0,0,0.27,0.2,0.02,0,0,0.02,0
11,57.9,6.2,0,0,1.38,0.3,0.05,0,0,0.03,0
12,44.6,5.32,42.4,5.19,0.463,0.3,0.04,0.3,0.06,0.03,1
13,34.7,4.96,38.6,4.74,0.462,0.3,0.04,0.3,0.06,0.03,1
14,53.5,7.34,65.6,9.21,0.816,0.3,0.05,0.4,0.08,0.03,1
15,9.17,0.988,11.2,1.42,0.154,0.1,0.02,0.1,0.04,0.02,1
16,75.9,7.62,65.3,6.33,0.335,0.4,0.05,0.3,0.08,0.03,1
17,59.2,7.29,55.4,7.07,0.325,0.3,0.05,0.3,0.07,0.03,1
18,14.6,1.82,13.1,1.98,0.0939,0.2,0.02,0.2,0.04,0.02,1
19,77.6,8.09,72.8,7.89,0.416,0.4,0.05,0.4,0.09,0.04,1
20,8.75,1.03,9.12,0.788,0.0461,0.1,0.02,0.1,0.03,0.01,1
21,16.1,1.54,12.5,1.74,0.0953,0.2,0.02,0.2,0.05,0.02,1
22,78.8,6.37,65.7,5.23,0.317,0.4,0.04,0.3,0.08,0.03,1
23,151,15.1,130,15,1.34,0.6,0.07,0.5,0.1,0.04,1
24,96.2,11.2,99.1,12.4,1.26,0.5,0.06,0.5,0.09,0.03,1
25,79.6,9.61,72.2,9.74,0.898,0.4,0.06,0.4,0.08,0.03,1
26,100,13.3,93.6,13.5,1.47,0.5,0.07,0.4,0.1,0.04,1
27,26.6,2.43,26.4,2.83,0.269,0.2,0.03,0.2,0.05,0.02,1
28,33.9,3.8,31.6,3.58,0.534,0.3,0.03,0.2,0.06,0.02,1
29,9.18,0.907,9.45,1.17,0.321,0.1,0.02,0.1,0.04,0.02,1
30,29.8,2.29,30.7,2.38,0.646,0.2,0.03,0.2,0.06,0.03,1
31,16.8,2.09,18.4,1.8,0.24,0.2,0.03,0.2,0.05,0.02,1
32,5.83,0.403,4.79,0.341,0.21,0.1,0.02,0.1,0.03,0.01,1
\end{filecontents*}
\csvreader[tabular=l|c|c|c|c|c,
    table head=\hline \hline Clump & \twelveCO(1-0) & \twelveCO(2-1) & \thirtCO(1-0) & \thirtCO(2-1) & CS(2-1) \\ \hline]
    {LineStrengthTable.csv}
    {ClumpNames=\name,W12CO10=\wco, W13CO10=\wtco, W12CO21=\wcot, W13CO21=\wtcot, WCS21=\wcs, W12CO10err=\ewco, W13CO10err=\ewtco, W12CO21err=\ewcot, W13CO21err=\ewtcot, WCS21err=\ewcs, test=\test}
    {\name 
    & \wco$\pm$\ewco 
    & \ifthenelse{\test>0}{\wcot$\pm$\ewcot}{---} 
    & \wtco$\pm$\ewtco 
    & \ifthenelse{\test>0}{\wtcot$\pm$\ewtcot}{---} 
    & \wcs$\pm$\ewcs
    }\\
    \textbf{Note:} All line fluxes are integrated over the whole clump in units of 10$^3$ K km s$^{-1}$ arcsec$^2$. Clumps 3, 10, and 11 do not have corresponding APEX data. The uncertainty is a 10\% error from the absolute flux calibration plus the rms error added in quadrature.
    \label{tab:line fluxes}
\end{table*}

\subsection{Derived Clump Properties} \label{subsec:derived props}

\begin{table*}[]
    \centering
    \caption{Derived Clump Properties}
    \begin{filecontents*}{PropertyTable.csv}
clump, mass, masslow, massup, meann, meannlow, meannup, meanT, meanTlow, meanTup, sigv, errsigv, Rell, errRm, alphavir, erralphavir, Nyso, Myso, Mysoerr
2,9.5,0.62,2.5,2.1,0.75,2,17,6,15,2.3,0.034,6.6,3.7,7.9,4.6,2,8.5,1.2
4,22,0.54,0.63,3.7,0.81,2.2,18,5.1,5.6,1.8,0.017,8,4.5,2.7,1.5,3,47,4.9
5,19,0.61,0.83,2,0.53,1.2,16,4.5,5.1,1.9,0.024,7,2.6,3.1,1.2,0,0,0
6,19,0.39,1.8,1.6,0.59,0.94,23,8.6,13,1.8,0.014,7.4,2.4,3,0.98,0,0,0
7,18,0.5,0.95,1.7,0.26,0.59,19,3.3,4.9,1.5,0.0076,8.9,3.8,2.6,1.1,1,6.5,0.87
8,36,2.1,4.5,0.65,0.35,0.46,32,10,21,2.2,0.0096,9.3,2.2,2.7,0.68,0,0,0
12,10,0.34,0.42,3.8,1.3,2.6,18,5.2,5.6,0.97,0.0057,6.7,0.6,1.4,0.13,2,13,1.5
13,10,0.83,1.2,2.4,0.88,1.8,20,6.9,8.8,1.2,0.007,7.3,1.3,2.2,0.47,0,0,0
14,15,0.25,0.35,3.9,1.1,2.2,20,6,6.6,1.6,0.012,7.8,0.96,3,0.37,5,52,3.8
16,16,1.3,2.1,1.3,0.44,0.77,23,9.9,18,2,0.035,6.7,0.91,3.7,0.66,0,0,0
17,15,0.23,0.43,2.5,0.52,1.1,19,5.9,7.3,1.3,0.0087,8,0.48,1.9,0.12,2,5.9,0.83
18,3.6,0.16,0.22,3.2,0.94,3.6,13,3.5,4,1.1,0.029,4.8,0.84,3.5,0.66,1,12,1.5
19,14,0.58,0.73,2.7,0.95,2,17,5,5.6,1.8,0.025,10,5,5.3,2.6,0,0,0
22,11,0.4,1.1,1.2,0.79,1.4,23,9.2,9,1.8,0.013,7.9,1.6,5.1,1.1,0,0,0
23,30,2.4,2.1,1.7,0.53,0.88,36,16,22,1.7,0.0078,8.5,2.5,1.7,0.53,0,0,0
24,22,1.1,1.2,2.6,0.8,1.2,29,10,13,1.6,0.0088,7.7,1.5,2.1,0.43,2,29,3.1
25,19,0.89,0.95,2.4,0.63,1,25,8.2,10,1.2,0.01,7.2,2.5,1.2,0.42,1,5,1.7
26,26,0.89,1.4,2.5,0.8,1.2,24,8.2,11,1.4,0.011,6.9,2.2,1.2,0.39,5,45,4
27,4.1,0.22,0.28,1.8,0.68,2,17,5,6.1,0.97,0.012,6.1,1.1,3.1,0.59,1,4.9,0.65
28,8.1,1.2,2.3,3,1.2,3.9,21,11,23,2.1,0.037,5.1,0.28,6.2,1.3,1,3.2,1.1
31,3.4,0.21,0.28,2.6,0.75,2.1,16,4.8,5.9,1.1,0.02,4.9,0.43,3.7,0.43,0,0,0
\end{filecontents*}
\csvreader[tabular=l|c|c|c|c|c|c|c|c,
    table head=\hline \hline Clump\footnote{CO clump identifying numbers are shown in Figure\,\ref{fig:clump IDS}. Clumps 3, 10, and 11 are not included because they do not have corresponding APEX data. Clumps 1, 20, 29, 30, and 32 are not included because more than 75\% of the pixels had poor \radex fits for \nh. Clumps 9, 15, and 21 are not included because the \radex fits had unphysical variations between pixels. } & Mass\footnote{Result of \radex fitting.}\footnote{Calculation described in \S\ref{subsec:derived props}.} & \nh$^\text{bc}$ & \tkin$^\text{bc}$ & $\sigma_v^\text{c}$ & $\sigma_R^\text{c}$ & $\alpha_\text{vir}$\footnote{Calculated with Equation\,\ref{eq:alphavir}.} & Number of & Total M\\ 
     & ($10^3$ M$_\odot$) & ($10^3$ cm$^{-3}$) & (K) & (km/s) & (pc) & & YSOs\footnote{Identifying YSOs, matching them to associated CO clumps, and fitting YSO masses is described in \S\ref{sec:ysos}.} & YSOs$^\text{e}$ ($M_\odot$) \\ \hline 
     ]
    {PropertyTable.csv}
    {clump=\num, mass=\m, masslow=\ml, massup=\mup, meann=\n, meannlow=\nl, meannup=\nup, meanT=\t, meanTlow=\tl, meanTup=\tup, sigv=\sv, errsigv=\esv, Rell=\R, errRm=\eR, alphavir=\av, erralphavir=\eav, Nyso=\Nyso, Myso=\Myso, Mysoerr=\eMyso}
    {\num & \m$^{+\mup}_{-\ml}$ & \n$^{+\nup}_{-\nl}$ & \t$^{+\tup}_{-\tl}$ & \sv$\pm$\esv & \R$\pm$\eR & \av$\pm$\eav & \Nyso & \ifthenelse{\Nyso>0}{\Myso$\pm$\eMyso}{---}}
    \label{tab:derived quantities}
\end{table*}

For each clump, we calculated the mass by summing the \NCO within the clump, then multiplying by the area of each pixel in cm$^{2}$, the ratio of H$_2$/CO, and a factor of 1.3 to convert from H$_2$ mass to total mass based on cosmic abundances. The H$_2$/CO ratio is based on the values of $R_{13}$ for the map (either 50 or 100), and H$_2$/\thirtCO\ in the outer Milky Way, where the metallicity is similar to the LMC ($\sim1/3$ of solar). This has been measured to be between H$_2$/\thirtCO$=10^6$ \citep{Heyer01} and H$_2$/\thirtCO$=3\times10^6$ \citep{Brand95}, so we adopt H$_2$/\thirtCO$=2\times10^6$ in this work.  Previous works in the LMC have used similar values: 2--6$\times$10$^6$ in \citet{Heikkila99}, and 3$\times$10$^6$ in \cite{Wong19}.
We keep H$_2$/\thirtCO\ constant rather than H$_2$/CO since \thirtCO\ is optically thinner; \twelveCO\ is usually optically thick, so the column density is more correlated with the \thirtCO\ lines. Keeping H$_2$/\thirtCO\ constant minimizes our dependence on the value assumed for $R_{13}$ and instead leaves us with the value of H$_2$/\thirtCO\ as the major systematic uncertainty.

The final mass estimate for the clump is the mean of the masses from assuming $R_{13}$=50 and $R_{13}$=100 and assuming a filling factor of 10\% and 20\%. The upper and lower errors come from the full range included in the upper and lower errors on the two masses, as described in \S\ref{subsec:getting maps}. When a single error is reported, it is the geometric mean of the upper and lower errors. To get temperatures and densities for each clump, we took the mass-weighted average of all pixels in the clump, with the upper and lower errors propagated through separately. The resulting clump masses are in the range $(3.4-35.5)\times10^3$~M$_\odot$, temperatures are in the range 13--36~K, and densities are in the range 650--3940~cm$^{-3}$.

To measure the linewidths, we found the mass-weighted mean line profile using the map of \NCO. We then fit a Gaussian to this average line profile and report $\sigma_v$, not FWHM. The resulting linewidths are in the range 1.2--2.1~km~s$^{-1}$. Shuffling the line profiles to a common central velocity first would change the linewidths by $\sim$10\%. 


To get the radius of the clumps, we fit ellipses to the half-light contour, giving us major and minor axes for the clump. We convert these FWHM values to $\sigma$ of a gaussian profile (FWHM = 2.35$\times\sigma$). 
The reported radius is the geometric mean of the major and minor axes.

We also compare the radii from the fitted ellipse to two other methods. One is finding the area of the clump within the half-light contour and then finding the radius of a circle with equivalent area. This is taken to be an effective FWHM, which is then converted to  $\sigma$ of a Gaussian profile. 
This results in an effective radius that is usually almost identical to the geometric mean of the fitted ellipse, differing by a factor of 0.99 on average.

The other method is to take the spatial second moment of the clump projected along the axes of the fitted ellipse. This results in $\sigma$ for the major and minor axes of the clump. 
The geometric mean of these major and minor radii is a factor of 0.72 lower on average than that fitted by the ellipse. 

In the rest of this paper, we use the radius from the ellipse fitting method, and the error in the radius takes into account how non-circular the clump boundary is. The resulting $\sigma_r$ are in the range 5--10 pc.

The values of all derived properties for each of the clumps with \radex fits are given in Table\,\ref{tab:derived quantities}.

\section{Associated YSOs} \label{sec:ysos}

Current star formation in a molecular clump is expected to affect the gas in that clump - heating and changes in optical depth due to bulk gas motions and photodissociation will change the excitation conditions.  In this section we describe the young stellar objects in the region, and how they are associated with CO clumps.  In the next section, we will compare our non-LTE fitting to other techniques of determining gas properties, and will find that the presence of YSOs does not appear to affect one’s ability to calculate gas properties using those different methods.  In section 7, we will address the different question of whether the presence of the YSOs is correlated with changes in the physical properties.

\begin{figure}
    \centering
    \includegraphics[width=0.5\textwidth]{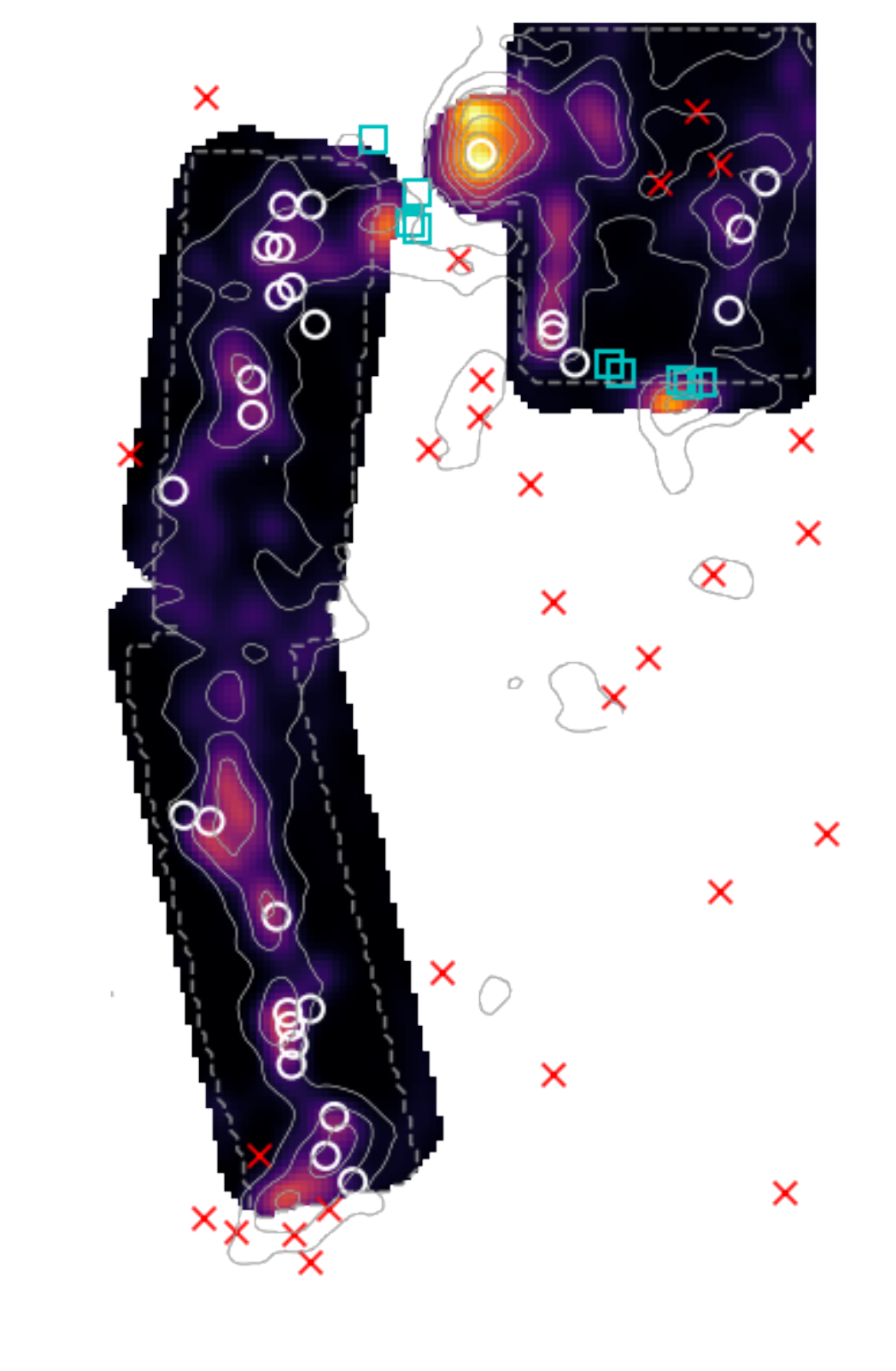}
    \caption{The 28 YSOs that were matched to CO clumps with \radex fits are shown as white circles, the 9 YSOs that were matched to CO clumps that could not be fitted are shown as cyan squares, and the YSOs with no associated clump in \thirtCO(1-0) are shown as red Xs. The color image is the integrated intensity of \thirtCO(1-0) (same maps as Figure\,\ref{fig:lowres 13CO}), which is the map that was used to assign clumps with \texttt{quickclump}. The contours are the integrated intensity of \twelveCO(1-0) as in Figure\,\ref{fig:moments}.}
    \label{fig:YSO matching}
\end{figure}

\subsection{YSO Selection} \label{subsec: yso selection}

Complete infrared surveys of the LMC with Spitzer \citep[SAGE][]{Meixner06} and Herschel \citep[HERITAGE][]{Meixner13} enabled the uniform selection of massive young stellar objects (MYSOs) across the entire galaxy. \citet[][henceforth W08]{Whitney08} used PSF-fit photometry from the SAGE legacy catalog and color selections between 1 and 10$\mu$m, chosen to include MYSO models and exclude evolved and main sequence stars.  \citet[][henceforth GC09]{GC09} used aperture photometry of SAGE images and the 4.5-8.0$\mu$m color plus manual examination of images and source environment. \citet[][henceforth S14]{Seale14} used 250$\mu$m PSF-fit photometry from the HERITAGE legacy catalog and 24$\mu$m emission to include MYSOs and resolved morphology to exclude background galaxies.

We combined the three existing MYSO catalogs in the Ridge region, first by matching all GC09 and W08 sources within 10\arcsec\ of each S14 source.  This resulted in the same associations that S14 published with a 5\arcsec\ matching radius, with the addition of a single matched source J85.202523-70.17060.  Visual examination of 8.0-24-250$\mu$m 3-color images prompted the removal of three S14 sources which were blended with other S14 sources - the guiding principle being to identify the sources that could be reliably photometered over a wide wavelength range.

The resulting list of 109 sources contains 24 MYSO candidates identified only from the shorter-wavelength lists GC09 and W08, 45 identified only from the longer-wavelength list S14, and 40 sources identified in multiple lists and matched. Next, we generated cutout images from 1.0-500$\mu$m and calculated aperture photometry at all bands.

The datasets used are 2MASS J, H, and K$_s$ (1.2$\mu$m, 1.6$\mu$m, 2.1$\mu$m, angular resolution $\sim$2$\arcsec$, aperture radius 3$\arcsec$) \citep{Skrutskie06}, SAGE IRAC bands 1-4 (3.6, 4.5, 5.8, and 8.0$\mu$m, resolution $\sim$2$\arcsec$, aperture radius 3$\arcsec$) and MIPS 24$\mu$m and 70$\mu$m (resolution 6$\arcsec$, 18$\arcsec$, aperture radius 9$\arcsec$, 18$\arcsec$), HERITAGE PACS 110 and 170$\mu$m (resolution 8$\arcsec$, 13$\arcsec$, aperture radius 11$\arcsec$, 18$\arcsec$), and HERITAGE SPIRE 250,350, 500$\mu$m (resolution 18$\arcsec$, 25$\arcsec$, 36$\arcsec$, aperture radius 18$\arcsec$, 25$\arcsec$, 37$\arcsec$). A local annular background was subtracted from each flux, and the uncertainty of the flux measurement is calculated from the standard deviation of values in the annulus.

We combined the published photometry with the new aperture photometry by visually inspecting the cutout images and spectral energy distribution (SED) of the two. 
For most sources, the published photometry agreed with the new aperture photometry within uncertainties, so the average was used.  For sources and bands in which the published photometry was lacking, the aperture photometry smoothly filled in the SED. 

The Spitzer MIPS 70$\mu$m resolution of 18\arcsec\ is significantly worse than the neighboring points in our spectral energy distributions (Spitzer MIPS 24$\mu$m and Herschel PACS 100$\mu$m at $\sim$6\arcsec). Consequently, the 70$\mu$m flux density was clearly affected by blending for many sources, so was used as an upper limit in the subsequent SED fitting.  If the image cutouts were confused, or the aperture and published photometry were very discrepant for a given source and band, we either eliminated that band from the fitting or used the largest photometric value as an upper limit.

\subsection{YSO Fitting} \label{subsec: yso fitting}

We fit the YSO SEDs with the \citet{Robitaille17} grid of single-YSO dust radiative transfer models \footnote{\url{https://zenodo.org/record/166732}} and the \citet{Robitaille07} $\chi^2$ fitting code\footnote{\url{https://sedfitter.readthedocs.io/en/stable/index.html}}.  Following many other studies \citep[e.g.][]{Carlson12,Chen10}, we used a minimum uncertainty in each band of 10\%, and calculated $\chi^2$ for every model.  Central sources in the \citet{Robitaille17} model grid are parameterized by the radius and temperature of the YSO, $R_\star$ and $T_\star$.  We interpolated the Z=0.004 PARSEC stellar evolutionary models \citep{Bressan12} that include the PMS phase\footnote{\url{http://stev.oapd.inaf.it/cgi-bin/cmd}} to determine the mass of the YSO, $M_\star$, and age for each Robitaille YSO model.
The Robitaille circumstellar dust distribution is that of a rotating flattened toroid in analytical form \citet{Ulrich76} parameterized by centrifugal radius $R_C$ and scaling density $\rho_0$.  Using the central source mass assigned to each model, the envelope accretion rate $\dot{M}$ is a function of $M_\star$, $\rho_0$, and $R_C$ \citep[][equation 5]{Robitaille17}.  The envelope accretion rate relative to the central source mass is a measure of evolutionary stage under the assumption that mean accretion rate decreases with time for protostars.

Finally, we assigned $M_\star$ and $\dot{M}$ to each source in the Ridge by marginalizing the model probability distributions over all other model parameters and measuring the first and second moments of each 1D probability distribution. We visually inspected the 2D probability distribution of $M_\star$ and $\dot{M}$ for each source and verified that the second moments that we are using as uncertainties for those parameters do indeed span the range of well-fitting models, even in the minority of cases where the probability distribution is not single-peaked. A table of $M_\star$ and $\dot{M}$ for each YSO that was matched with a CO clump is given in Table\,\ref{tab:yso properties}.

It is important to recognize that despite the use of a specific set of models, fundamentally the quantities being measured and parameterized are the total luminosity of the central source, which is tightly correlated with the derived $M_\star$, and the amount of dust extinction around that central source, which is highly correlated with the envelope accretion rate $\dot{M}$ (except for more evolved sources that have little envelope in which case the disk has more of an effect on the SED).

Comparing the detection limits of the SAGE and Herschel surveys to all of the \citet{Robitaille17} models, we expect that all protostars with a mass above 6 M$_\odot$ will be detected. At 2.5 M$_\odot$, half of all protostars would be detected, though it is possible to detect some protostars down to 1 M$_\odot$, depending on their evolutionary state.

\subsection{YSO Matching} \label{yso matching}

We then matched the YSOs to CO clumps by eye, assigning them to a clump only if their positions coincided with CO emission (see Figure\,\ref{fig:YSO matching}). In cases where the YSO overlapped with strong emission from more than one clump, it was assigned to the clump that was brightest in its location. This process resulted in 37 of the YSOs being matched to CO clumps in the Molecular Ridge. Of these 37, nine were associated with the three clumps that do not have data for \twelveCO(2-1) and \thirtCO(2-1) from APEX (clumps 3, 10, and 11) and so could not be included in all parts of the analysis. Table\,\ref{tab:yso properties} includes the number of the clump assigned to each of these 37 YSOs. 

\begin{table}[]
    \centering
    \caption{Fitted YSO Properties}
    \begin{filecontents*}{YSOPropsTable.csv}
name, m, merr, logmdot, logmdoterr, clump
J84.636381-70.144787,	4.9,	0.8,	-5.2,	0.3,	1
J84.677275-70.194884,	4.8,	1.0,	-5.2,	0.3,	2
J84.663696-70.163174,	3.7,	0.5,	-5.2,	0.3,	2
J84.722292-70.227647,	4.9,	0.6,	-4.6,	0.3,	3
J84.739529-70.22909,	15.4,	2.0,	-6.8,	0.3,	3
J84.815252-70.224064,	9.2,	2.4,	-8.0,	1.1,	3
J84.746459-70.227471,	8.6,	1.2,	-4.6,	0.3,	3
J84.829777-70.220668,	11.5,	1.6,	-11.3,	0.4,	3
J84.879808-70.20452,	21.8,	3.9,	-7.4,	0.3,	4
J84.879594-70.200559,	20.5,	2.7,	-9.1,	0.3,	4
053925.03-701255.0,	4.6,	1.3,	-6.1,	0.5,	4
J84.962153-70.133828,	6.5,	0.9,	-4.6,	0.3,	7
J85.101423-70.133438,	11.2,	1.8,	-8.6,	0.4,	10
J85.057077-70.165809,	15.1,	2.7,	-10.1,	0.7,	11
J85.050246-70.168067,	15.4,	2.0,	-11.6,	0.7,	11
054012.00-700916.0,	15.3,	2.1,	-6.3,	0.3,	11
054037.80-700914.0,	6.5,	0.9,	-8.7,	0.5,	12
J85.18818-70.154154,	6.2,	1.2,	-8.0,	0.8,	12
J054036.65-701201.1,	4.2,	1.3,	-8.1,	1.4,	14
J85.179372-70.186003,	15.4,	2.0,	-7.9,	0.3,	14
J85.193443-70.189071,	15.4,	2.1,	-6.8,	0.3,	14
J85.207422-70.169937,	8.1,	1.5,	-4.6,	0.3,	14
J85.192755-70.17054,	8.5,	1.3,	-4.6,	0.3,	14
J85.225346-70.235587,	4.7,	0.8,	-5.2,	0.3,	17
054054.33-701318.8,	1.2,	0.1,	-7.8,	0.5,	17
J85.317317-70.265032,	11.5,	1.5,	-9.1,	0.3,	18
J85.276729-70.393765,	20.5,	2.7,	-9.1,	0.3,	24
054113.61-702329.1,	9.0,	1.5,	-7.8,	0.5,	24
J054047.85-702551.1,	5.0,	1.7,	-8.0,	1.5,	25
J85.18708-70.468313,	20.5,	2.7,	-6.8,	0.3,	26
J85.180541-70.480586,	6.5,	0.9,	-4.6,	0.3,	26
054044.25-702824.7,	6.9,	2.0,	-5.8,	0.4,	26
054043.94-702918.5,	4.6,	1.7,	-7.9,	1.6,	26
054038.61-702800.5,	6.6,	1.0,	-6.9,	0.3,	26
J85.143381-70.523952,	4.9,	0.6,	-5.2,	0.3,	27
J85.133273-70.508695,	3.2,	1.1,	-5.2,	0.3,	28
J054026.85-703202.9,	2.1,	0.3,	-11.3,	0.3,	29
\end{filecontents*}
\csvreader[tabular=l|c|c|c,
    table head=\hline \hline Name & $M_*$ & $\log{(\dot{M})}$ & Clump \\ 
    & ($M_\odot$) & ($M_\odot$/yr) & Assignment \\\hline
     ]
    {YSOPropsTable.csv}{name=\name, m=\m, merr=\merr, logmdot=\mdot, logmdoterr=\emdot, clump=\clump}
    {\name & \m$\pm$\merr & \mdot$\pm$\emdot & \clump}
    \label{tab:yso properties}
\end{table}

\section{Comparing \radex Fitting to Other Methods} \label{sec:Comparisons}

\begin{table}
    \caption{Comparison of methods}
    \begin{center}
    \begin{tabular}{c|c c }
        \hline
        \hline
         & $M/M_\text{RF}$ & $T_\text{ex}$/$T_{\text{kin,RF}}$ \\ 
         \hline
         LTE (1-0)     & 1.66$\pm$0.19 & 0.71$\pm$0.10 \\
         LTE (2-1)     & 0.55$\pm$0.10 & 0.78$\pm$0.15 \\
         \XCO          & 1.20$\pm$0.33 & NA            \\
    \end{tabular}
    \end{center}
    Comparison of different methods of determining mass (from \NCO) and \tkin or $T_\text{ex}$. These values are the average of the ratios for all the clumps that were fit with these methods and the standard deviation among the clumps. $M_\text{RF}$, $T_{\text{kin,RF}}$ are the values derived from the \radex fitting as described in \S\ref{subsec:getting maps}.  
    \label{tab:fitting comparison}
\end{table}

\subsection{LTE Method} \label{subsec: LTE}

We compared the clump masses derived by the \radex fitting to the results of LTE assumptions with both the (1-0) lines and the (2-1) lines. For LTE, we use the peak of the \twelveCO\ line for each pixel, divided by a beam filling factor of 10\% or 20\% that we used in the \radex fitting (results for each were combined after as described in \S\ref{subsec: fitting method}), to get $T_\text{ex}$:

\begin{equation}
    T_B = T_{ul}(1 - e^{-\tau_\nu})\left[\frac{1}{e^{T_{ul}/T_{ex} } - 1} - \frac{1}{e^{T_{ul}/T_{bg}} - 1}\right],
\end{equation}

where we assume optically thick \twelveCO, so $e^{-\tau_\nu} \approx 0$. $T_B$ is the peak brightness temperature for the pixel divided by the filling factor, $T_{ul}=5.532$~K for \twelveCO(1-0), $T_{ul}=11.06$~K for \twelveCO(2-1), and $T_{bg} = 2.73$~K. Then, we calculate the optical depth of \thirtCO\ from

\begin{equation}
    \tau_\nu = -\ln{\left[1 - \frac{T_B}{T_{ul}}\left[\frac{1}{e^{T_{ul}/T_{ex} } - 1} - \frac{1}{e^{T_{ul}/T_{bg}} - 1}\right]^{-1}\right]}
\end{equation}

where now $T_B$ is from \thirtCO(1-0) divided by the filling factor of 10\%, $T_{ul} = 5.289$~K for \thirtCO(1-0), and $T_{ul} = 10.58$~K for \thirtCO(2-1). Then, the column density of \thirtCO\ is

\begin{equation}
    N_{13} = \frac{8\pi\nu_0^2}{c^2 A_{ul}}\frac{Q}{g_u}\frac{1}{1 - e^{-T_{ul}/T_{ex}}}\int{\tau_\nu d\nu}
\end{equation}

where we use $Q = \frac{T_{ex}}{B_0} + \frac{1}{3}$ and $B_0 = 2.644$~K for \thirtCO. We adopt H$_2$/\thirtCO$=2\times10^6$ and sum over all pixels in the clump to get masses from the LTE assumptions. A comparison of the results using the (1-0) and (2-1) lines against our \radex-fitting method is given in Table\,\ref{tab:fitting comparison}. 

Using (2-1) lines with the LTE assumptions results in a much lower mass estimate and a lower $T_\text{ex}$ estimate than the \radex-fitted \tkin (by average factors of 0.55$\pm$0.10 and 0.78$\pm$0.15, respectively). A low mass estimate from LTE calculations is expected if \thirtCO\ is sub-thermally excited and $T_\text{ex}^{13} < T_\text{ex}^{12}$: The $T_\text{ex}$ derived from \twelveCO\ is too high for \thirtCO, which makes the optical depth of \thirtCO\ underestimated, and so the column density and mass are underestimated \citep[e.g.][]{Castets90,Padoan00,Heyer15}. The temperature estimate being lower than the non-LTE \tkin could mean that \twelveCO\ is also sub-thermally excited, but less so than \thirtCO. 

The mass estimate when using the (1-0) lines is higher than our fitted \radex (by an average factor of 1.66$\pm$0.19), which is harder to explain though not unprecedented \citep{Indebetouw20}. One way this could happen is if the \twelveCO\ is not actually optically thick, which would make the LTE $T_\text{ex}$ estimate too low and shift the optical depth estimate higher. This could be exacerbated if the beam filling factor we use is too large, which would also artificially lower the measured excitation temperature. The non-linearity of the LTE equations means that an underestimated $T_\text{ex}$ due to a high filling factor would cause a larger shift to high column density than is corrected for with the filling factor being multiplied back in at the end. This filling factor was also used in the \radex fitting, but the non-linearity of that is even more extreme, making it difficult to predict how much it would affect our results relative to the LTE calculations.

\subsection{\XCO Method} \label{subsec: xco}

We also include in Table\,\ref{tab:fitting comparison} a comparison of the mass from a typical Milky Way value of \XCO$=2\times10^{20}$ cm$^2$/(K km s$^{-1}$) \citep{Bolatto13}. These masses are on average slightly higher than those fit with \radex---though they are consistent within the deviation between clumps (the average factor is 1.20$\pm$0.33).

\begin{figure}
    \centering
    \includegraphics[width=0.45\textwidth]{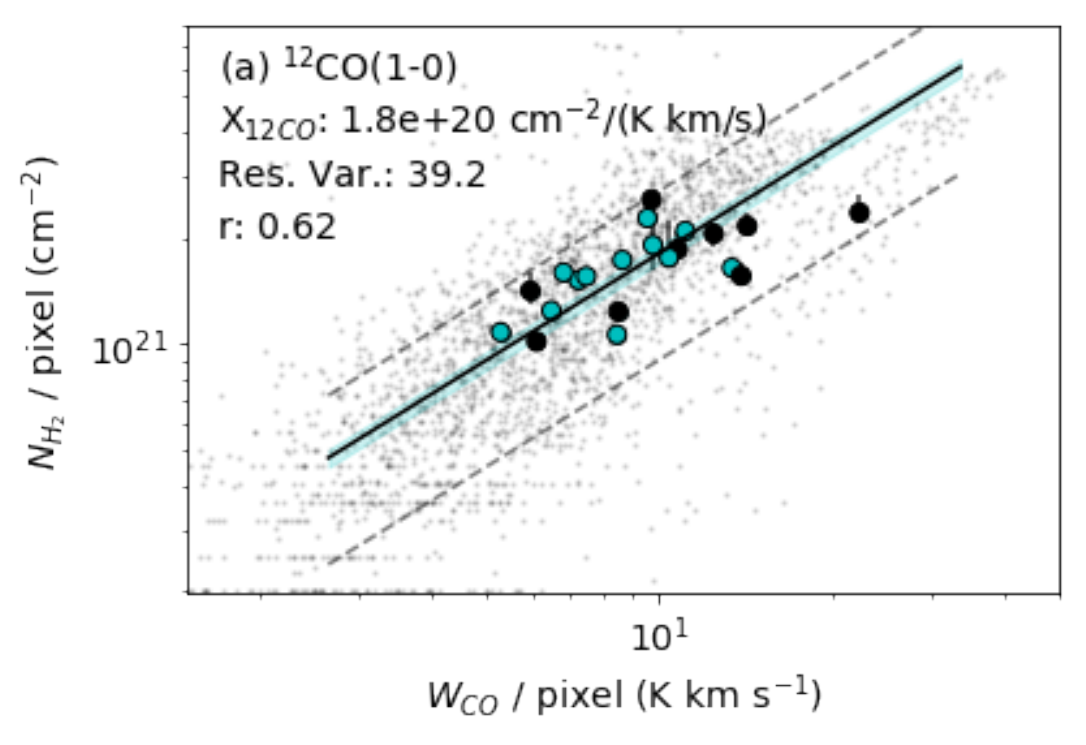}
    \includegraphics[width=0.45\textwidth]{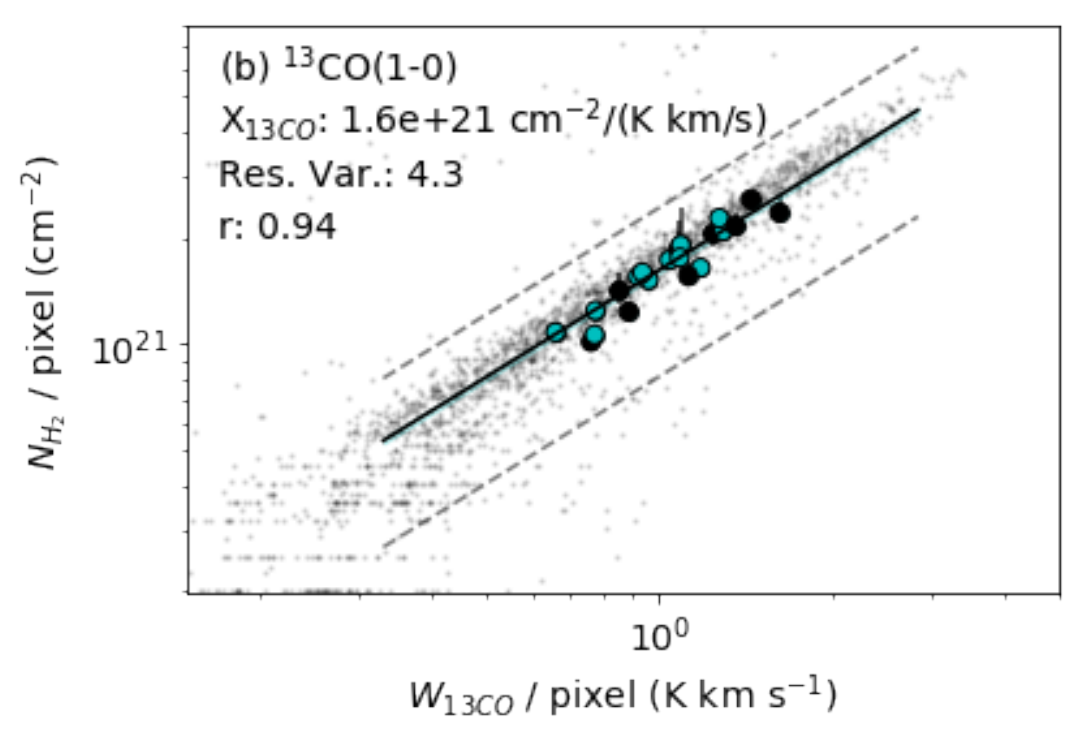}
    \caption{Fitted values of $N_{H_2}$ per pixel against the integrated flux per pixel of \twelveCO(1-0) (top) and \thirtCO(1-0) (bottom). Blue circles indicate clumps with at least one associated YSO, while the black circles are clumps without any associated YSOs. The small black points show the values of individual pixels. The fitted linear trend (fitted linearly, not in log space as is plotted here) corresponds to a value of \XCO$=(1.8\pm0.1)\times10^{20}$ cm$^{-2}$/(K km s$^{-1}$) and \XthCO$=(1.6\pm0.03)\times10^{21}$ cm$^{-2}$/(K km s$^{-1}$), respectively. The dotted lines on either side show the systematic range from the uncertainty in the ratio  H$_2$/\thirtCO\ used to get $N_{H_2}$. \thirtCO\ has a tighter trend with $N_{H_2}$ with a residual variance of 4.3 and correlation coefficient of 0.94 while \twelveCO\ has a residual variance of 39 and correlation coefficient of 0.62, suggesting that the \XthCO factor may be a better tool for estimating mass than the typical \XCO.}
    \label{fig:XCO plots}
\end{figure}

We fit the linear relation between the \radex-fitted total $N_{H_2}$ and the summed \twelveCO(1-0) flux of the clump in units of K km s$^{-1}$ ($W_\text{CO}$), both divided by the number of pixels in the clump with $N_{H_2}$ solutions (Figure\,\ref{fig:XCO plots}a). The linear trend was fit to the linear values, not in log space. Values of individual pixels within the clumps with $N_{H_2}$ solutions are also shown behind the clump average points. The slope of this trend is the \XCO conversion factor ($N_{H_2} = X_\text{CO} \times W_\text{CO}$). 
%
%
This relation is subject to the systematic error in the ratio H$_2$/\thirtCO\ used in \S\ref{subsec:derived props}. We use  H$_2$/\thirtCO$=2\times10^6$ with a range of $(1-3)\times10^6$ \citep{Brand95,Heikkila99,Heyer01}. A higher H$_2$/\thirtCO\ value results in higher derived masses and a larger derived \XCO.

Fitting a linear relation results in a value for \XCO of $(1.81\pm0.1)\times10^{20}$ cm$^{-2}$/(K km s$^{-1}$), and taking into account the full systematic range due to the uncertainty in H$_2$/\thirtCO\ and the error in the fitted \XCO results in a value of \XCO in the range $(0.85-2.87)\times10^{20}$ cm$^{-2}$/(K km s$^{-1}$), shown in Figure\,\ref{fig:XCO plots}a. This is lower than we expect for the Ridge since the LMC has a metallicity of 1/3 solar and has been estimated to have a value closer to \XCO$\sim4\times10^{20}$ cm$^{-2}$/(K km s$^{-1}$) \citep{Hughes10,Bolatto13}. This discrepancy could be because the non-LTE fitting is more accurate than previous measures or could be due to an underestimate of the H$_2$/\thirtCO\ ratio; a ratio of H$_2$/\thirtCO$\sim4\times10^6$ would make our fitted \XCO consistent with other estimates for the LMC. 

We also perform this fitting with the summed \thirtCO(1-0) intensities to get a value of \XthCO (Figure\,\ref{fig:XCO plots}b). We find a value of \XthCO$ = (1.62\pm0.03)\times10^{21}$ cm$^{-2}$/(K km s$^{-1}$), and taking into account the systematic range of H$_2$/\thirtCO$=(1-3)\times10^6$ and the error in the fitted \XthCO results in a \XthCO range of $(0.80-2.48)\times10^{21}$ cm$^{-2}$/(K km s$^{-1}$). 

The relation between $N_{H_2}$ and \thirtCO\ emission is tighter than the relation of $N_{H_2}$ with \twelveCO. The fitted \XCO has a 5.8\% error and a residual variance of 39, while the fitted \XthCO has a 1.9\% error and a residual variance of 4.3. Calculating a Pearson correlation coefficient for the two trends results in a value of $r=0.62$ for \XCO and $r=0.94$ for \XthCO. This shows that using a \XthCO would be a more precise method of determining mass than \XCO, which makes sense since \thirtCO\ is more optically thin and so can better trace the quantity of gas, while \twelveCO\ is usually optically thick. 

\begin{figure*}
    \centering
    \includegraphics[width=0.45\textwidth]{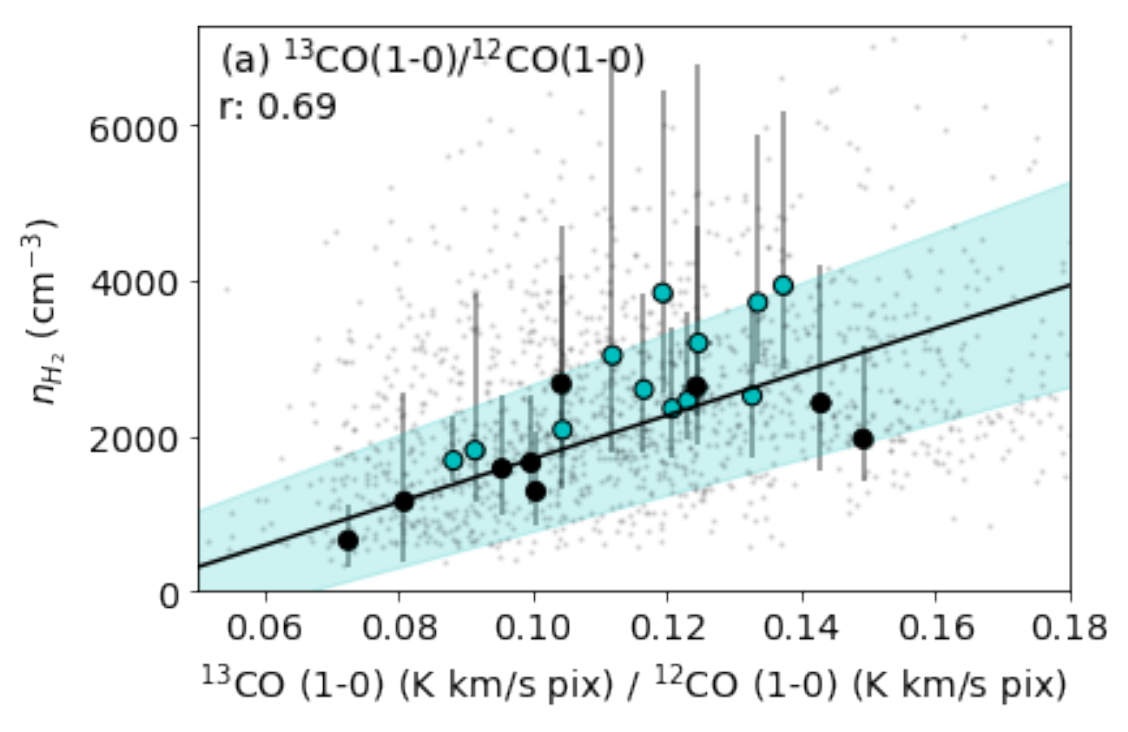}
    \includegraphics[width=0.45\textwidth]{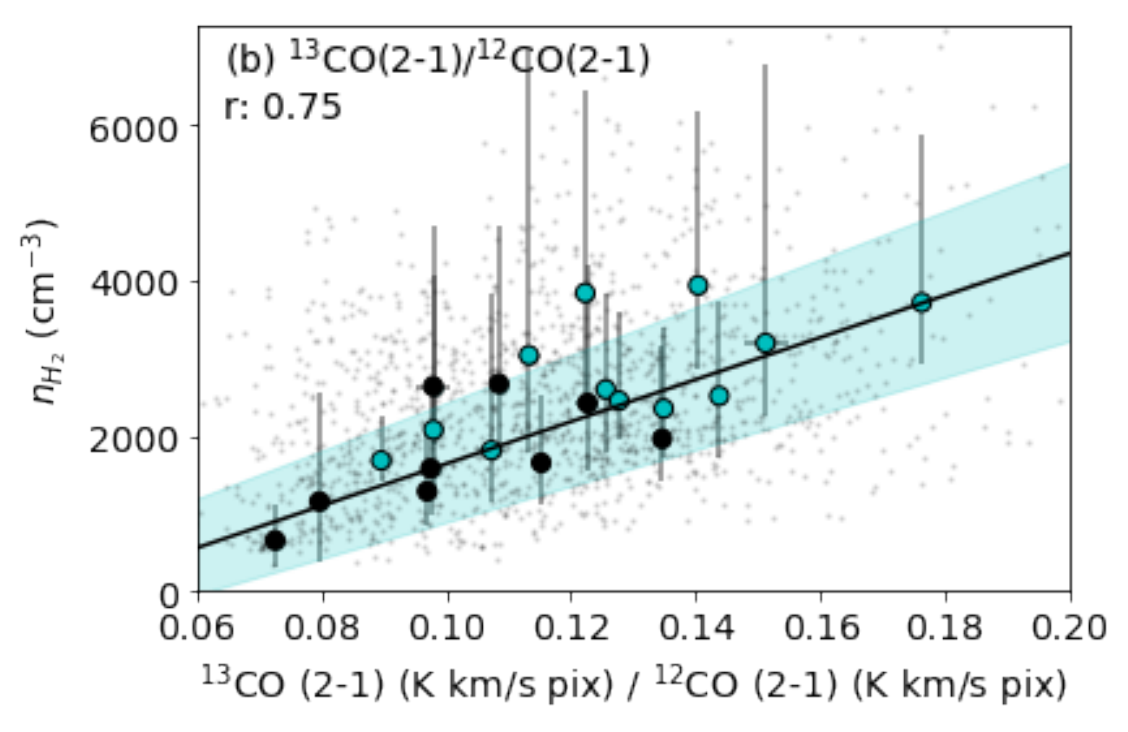}
    \includegraphics[width=0.45\textwidth]{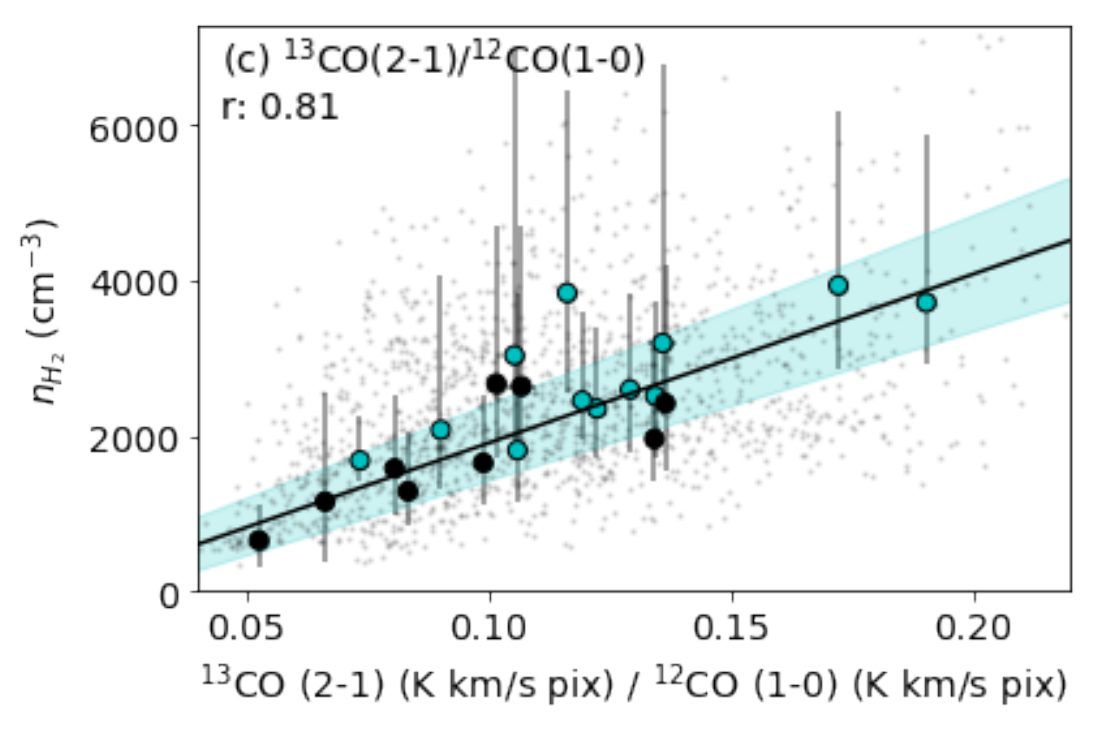}
    \includegraphics[width=0.45\textwidth]{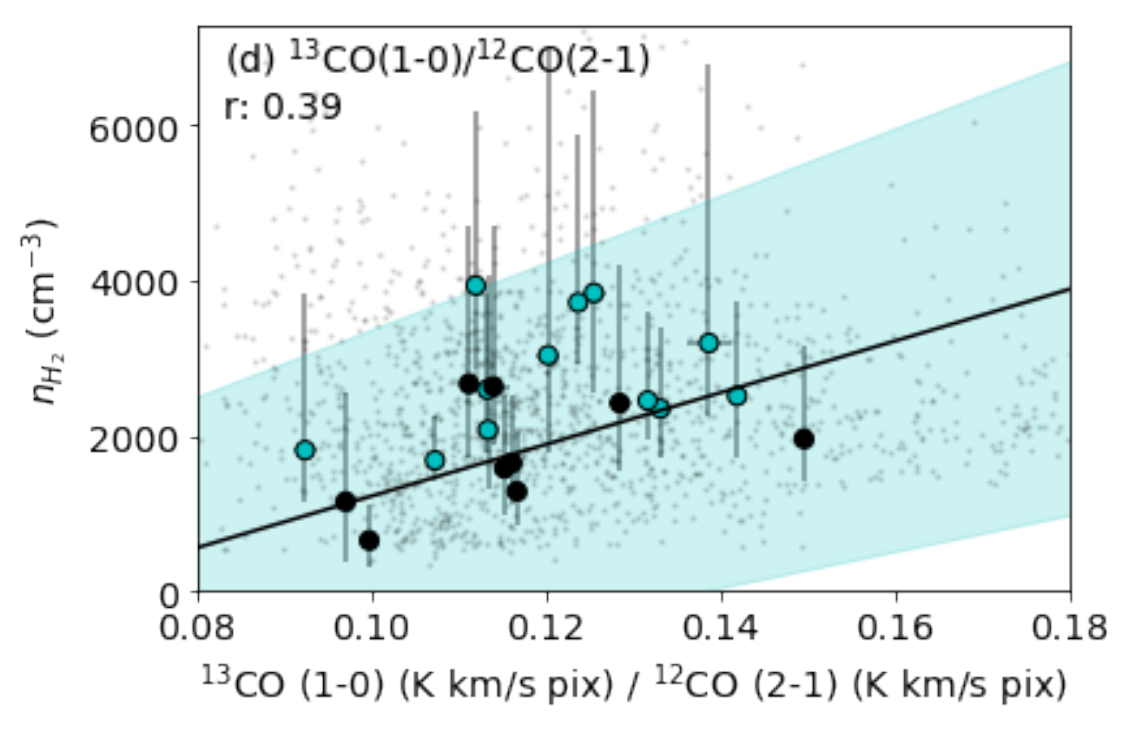}
    \caption{Volume density fitted by \radex plotted against ratios of \thirtCO\ to \twelveCO. The ratio with the strongest correlation is \thirtCO(2-1)/\twelveCO(1-0) in panel (c), with a residual variance of 0.18 and Pearson correlation coefficient of $r=0.81$. The ratios \thirtCO(1-0)/\twelveCO(1-0) and \thirtCO(2-1)/\twelveCO(2-1) both also show strong trends with residual variances of 0.32 and 0.24, respectively, and $r=0.69$ and $r=0.75$, respectively. The ratio of \thirtCO(1-0)/\twelveCO(2-1) shows very little trend with a residual variance of 0.57 and $r=0.39$. Plot symbols are the same as in Figure\,\ref{fig:XCO plots}. The blue points indicate clumps with at least one associated YSO, demonstrating that YSO presence does not appear to affect the trends at all.}
    \label{fig:n 13/12 trend}
\end{figure*}

Another possible source of scatter for \twelveCO\ could be the faint pixels that do not have corresponding \thirtCO\ emission. Since detections in three of the four lines were required before the fitting proceeded, there are several pixels that have \twelveCO(1-0) emission above 5$\sigma$ but no emission in two of the other lines. These are the faintest of the pixels and so would have correspondingly low \NCO, so including them would not make large changes to the reported $N_{H_2}$ or $W_\text{CO}$, but could account for some of the scatter.

We also show in both panels of Figure\,\ref{fig:XCO plots} the values of $N_{H_2}$ and $W_\text{CO}$ for individual pixels. In both cases, the pixels have shallower slopes than the clump-averaged values and fitting the pixels instead would result in a value of \XCO$ = (1.13\pm0.01)\times10^{20}$ cm$^{-2}$/(K km s$^{-1}$) and  \XthCO$ = (1.32\pm0.01)\times10^{21}$ cm$^{-2}$/(K km s$^{-1}$). For \XCO, the pixels show a stronger correlation between $W_\text{CO}$ and $N_{H_2}$ than the clump-averaged values, with a correlation coefficient of $r=0.74$. In contrast, the clump-averaged values for \XthCO have a stronger trend than the pixels, which have a correlation coefficient of $r=0.89$.
We would expect the clump-averaged values to show a stronger trend since the \XCO method works best when integrating over variations in physical conditions \citep{Bolatto13}.

\subsection{Diagnostic Line Ratios} \label{subsec: diagnostics}

\subsubsection{Isotopologues} \label{subsubsec: isotopologue ratios}

Ratios of isotopologues (e.g. \thirtCO/\twelveCO) can trace volume density in the case where one line is optically thick and the other line is sub-thermally excited \citep{Nishimura15}. We examine how well the ratios of the \thirtCO\ and \twelveCO\ lines' total fluxes in K km s$^{-1}$ predict the fitted \radex volume density, \nh. Figure\,\ref{fig:n 13/12 trend} shows that the strongest combination of lines to predict \nh is \thirtCO(2-1)/\twelveCO(1-0), with a residual variance of 0.18 and Pearson correlation coefficient of $r=0.81$. This makes sense since we expect \thirtCO(2-1) to be the most sub-thermally excited of the lines, especially from the LTE comparison in \S\ref{subsec: LTE}, and we expect \twelveCO(1-0) to be the most optically thick.

The weakest predictor of \nh is \thirtCO(1-0)/\twelveCO(2-1), with a residual variance of 0.57 and $r=0.39$. \thirtCO(1-0)/\twelveCO(1-0) and \thirtCO(2-1)/\twelveCO(2-1) have residual variances of 0.32 and 0.24, respectively, and $r=0.69$ and $r=0.75$, respectively. All of these combinations do show a positive trend, indicating that \thirtCO/\twelveCO\ is indeed a good diagnostic of gas density in this physical regime, where the \thirtCO\ is sub-thermally excited. In areas of higher temperature and density where all lines become thermalized, this relation would likely no longer hold. 

\begin{figure*}
    \centering
    \includegraphics[width=0.45\textwidth]{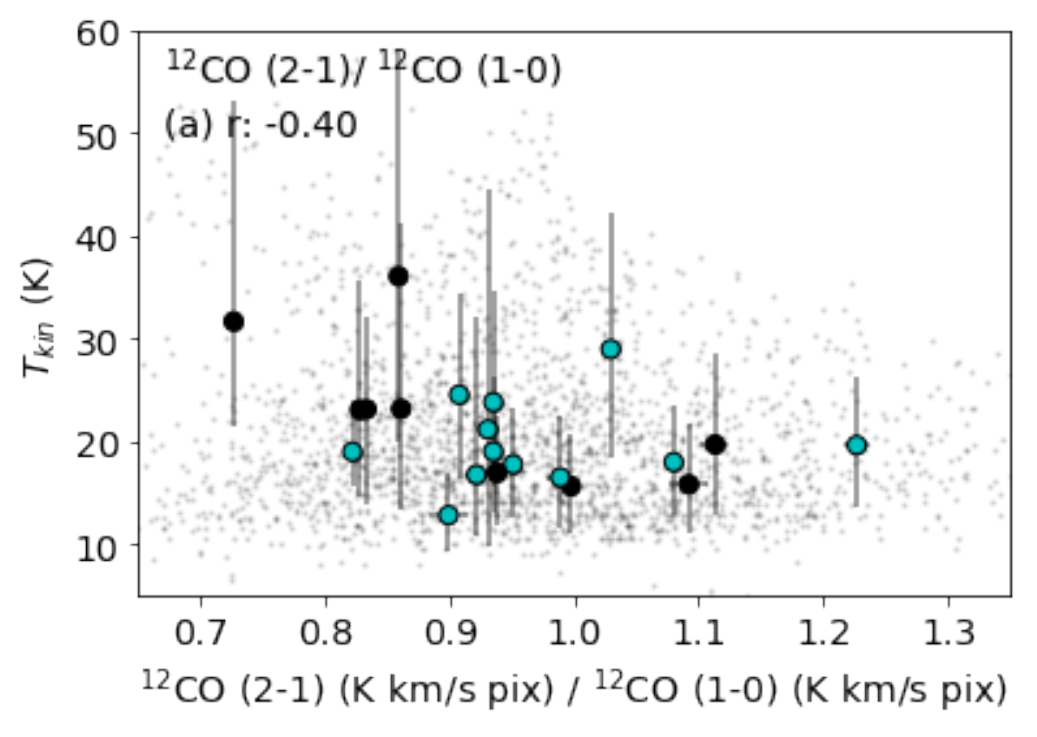}
    \includegraphics[width=0.45\textwidth]{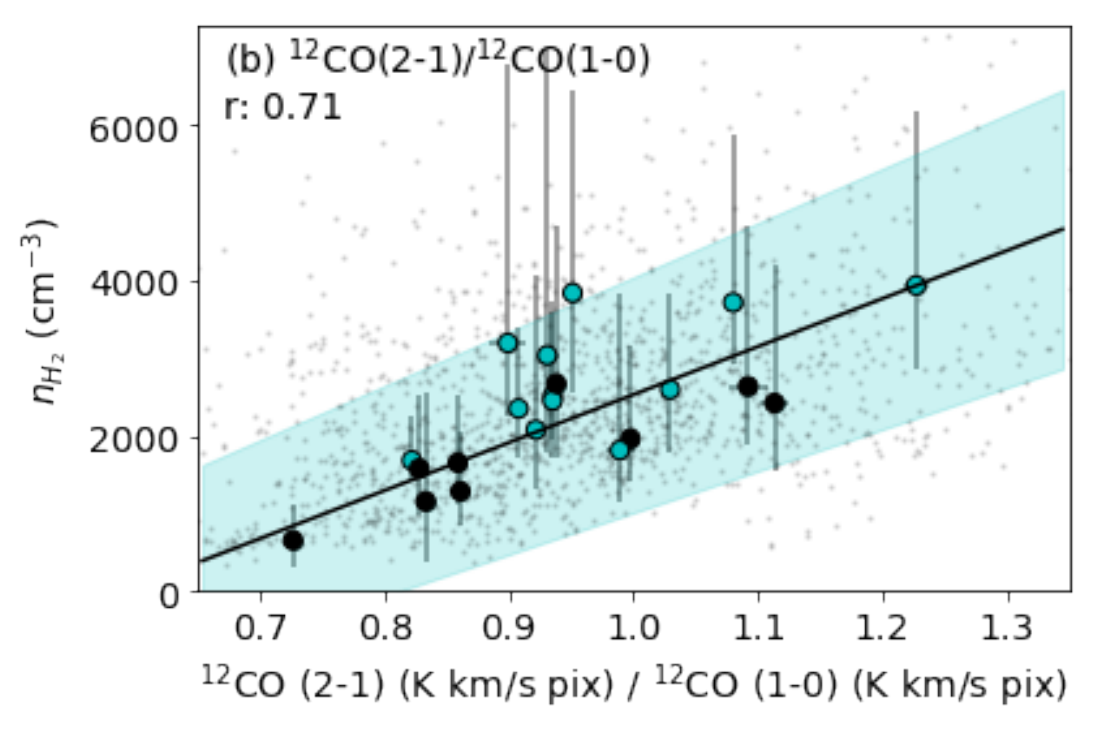}
    \includegraphics[width=0.45\textwidth]{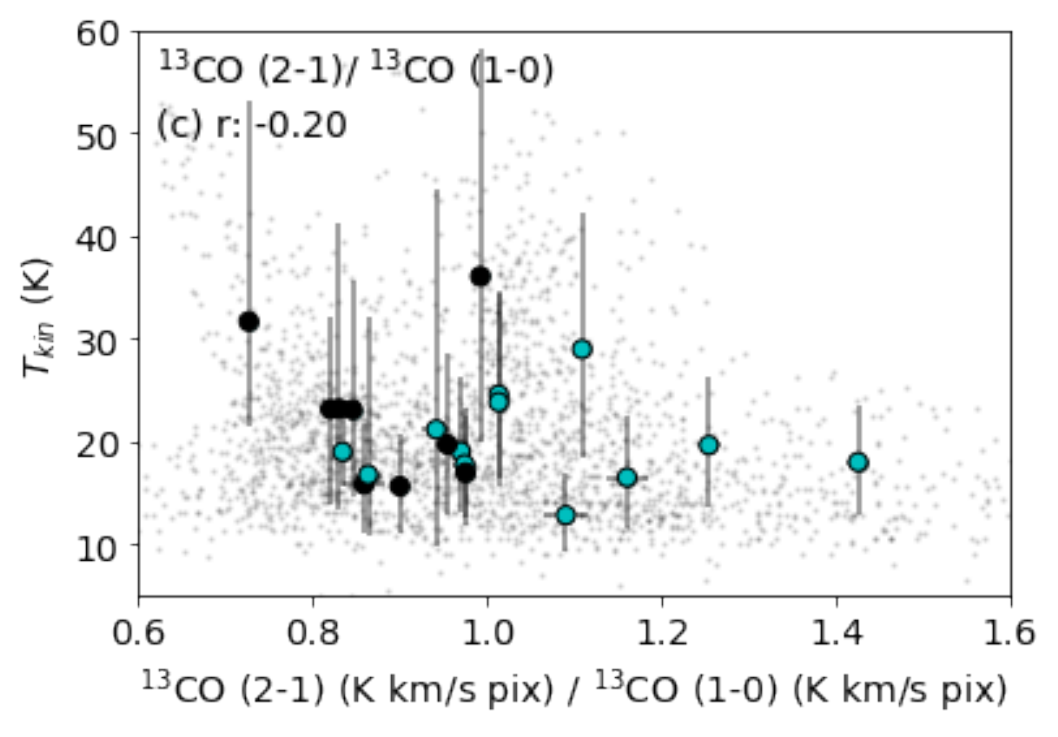}
    \includegraphics[width=0.45\textwidth]{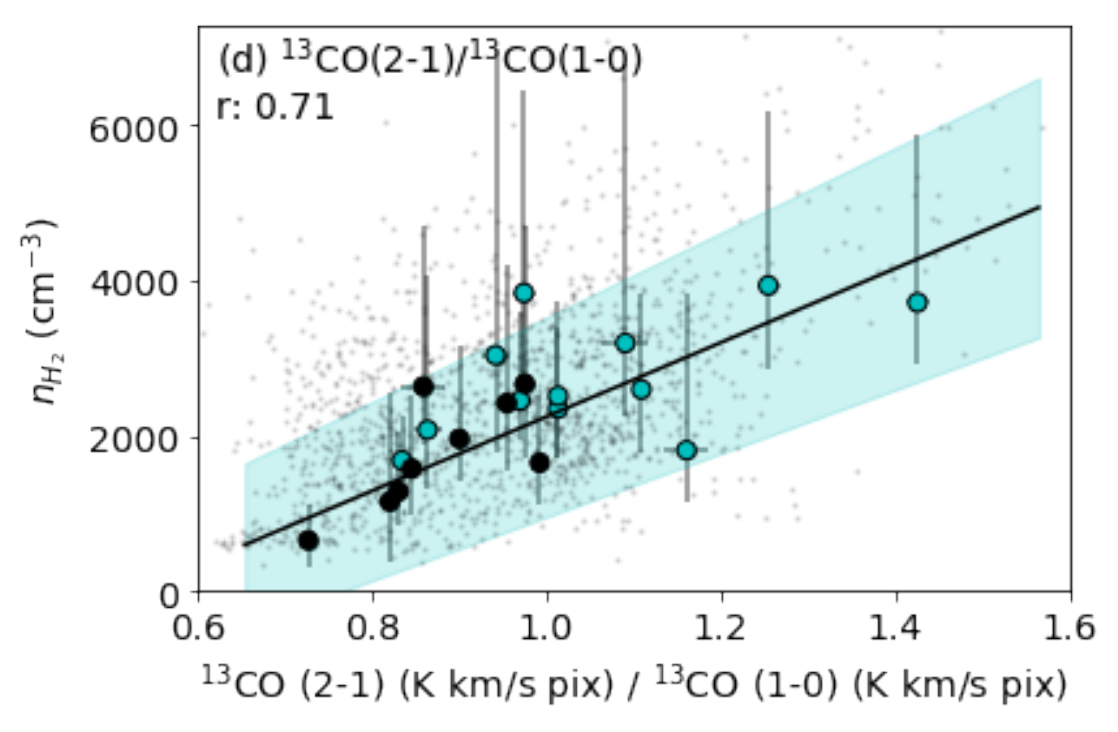}
    \caption{Kinetic temperature (left column) and \radex-fitted density (right column) against ratios of upper and lower transitions (\twelveCO(2-1)/\twelveCO(1-0) above and \thirtCO(2-1)/\thirtCO(1-0) below). There is not a strong correlation with \tkin for either \twelveCO\ or \thirtCO, suggesting that such line ratios are not a good diagnostic of kinetic temperature in this physical regime. They do, however, show a correlation with density, which has been fitted with a linear trend. The two ratios have a similar correlation, both with $r=0.71$ and residual variances of 0.22 and 0.25, respectively, though this is not as strong as the correlation of \nh with \thirtCO(2-1)/\twelveCO(1-0) (Figure\,\ref{fig:n 13/12 trend}c). Plot symbols are the same as in Figure\,\ref{fig:XCO plots}. Blue points indicate clumps with at least one associated YSO, which shows that YSO presence seems to have no affect on these trends, except that all clumps that have \thirtCO(2-1)/\thirtCO(1-0)$>1$ also have at least one associated YSO.}
    \label{fig:T 21/10 trend}
\end{figure*}

The fitted \nh has asymmetrical error bars, and when fitting the trends in Figure\,\ref{fig:n 13/12 trend}, we simply use the mean of the upper and lower errors. This results in overestimated errors \citep{Barlow04}, but a proper treatment of the errors would require a computationally rigorous analysis to properly account for the non-gaussian nature of the \nh probability distribution. We decided to report the simpler analysis with the acknowledgement that the errors are overestimated. Since the mean errors are usually lower than the upper errors and larger than the lower errors, the fit is also likely biased towards lower values of \nh. 

In all plots, we indicate the clumps that have at least one associated YSO as blue instead of black. Although YSOs could potentially affect the excitation, we do not see any indication that the presence of YSOs affects our ability to recover and understand that excitation, using different methods. 

We also show in Figure\,\ref{fig:n 13/12 trend} the values for individual pixels. In all panels, the correlations of the pixels is much weaker than that of the clump-averaged values, having correlation coefficients of 0.35, 0.43, 0.53, and 0.11, respectively, for panels (a) through (d). This indicates that these trends are most accurate when averaged over the whole clump.

\subsubsection{Excitation Levels} \label{subsubsec:excitation ratios}

Ratios of upper to lower excitation levels of CO (e.g. \twelveCO(2-1)/\twelveCO(1-0)) scale with excitation temperature and density when both lines are optically thin, and the ratio approaches unity as the lines get increasingly optically thick \citep{Sakamoto94,Nishimura15,Penaloza17}. Figure\,\ref{fig:T 21/10 trend} show the measured ratios of the total flux in K km s$^{-1}$ pix for \twelveCO(2-1)/\twelveCO(1-0) and for \thirtCO(2-1)/\thirtCO(1-0) plotted against the \radex-fitted \tkin and \nh.

The plots of \tkin against \twelveCO(2-1)/\twelveCO(1-0) and \thirtCO(2-1)/\thirtCO(1-0) in Figure\,\ref{fig:T 21/10 trend} do not show a strong linear trend ($r=-0.40$ and $r=-0.20$, respectively). This could be because the line ratios do not correlate with the excitation temperature, or because the excitation temperature is not correlated with the kinetic temperature in these clumps. We have seen in \S\ref{subsec: LTE} and \S\ref{subsubsec: isotopologue ratios} that the lines are likely sub-thermally excited, which would be consistent with the excitation temperature not tracing the kinetic temperature as well.

The ratios of excitation levels seem much more correlated with \nh than \tkin. For these, we were able to fit a linear relation as shown in Figure\,\ref{fig:T 21/10 trend}.  The two ratios have a similar correlation, both with $r=0.71$ and residual variances of 0.22 and 0.25, respectively, though neither are as good a tracer as the \thirtCO(2-1)/\twelveCO(1-0) relation from Figure\,\ref{fig:n 13/12 trend}c. 

The ratios of (2-1)/(1-0) shown in Figure\,\ref{fig:T 21/10 trend} are greater than unity for many of the clumps. This could happen if the lines are optically thin and the gas is hot, allowing the (2-1) lines to be excited. This is consistent with the results of the LTE mass estimate with (1-0) lines being an overestimate, which would require optically thin lines. The clumps with the high (2-1)/(1-0) ratios may have an embedded source of internal heating.

As in Figure\,\ref{fig:n 13/12 trend}, we indicate the clumps that have at least one associated YSO as blue instead of black. Again in these plots, the presence of YSOs seems to have no strong relation to the trends, except that all clumps that have \thirtCO(2-1)/\thirtCO(1-0)$>1$ do have at least one associated YSO. This is not the case for \twelveCO(2-1)/\twelveCO(1-0).

We also show in Figure\,\ref{fig:T 21/10 trend} the values for individual pixels. In all panels, the correlations of the pixels are much weaker than that of the clump-averaged values, having correlation coefficients of -0.15, 0.49, -0.01, and 0.52, respectively, for panels (a) through (d). This indicates that these trends are most accurate when averaged over the whole clump.

\subsubsection{Dense Gas Tracer, CS(2-1)} \label{subsubsec: CS ratios}

\begin{figure}
    \centering
    \includegraphics[width=0.45\textwidth]{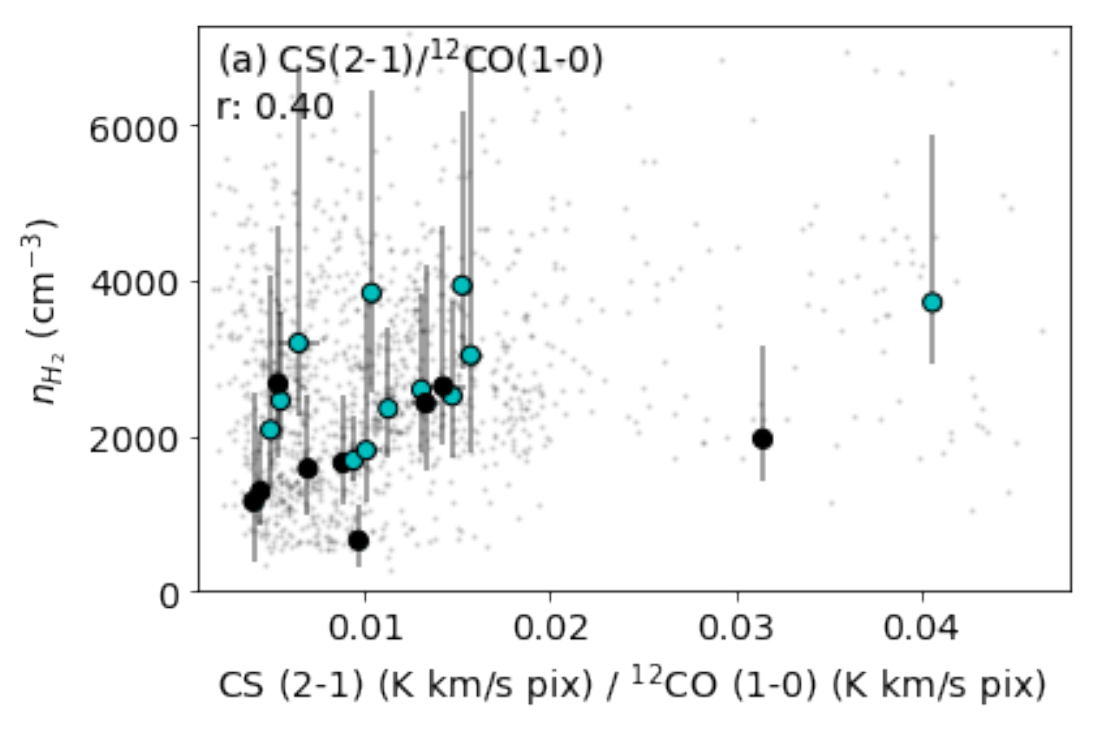}
    \includegraphics[width=0.45\textwidth]{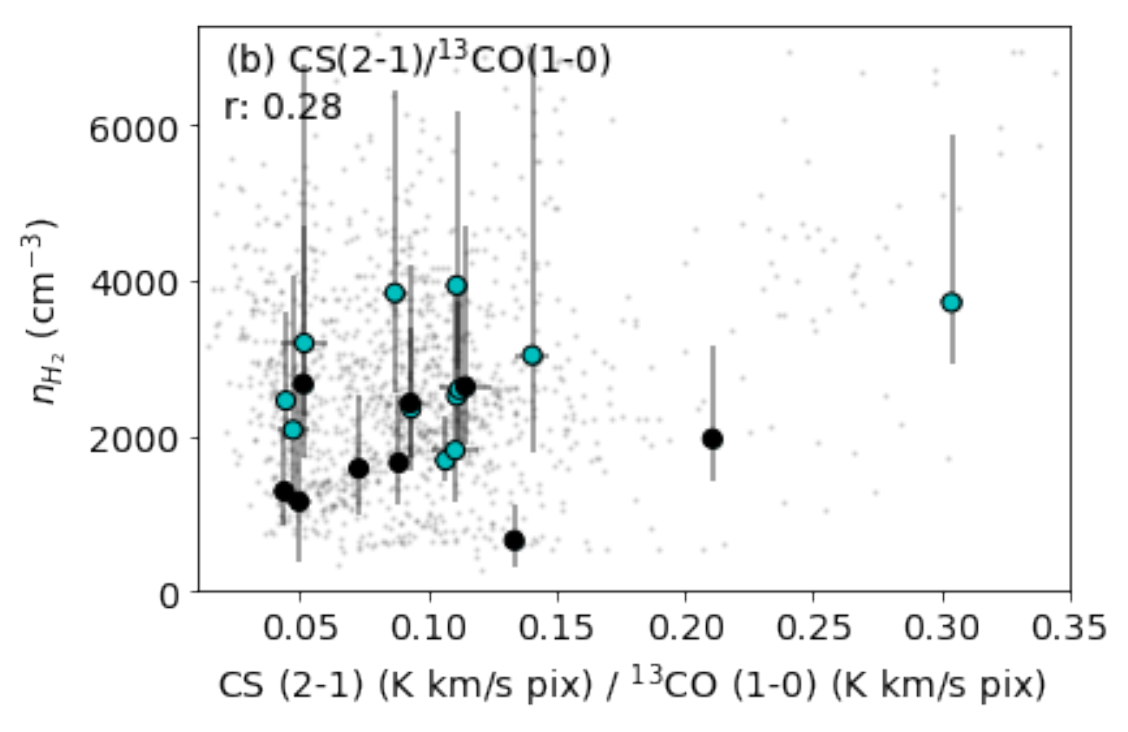}
    \caption{Density fitted by \radex against ratios of CS(2-1) to \twelveCO(1-0) \emph{(top)} and to \thirtCO(1-0) \emph{(bottom)}. CS(2-1)/\twelveCO(1-0) has a stronger correlation with density than CS(2-1)/\thirtCO(1-0) with  correlation coefficients of $r=0.40$ and $r=0.28$, respectively. The small black points show the values of individual pixels, which have slightly weaker trends than the clump-averaged values. Despite the high critical density of CS, neither of the ratios are as strong as most of the trends of density with CO line ratios shown in Figures\,\ref{fig:n 13/12 trend} and \ref{fig:T 21/10 trend}. This could be due to molecular abundance variations or because of the scales traced by the CO in the \radex fitting. Plot symbols are the same as in Figure\,\ref{fig:XCO plots}.}
    \label{fig:CS/CO}
\end{figure}

CS is a commonly observed dense gas tracer in molecular clouds with an optically thin critical density of $10^5$~cm$^{-3}$ at 20 K for CS(2-1) \citep{Shirley15}. We examine here how the ratios of CS(2-1) to CO correlate with the \radex-fitted density.

Figure\,\ref{fig:CS/CO} shows the ratio of the integrated intensities across a clump in units of K km s$^{-1}$ pix of CS(2-1) to \twelveCO(1-0) and CS(2-1) to \thirtCO(1-0). For both ratios, there is a weak linear trend with \nh, with correlation coefficients of $r=0.40$ and $r=0.28$ for the ratio with \twelveCO(1-0) and \thirtCO(1-0), respectively. This demonstrates that CS(2-1)/\twelveCO(1-0) may be a slightly better indicator of density than CS(2-1)/\thirtCO(1-0).

We also show in Figure\,\ref{fig:CS/CO} the values of individual pixels, which have slightly weaker correlations with $r=0.35$ and $r=0.26$, respectively. We also indicate clumps that have at least one associated YSO as blue instead of black circles, which shows that YSO presence appears to have no effect on these trends.

Despite the high critical density of CS, Figure\,\ref{fig:CS/CO} suggests that CS(2-1) cannot predict the density as strongly as most ratios of \thirtCO\ to \twelveCO\ shown in Figure\,\ref{fig:n 13/12 trend} or the ratios of excitation levels shown in Figure\,\ref{fig:T 21/10 trend}b and \ref{fig:T 21/10 trend}d. This could be because of variations in molecular abundance that influence the strength of the CS emission in addition to the density. This could also be because the \radex fitting is primarily tracing the CO density at scales of 1-2 pc, while the CS emission may be coming from more compact cores within the clumps. With higher resolution multiline observations and parameter fitting, we may begin to see more correlation between the fitted density and the CS emission.

\section{Trends with Star Formation} \label{sec: SF trends}

\subsection{Trend with Fitted Density} \label{subsec: density trend}

We looked for correlations between any of the derived quantities from \S\ref{subsec:derived props} and star formation. Our measures of star formation for a clump are the number of YSOs associated with it and their masses (\S\ref{sec:ysos}). We use a Pearson correlation coefficient, $r$, to evaluate how strong a relation there is between the two variables, though we acknowledge a linear trend may not best describe the expected relationship between the variables. For that reason, we are more interested in the relative values of the correlation coefficients than the absolute values.

\begin{figure}
    \centering
    \includegraphics[width=0.5\textwidth]{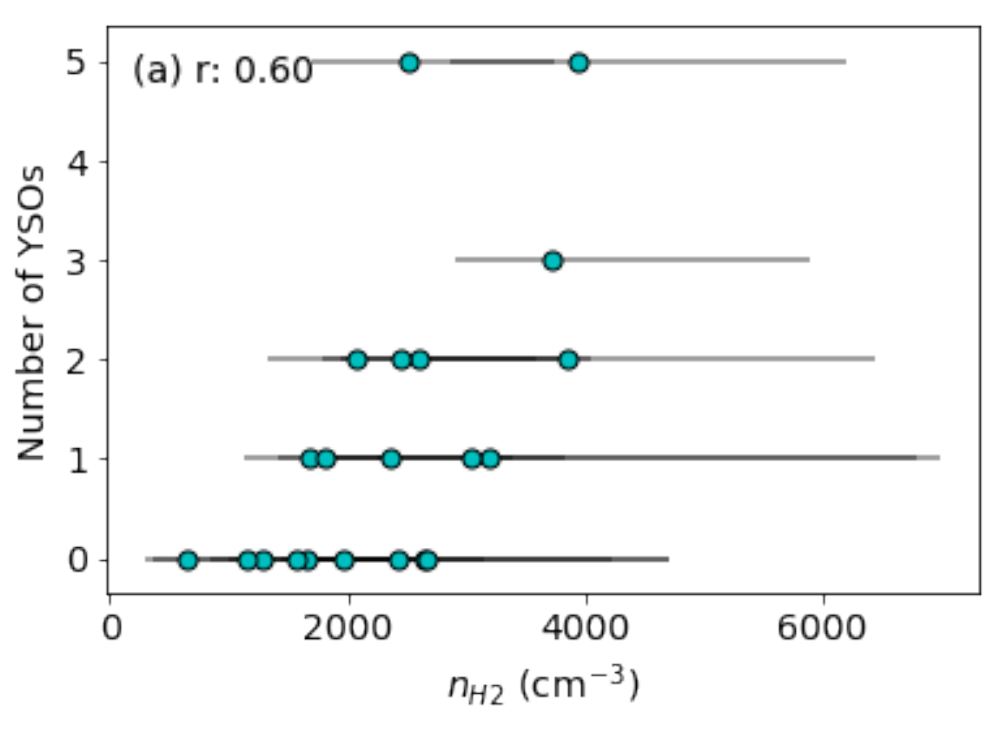}
    \includegraphics[width=0.5\textwidth]{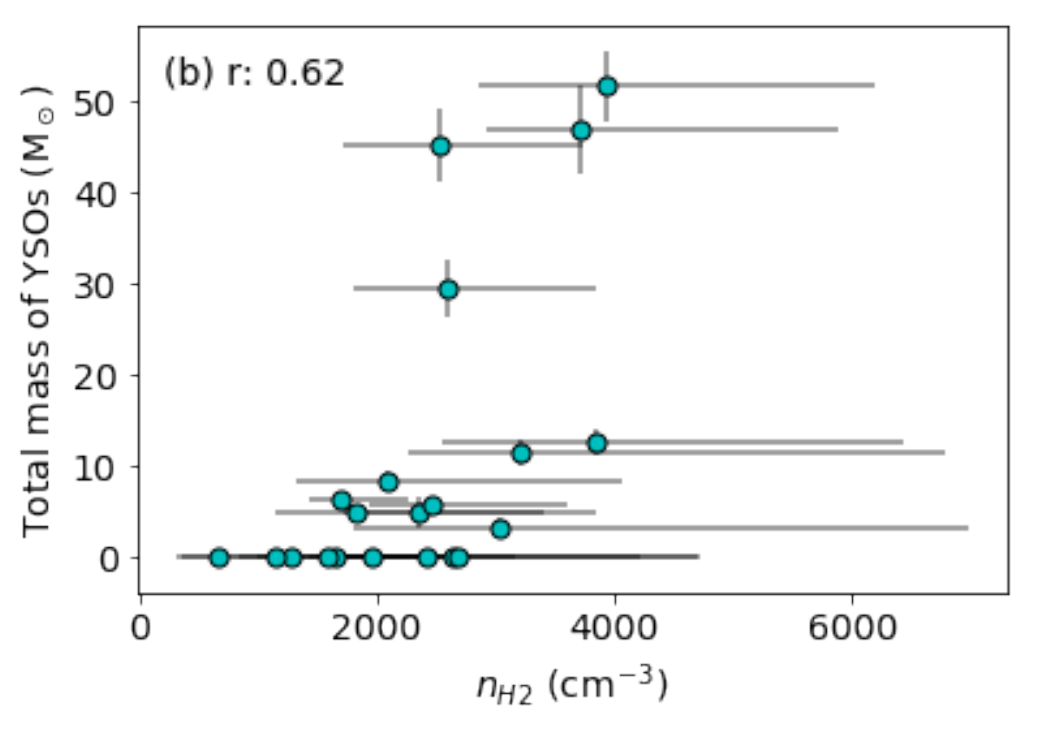}
    \includegraphics[width=0.5\textwidth]{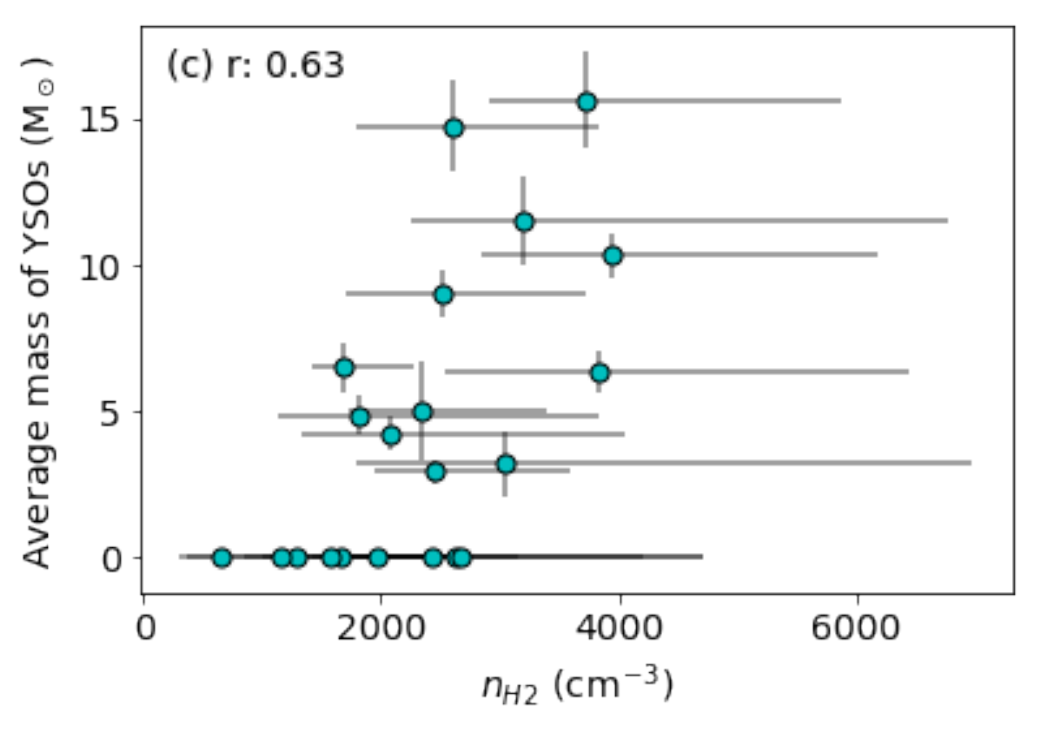}
    \caption{The number of YSOs associated with a given clump (\emph{top}), their total mass (\emph{middle}), and their average mass (\emph{bottom}) plotted against the \radex-fitted volume density, \nh. These three trends have correlation coefficients of 0.60, 0.62, and 0.63, respectively. The \radex-fitted volume density had the strongest correlation with YSO presence of any parameter we considered, including the properties derived in \S\ref{subsec:derived props} and several common star formation tracers. }
    \label{fig:n trends}
\end{figure}

By far the strongest correlation of YSO presence is with the mass-weighted \radex-fitted volume density, \nh. The density appears correlated with the number of YSOs associated with each clump ($r=0.60$), the total mass of all YSOs associated with the clump ($r=0.62$), and even the average mass of YSO associated with the clump ($r=0.63$). Figure\,\ref{fig:n trends} shows that clumps fit with a higher \nh have more associated YSOs, and more massive associated YSOs, and that clumps with lower values of \nh have no associated YSOs, or less massive YSOs. 

\subsection{Other Common Star Formation Tracers} \label{subsec: other SF tracers}

We investigated whether or not a threshold or trend appears in other, more commonly or easily observed properties: the mean volume density ($\Bar{n} = 3M / 4\pi R^3 \mu m_{H_2}$, corrected to match the filling factor scale of \nh by dividing $\Bar{n}$ by $f^{3/2}$), the ratio of the integrated intensities of CS(2-1) to \twelveCO(1-0), the gas surface density ($\Sigma = M/$Area), and the virial parameter, $\alpha_\text{vir}$. The gas surface density is directly proportional to $N_{H_2}$ and $A_V$, which are cited as showing star formation thresholds \citep[i.e.][]{Kennicutt98,Lada10}. The relation between these four parameters and the total mass of associated YSOs, as well as with the fitted density, \nh, are shown in Figure\,\ref{fig:other trends}.

\subsubsection{Mean Density} \label{subsubsec: SF mean density}

The mean density (Figures\,\ref{fig:other trends}a and \ref{fig:other trends}b) does not show much of a trend with the total mass of associated YSOs ($r=0.37$) and shows almost no trend with the fitted density ($r=0.06$). Figure\,\ref{fig:other trends}b shows a line indicating a one-to-one correlation between mean density and fitted density, and the fitted density is almost always larger than the mean density. This is expected since the mean density is an average over the whole clump, while the fitted density is a mass-weighted average. Since more of the clump's mass is in the denser regions, the fitted density is higher. The mass weighting also means the fitted density depends on the internal structure and density profile of the clump, while the mean density contains none of that information.

\subsubsection{Dense Gas Ratio} \label{subsubsec: SF dense gas}

The ratio of dense gas tracers with critical densities above $10^4$~cm$^{-3}$ (e.g. HCN, HCO$^+$, CS) to CO is often used as a tracer of the star formation rate in galaxies \citep{GaoSolomon04,Wang11,Zhang14,Li21}. In \S\ref{subsubsec: CS ratios}, we examined how ratios of CS(2-1) to CO correlated with \radex-fitted density and found that CS(2-1)/\twelveCO(1-0) shows a slightly stronger correlation with \nh than CS(2-1)/\thirtCO(1-0), though neither ratio is as good a predictor of \nh as most of the ratios between CO lines (\S\ref{subsubsec: isotopologue ratios} and \S\ref{subsubsec:excitation ratios}). 

In Figures\,\ref{fig:other trends}c and \ref{fig:other trends}d, we see that the ratio of CS(2-1)/\twelveCO(1-0) shows the strongest correlation with the \radex-fitted density of the star formation-tracing parameters ($r=0.40$) and an even stronger correlation with associated YSO presence ($r=0.47$), although this is still not as strong a correlation as with fitted density (Figure\,\ref{fig:n trends}). This measurement is difficult to relate to other observations since it cannot be directly compared with dense gas ratios of other common dense gas tracers, such as HCN or HCO$^+$. 

The two clumps in Figure\,\ref{fig:other trends}c and \ref{fig:other trends}d that have much larger CS(2-1)/\twelveCO(1-0) ratios than the rest are clumps 4 and 5, which have particularly bright CS(2-1) emission in Figure\,\ref{fig:moments}. It is not immediately clear what is causing this enhanced CS(2-1) emission in the region, since it does not appear particularly unique in any of the fitted physical conditions (see Figure\,\ref{fig:Ridge full maps}). The enhanced CS emission could be related to variations in the molecular abundance of CS across the region. It also could be that the region would appear more unique at higher resolutions where dense cores would be detected and not convolved with the surrounding more diffuse gas.

\subsubsection{Surface Density} \label{subsubsec: SF surface density}

Gas surface density, $\Sigma_\text{gas}$, is one of the most commonly used star formation threshold measures. It is correlated with star formation rate via the Kennicutt-Schmidt Relation on several scales \citep[see reviews in][]{Elmegreen18,KennicuttEvans12}. One of the most commonly cited star formation thresholds is $A_V > 8$ from \citet{Lada10}, which corresponds to $\Sigma_\text{gas} > 116$ M$_\odot$ pc$^{-2}$ in their sample of Milky Way clouds.  

In the LMC, \citet{Dobashi08} found a global relation of $\frac{A_V}{N_H} = 1.7\times10^{-22}$ mag / H cm$^2$, although this value varies from  $2.5\times10^{-22}$ mag / H cm$^2$ near 30 Dor, to  $0.63\times10^{-22}$ mag / H cm$^2$ near the outskirts of the LMC. Assuming that all the hydrogen in the Ridge is molecular, a threshold of $A_V>8$ corresponds to $N_{H_2} > 2.4\times10^{22}$ cm$^{-2}$, based on the global estimate, and this in turn corresponds to a gas surface density of $\Sigma_\text{gas} > 490$~M$_\odot$~pc$^{-2}$, with a lower limit of $\Sigma_\text{gas} > 330$~M$_\odot$~pc$^{-2}$ based on $\frac{A_V}{N_H}$ measured in the outskirts of the LMC. 

The threshold of $A_V>8$ from \citet{Lada10} was measured on $\sim0.1$ pc scales, so at 45\arcsec\ (11 pc) resolution, a threshold of 490~$M_\odot$~pc$^{-2}$ would correspond to $\sim4.5 M_\odot$ pc$^{-2}$ if all of the emission is coming from compact sources with no diffuse component. The smallest gas surface density measured for a clumps in the Ridge with an associated YSO is $\Sigma = 10$ M$_\odot$ pc$^{-2}$, and since a diffuse envelope is likely, this would be consistent with the $A_V > 8$ star formation threshold on smaller scales. 

The surface density shows the weakest correlation with the presence of YSOs ($r=0.06$) and weak, surprisingly negative correlation with fitted density as well ($-0.31$) (Figures\,\ref{fig:other trends}e and \ref{fig:other trends}f). There are clumps with high surface density and no associated YSOs, and clumps with low surface density in the Ridge that do have associated YSOs. 

\subsubsection{Virial Parameter} \label{subsubsec: SF virial}

\begin{figure*}
    \centering
    \includegraphics[width=0.38\textwidth]{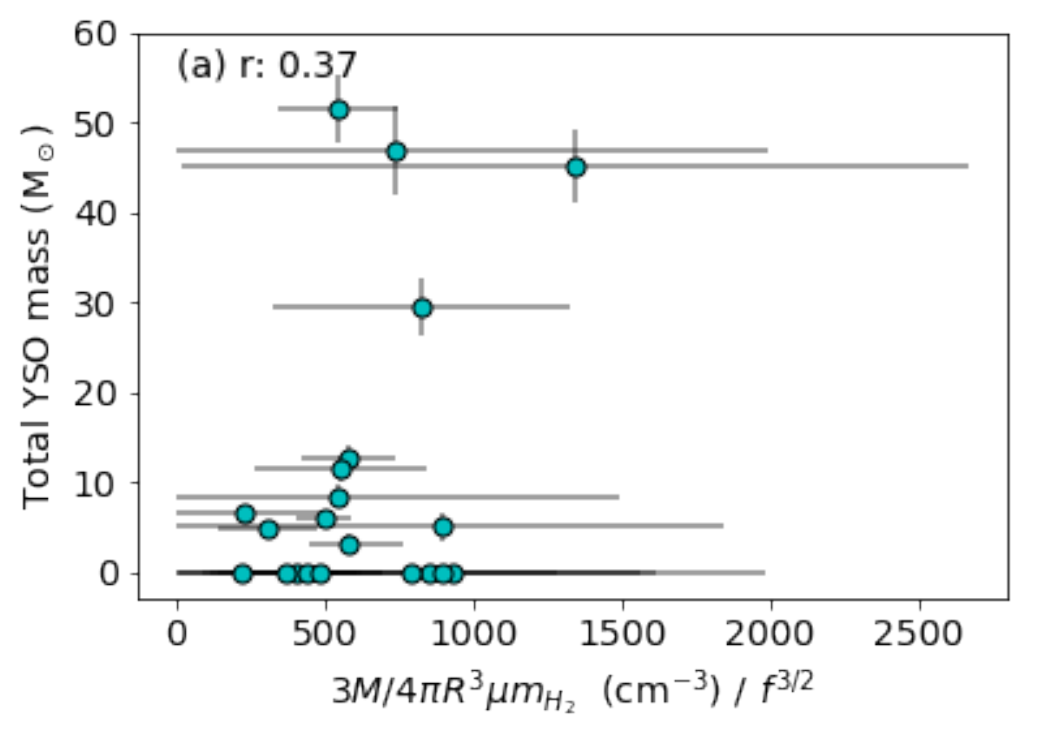}
    \includegraphics[width=0.4\textwidth]{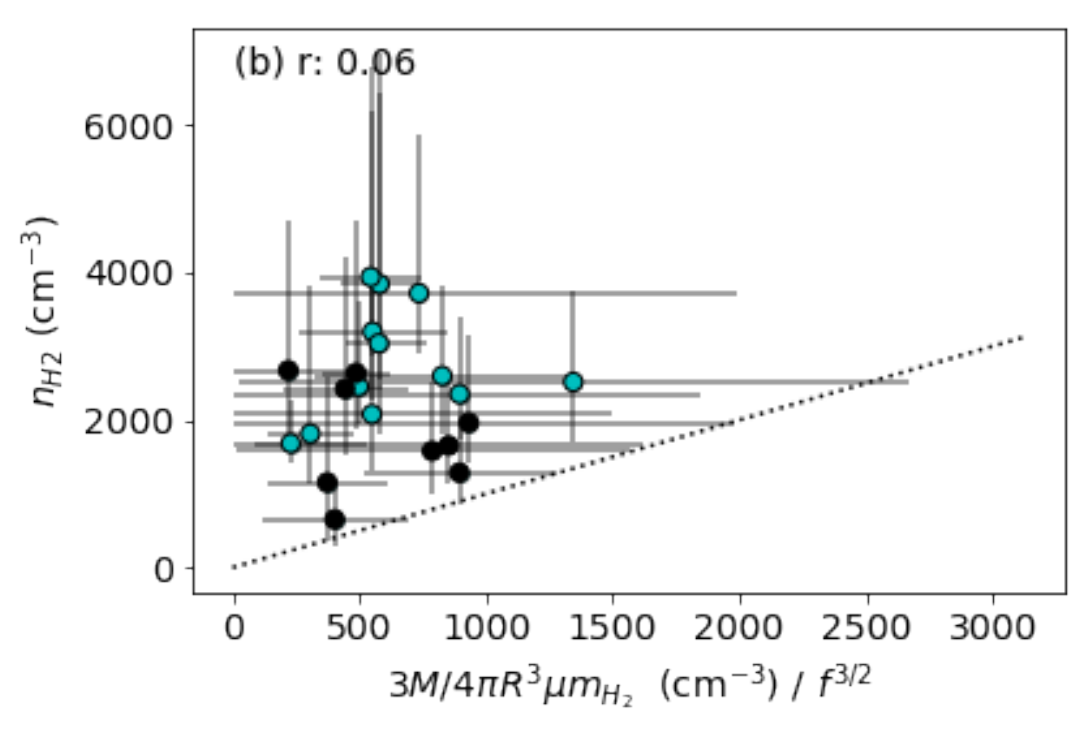}\\
    \includegraphics[width=0.38\textwidth]{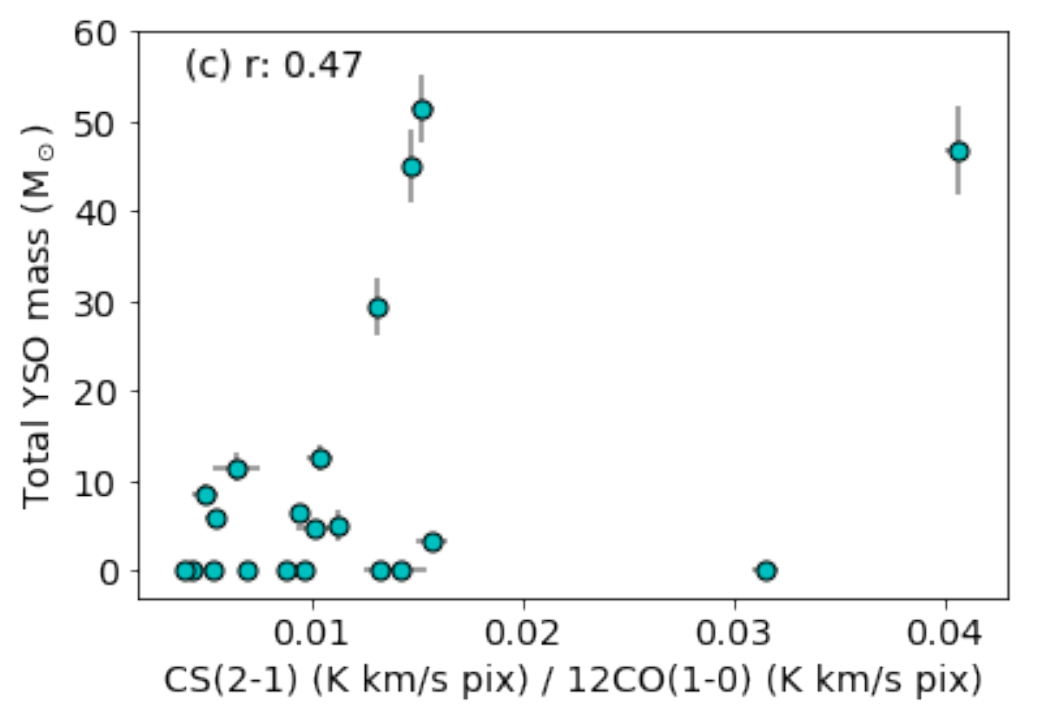}
    \includegraphics[width=0.4\textwidth]{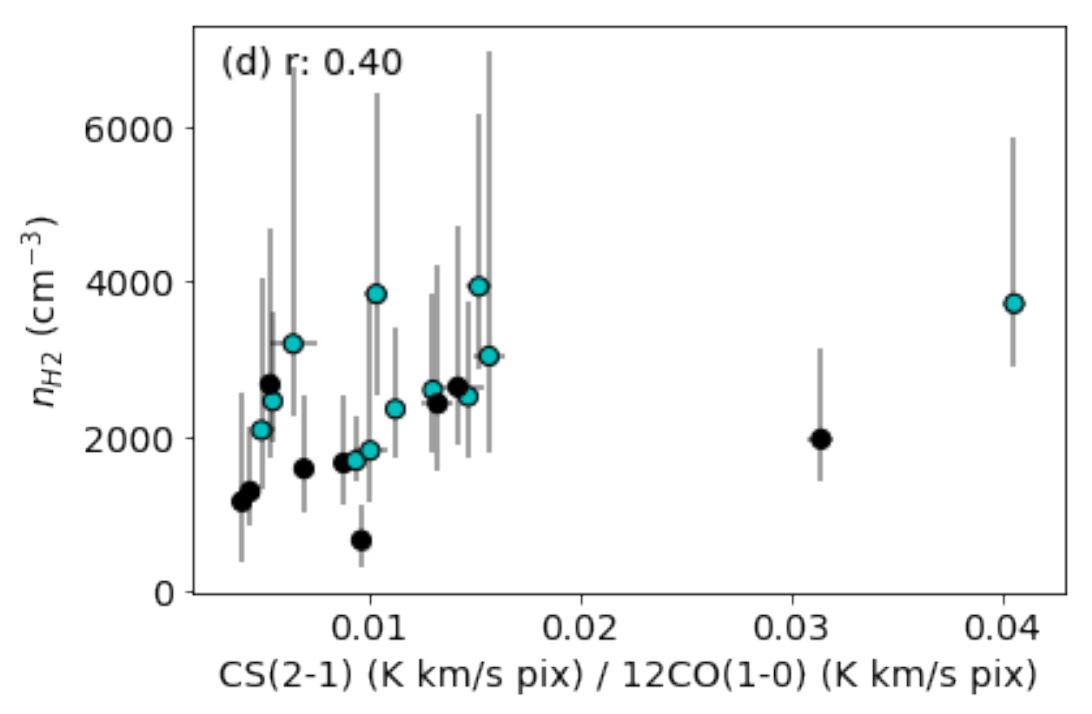}\\
    \includegraphics[width=0.38\textwidth]{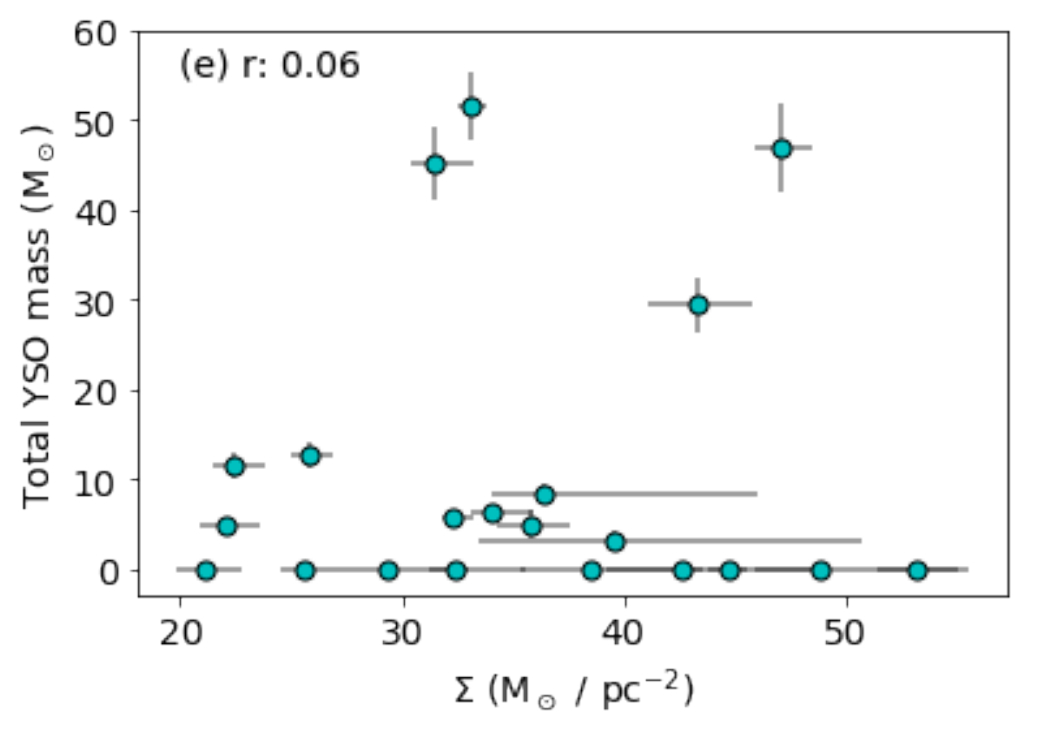}
    \includegraphics[width=0.4\textwidth]{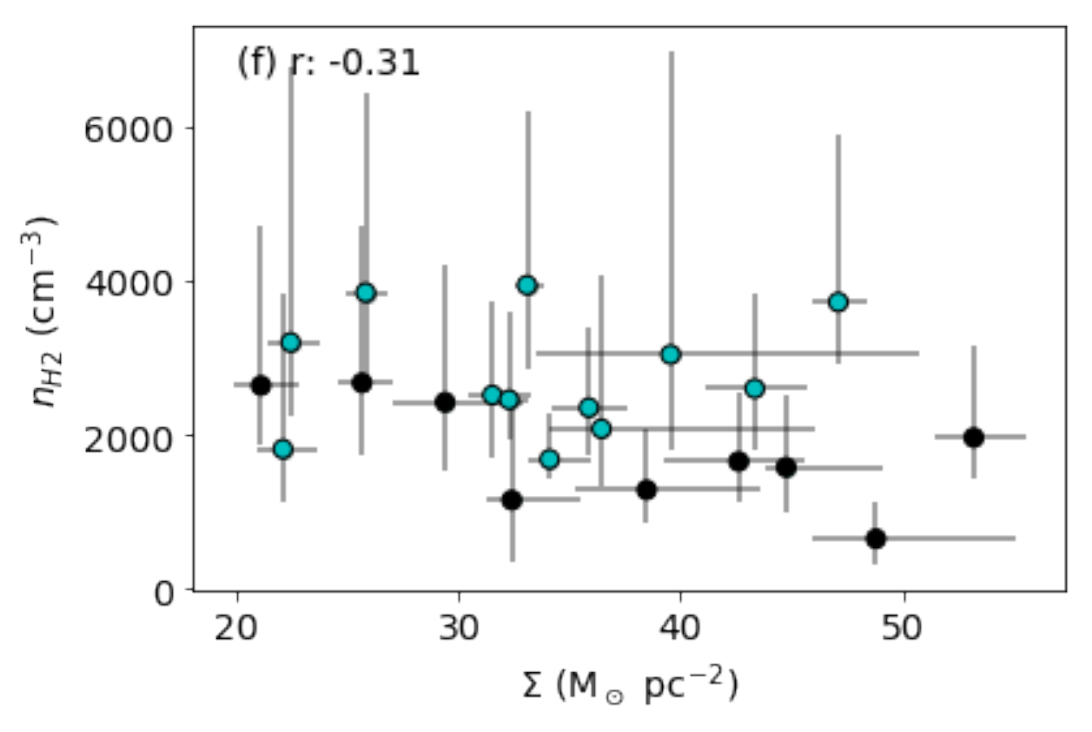}\\
    \includegraphics[width=0.38\textwidth]{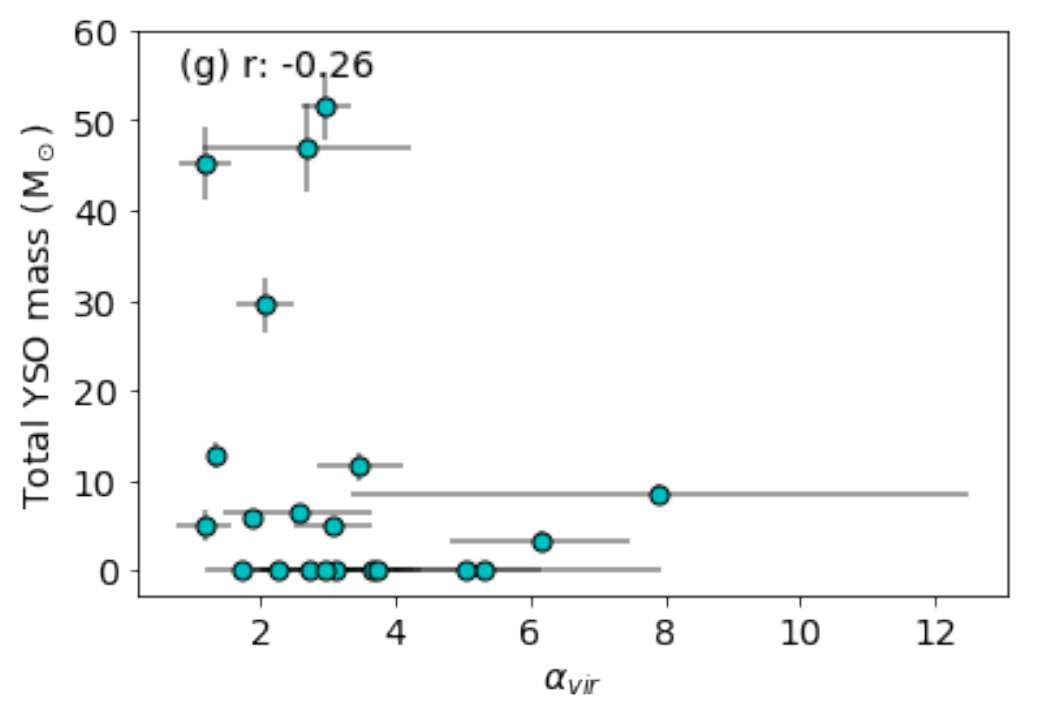}
    \includegraphics[width=0.4\textwidth]{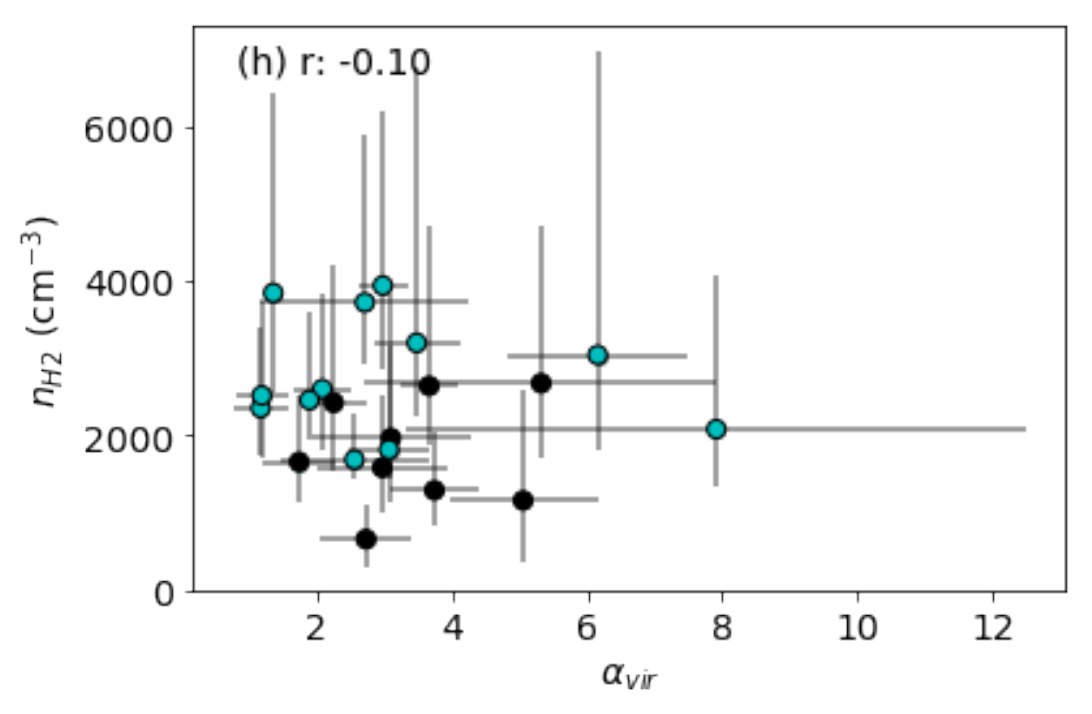}
    \caption{\emph{Left column: } The total mass of YSOs associated with a clump plotted against four parameters commonly used as indicators or thresholds for stars formation: the mean density, the dense gas fraction, the gas surface density, and the virial parameter $\alpha_\text{vir}$.  \emph{Right column: } The same four parameters, plotted against the fitted density, \nh. The density is the most correlated with total YSO mass, with a correlation coefficient of $r=0.62$ (Figure\,\ref{fig:n trends}). None of the other parameters show as strong a correlation, with CS(2-1)/\twelveCO(1-0) having the second-closest relation with $r=0.47$. In panel (b), the dotted line shows a one-to-one correlation between the fitted and mean densities. All clumps appear consistent with having a higher fitted density than mean density, as expected. The blue points indicate clumps that have at least one associated YSO, while the black points have no associated YSOs.}
    \label{fig:other trends}
\end{figure*}

The last metric we considered is the virial parameter, $\alpha_\text{vir}$, calculated as

\begin{equation} \label{eq:alphavir}
    \alpha_\text{vir} = \frac{5\sigma_v^2 R}{G M},
\end{equation}

where $R$ in this equation is 1.91 times the $\sigma_R$ reported in Table\,\ref{tab:derived quantities} to get an ``effective radius'' \citep{Solomon87}. This parameter indicates whether or not the clumps are in virial equilibrium. The correlation between $\alpha_\text{vir}$ and associated YSOs is weak, but negative as we would expect, since clumps with a high $\alpha_\text{vir}$ are less prone to collapse ($r=-0.26$; Figure\,\ref{fig:other trends}g). Clumps with a high $\alpha_\text{vir}$ have a few associated YSOs, but clumps with a low $\alpha_\text{vir}$ seem to have the most massive associated YSOs. Plotting $\alpha_\text{vir}$ against fitted density shows almost no trend ($r=-0.1$; Figure\,\ref{fig:other trends}h)). 

\makeatletter\onecolumngrid@push\makeatother
\clearpage
\makeatletter\onecolumngrid@pop\makeatother

\subsection{Discussion of Density Correlation} \label{subsec: densty discussion}

The correlation with \radex-fitted density and YSO presence could indicate a threshold density for star formation, which is commonly invoked in theories of star formation \citep[][and references therein]{Evans99}. Typically, density thresholds measured in the Milky Way are \nh$>10^4-10^5$ cm$^{-3}$ \citep{Evans99}, though these are often measured on scales of $\sim0.1$ pc. Since the densities in the Ridge are measured on scales of 4.5---9\arcsec\ (1-2 pc) depending on the filling factor, finding clumps with associated YSOs and densities of $\gtrsim10^3$ cm$^{-3}$ is not inconsistent with these Milky Way density threshold measurements. However, with a detection limit for YSOs around $\sim$2.5 M$_\odot$ and a trend with YSO mass, there could be lower mass, undetected YSOs associated with the lower density clumps as well. Also, the trends we see in Figure\,\ref{fig:n trends} appear more continuous, rather than the step-function that would be expected of a strict density threshold. 

To understand other star formation thresholds that have been observed, we need to relate the actual physical conditions of the gas to the observational measurements we use to describe them, which are usually only projections of those conditions. The strong correlation of our \radex-fitted density with the presence of YSOs may indicate that the fitting more directly measures the actual physical condition of the gas than any of the other parameters tested in Figure\,\ref{fig:other trends}. Unlike measurements of mean density or surface density, the fitting allows us to probe the conditions of the gas that is the source of the emission, without being as affected by line-of-sight effects or optical depth.

\citet{Khullar19} find in their simulations that while a high \nh is necessary for efficient star formation, typical star formation thresholds such as surface density do not actually correspond to that physical threshold. It seems like the virial parameter, $\alpha_\text{vir}$ ought to probe the actual physical conditions and be a good predictor of star formation and collapse; however, most measured $\alpha_\text{vir}$ use Equation\,\ref{eq:alphavir}, which depends on the total mass and estimations of the radius of clouds that are not symmetric or spherical. The resulting measurement is subject to the same projection effects and averaging as measuring the mean density instead of the fitted density. 

While the fitted density seems to be a better predictor of star formation than any of the other measurements, we cannot directly relate that to the conditions of the clouds when they actually formed the associated YSOs, or whether those are even the same clouds that we are observing now.  
Furthermore, there are many other factors that determine whether or not a cloud will form stars, and it seems unlikely that there is a single one-size-fits-all density threshold that guarantees the formation of stars, as discussed in \citet{Elmegreen18}. 

What we are seeing in Figure\,\ref{fig:n trends} is more likely an indicator of the local environment in the Ridge, and what density is required for the molecular gas to form stars above $\sim$2.5 M$_\odot$ when averaged over $\sim$1 pc. It would be interesting to test if the Ridge is forming fewer massive stars than regions to the north because the densities in the Ridge are lower, or if the threshold for forming stars is higher, making it more difficult to form stars than in other regions of the LMC.  If the Ridge is more turbulent or more magnetically supported, it could raise the density threshold compared to other regions. 

As shown in Figure\,\ref{fig:other trends}, we cannot accurately compare this fitted density by measuring the mean density, surface density, dense gas ratio, or $\alpha_\text{vir}$. However, as we saw in \S\ref{subsec: diagnostics}, ratios of \thirtCO\ to \twelveCO\ do show a strong correlation with the fitted gas density. We cannot easily say how those trends or specific numbers translate to other regions or size scales though, since the trend is likely dependent on the excitation and optical depth of the observed lines and a full non-LTE analysis like the one presented here would be necessary. To make a robust comparison of star formation in the Ridge to other massive-star-forming regions in the LMC, we would need to perform the full \radex fitting process described in this work in other regions of the LMC. Such a study could give insights into whether the Ridge has lower densities on average compared to those regions or whether stars in those regions are able to form at lower densities, suggesting that  the gas density required for star formation to occur may depend on galactic environment.

\citet{Indebetouw13} published fluxes of \twelveCO(2-1) and \thirtCO(2-1) at $<1$ pc resolution for 103 clumps they identified in the region of 30 Dor. The ratios of \thirtCO(2-1)/\twelveCO(2-1) for those clumps ranged from 0.05 to 0.46, where the clumps in the Ridge range from 0.06 to 0.14. Furthermore, \citet{Indebetouw13} flag six of those clumps as being associated with YSOs or clusters. These six clusters have \thirtCO(2-1)/\twelveCO(2-1) ratios that range from 0.14 to 0.22. These are higher than the ratios for most of the Ridge clumps, though they are not the highest ratios of the clumps measured in 30 Dor. If the numbers from the trend in Figure\,\ref{fig:n 13/12 trend} hold in 30 Dor and at smaller size scales, this could indicate that 30 Dor does indeed have higher densities than in the Ridge. To be sure how these two regions compare in densities though, we intend to perform our \radex fitting on observations at a similar resolution.

\section{Conclusions}\label{sec:conclusions}

We present new observations of the Molecular Ridge in the LMC, including \thirtCO(1-0) and CS(2-1) from ALMA at 16\arcsec\ resolution, and \twelveCO(2-1) and \thirtCO(2-1) from APEX at 30\arcsec\ resolution, as well as archival \twelveCO(1-0) from MAGMA at 45 \arcsec\ resolution. We analyzed these observations by fitting them to \radex models and assessed how well this fitting technique was able to recover \tkin, \NCO, and \nh from simulated line emission. The results are summarized below:

\begin{itemize}
    \item We were able to reliably recover \tkin, \nh, and \NCO from simulated line emission by using a combination of the 95\% confidence interval and 1.0 Bayesian interval. The performance of the fitting varied across the range of \tkin, \nh, and \NCO that we tested, and is dependent on the expected rms error in the line observations. We also determined that dropping one of the four lines---\twelveCO(1-0), which had the lowest resolution---would result in a significant loss of fitting sensitivity, especially in moderate values of \NCO. We found that it was unnecessary to account for sharp boundaries between clumps, as fitting Gaussian line profiles did not change the results significantly.
    
    
    
    \item LTE calculations from the (2-1) lines result in much lower clump masses than the \radex fitting, which implies that the lines are sub-thermally excited and the excitation temperatures of \twelveCO(2-1) and \thirtCO(2-1) are not equal. When calculating LTE masses from the (1-0) lines, the masses are higher than the \radex fitting. This could happen if the \twelveCO\ lines are actually optically thin rather than thick or if the adopted beam filling factor was too large, though the relative effects of the filling factor are nonlinear and difficult to predict.
    
    \item We calculated a value for \XCO in the Ridge based on the \radex fitted masses, getting \XCO $= 1.8\times10^{20}$ cm$^{-2}$ / (K km s$^{-1}$), which is lower than we would expect for the LMC with 1/3 solar metallicity. This could be because the non-LTE fitting is better tracing the molecular mass or because the abundance ratio of H$_2$/\thirtCO$=2\times10^6$ that we used was too low. We also calculated a value for \XthCO, where the total $N_{H_2}$ is related to the integrated flux of \thirtCO(1-0) rather than \twelveCO(1-0). We get \XthCO$= 1.6\times10^{21}$ cm$^{-2}$ / (K km s$^{-1}$), and the correlation is much tighter than it is with \twelveCO. This indicates that using \thirtCO(1-0) for mass estimates would be more accurate than using \twelveCO(1-0). 
    
    \item The ratio \thirtCO(2-1)/\twelveCO(1-0) had the tightest trend with the \radex-fitted gas density, \nh, though all ratios of \thirtCO\ to \twelveCO\ fluxes are diagnostic of the volume density to a lesser extent.
    Ratios of upper level transitions to lower (\twelveCO(2-1)/\twelveCO(1-0)) are also correlated with density and not kinetic temperature. These relations are likely due to the observed lines being sub-thermally excited, so the density of the gas is important for excitation and the excitation temperature is lower than the kinetic temperature. Six clumps had a \twelveCO(2-1)/\twelveCO(1-0) ratio greater than 1 in K km/s pix units, meaning that the \twelveCO\ may actually be optically thin and relatively hot in some clumps to allow \twelveCO(2-1) to be brighter than \twelveCO(1-0). The \thirtCO(2-1)/\thirtCO(1-0) ratios were above 1 for eight clumps, and all eight clumps have an associated YSO, which could be a source of heating to excite the (2-1) line. We also find that neither the ratio of CS(2-1)/\twelveCO(1-0) nor CS(2-1)/\thirtCO(1-0) show as strong a correlation with density as most of the \thirtCO/\twelveCO or (2-1)/(1-0) ratios, despite CS being commonly used as a dense gas tracer.
    
    \item  We find that no star formation parameter that can be calculated from simple mass estimates, like the mass based on an X-factor, showed a strong trend with star formation. Rather, the strongest predictor of the presence of YSOs associated with a clump was its \radex-fitted gas density, \nh. This fitted density is correlated with the number of associated YSOs, as well as the total and average mass of those YSOs. The simpler parameters we investigated were the mean density calculated from total mass and size, the ratio of the dense gas tracer CS(2-1) to \twelveCO(1-0), the surface density (which is directly related to $A_V$ and $N_{H_2}$), and the virial parameter, $\alpha_\text{vir}$. The correlation of \nh with YSO presence demonstrates that the \radex fitting may better probe the physical conditions of the gas on these scales, though the actual relationship between the fitted density and some critical ``threshold'' density required for star formation is uncertain.
    
    \item We hypothesize that the Molecular Ridge may not be actively forming massive stars as much as the northern regions either because its gas density is lower than those other regions or because it has a higher density threshold for stars to form. A higher density threshold could be due to turbulent or magnetic support for example. The results of this study show that the \radex-fitted volume density of the gas cannot be traced accurately by easily measured observables, such as surface density or a global mean density. While ratios of \thirtCO\ to \twelveCO\ are diagnostic of gas density, the scaling of this relationship is likely dependent on the local physical conditions and may not be accurate for determining relative gas density in other regions. To test these hypotheses, we will conduct a follow-up study of other active star-forming regions in the LMC with the \radex fitting method presented here. 
    
\end{itemize}

\begin{acknowledgements}

We thank the anonymous referee whose helpful comments improved this manuscript. This research is supported by NSF grants 1413231 and 1716335 (PI: K.~Johnson) and NSF AAG award AST 1312902 to U. Virginia (PI: R.~Indebetouw). This material is based upon work supported by the National Science Foundation Graduate Research Fellowship Program under Grant No. 1842490. Any opinions, findings, and conclusions or recommendations expressed in this material are those of the author(s) and do not necessarily reflect the views of the National Science Foundation. T. W. acknowledges support from the NSF through grant AST-2009849. This work was supported by NAOJ ALMA Scientific Research Grant Number 2016-03B.

This paper makes use of the following ALMA data: ADS/JAO.ALMA\#2012.1.00603.S., \\ADS/JAO.ALMA\#2015.1.00196.S, and \\ADS/JAO.ALMA\#2017.1.00271.S. ALMA is a partnership of ESO (representing its member states), NSF (USA) and NINS (Japan), together with NRC (Canada), NSC and ASIAA (Taiwan), and KASI (Republic of Korea), in cooperation with the Republic of Chile. The Joint ALMA Observatory is operated by ESO, AUI/NRAO and NAOJ. The National Radio Astronomy Observatory is a facility of the National Science Foundation operated under cooperative agreement by Associated Universities, Inc.

Based on observations with the Atacama Pathfinder EXperiment (APEX) telescope. APEX is a collaboration between the Max Planck Institute for Radio Astronomy, the European Southern Observatory, and the Onsala Space Observatory. Swedish observations on APEX are supported through Swedish Research Council grant No 2017-00648

\facility{APEX, ALMA}

\software{Pipeline-CASA51-P2-B v.40896 \citep{Davis21}, 
    CASA \citep[v.5.1.1-5, v.5.6.1; ][]{McMullin07}, 
    \radex \citep{radex}, 
    \texttt{quickclump} \citep{Sidorin17}, 
    Astropy \citep{astropy}, 
    Matplotlib \citep{matplotlib}, 
    NumPy \citep{numpy}, 
    SciPy \citep{scipy}
}

\end{acknowledgements}

\bibliographystyle{yahapj}
\bibliography{references.bib}

\begin{thebibliography}{}
\providecommand\natexlab[1]{#1}
\providecommand\JournalTitle[1]{#1}

\bibitem[{{Astropy Collaboration} {et~al.}(2013){Astropy Collaboration},
  {Robitaille}, {Tollerud}, {Greenfield}, {Droettboom}, {Bray}, {Aldcroft},
  {Davis}, {Ginsburg}, {Price-Whelan}, {Kerzendorf}, {Conley}, {Crighton},
  {Barbary}, {Muna}, {Ferguson}, {Grollier}, {Parikh}, {Nair}, {Unther},
  {Deil}, {Woillez}, {Conseil}, {Kramer}, {Turner}, {Singer}, {Fox}, {Weaver},
  {Zabalza}, {Edwards}, {Azalee Bostroem}, {Burke}, {Casey}, {Crawford},
  {Dencheva}, {Ely}, {Jenness}, {Labrie}, {Lim}, {Pierfederici}, {Pontzen},
  {Ptak}, {Refsdal}, {Servillat}, \& {Streicher}}]{astropy}
{Astropy Collaboration}, {Robitaille}, T.~P., {Tollerud}, E.~J., {et~al.} 2013,
  \href{http://dx.doi.org/10.1051/0004-6361/201322068}{\JournalTitle{\aap},
  558, A33}

\bibitem[{Barlow(2004)}]{Barlow04}
Barlow, R. 2004, Asymmetric Errors,
  \href{http://arxiv.org/abs/physics/0401042}{{\sffamily arXiv:physics/0401042
  [physics.data-an]}}

\bibitem[{{Bica} {et~al.}(1996){Bica}, {Claria}, {Dottori}, {Santos}, \&
  {Piatti}}]{Bica96}
{Bica}, E., {Claria}, J.~J., {Dottori}, H., {Santos}, J.~F.~C., J., \&
  {Piatti}, A.~E. 1996,
  \href{http://dx.doi.org/10.1086/192251}{\JournalTitle{\apjs}, 102, 57}

\bibitem[{{Bolatto} {et~al.}(2013){Bolatto}, {Wolfire}, \& {Leroy}}]{Bolatto13}
{Bolatto}, A.~D., {Wolfire}, M., \& {Leroy}, A.~K. 2013,
  \href{http://dx.doi.org/10.1146/annurev-astro-082812-140944}{\JournalTitle{\araa},
  51, 207}

\bibitem[{{Brand} \& {Wouterloot}(1995)}]{Brand95}
{Brand}, J., \& {Wouterloot}, J.~G.~A. 1995, \JournalTitle{\aap}, 303, 851

\bibitem[{{Bressan} {et~al.}(2012){Bressan}, {Marigo}, {Girardi}, {Salasnich},
  {Dal Cero}, {Rubele}, \& {Nanni}}]{Bressan12}
{Bressan}, A., {Marigo}, P., {Girardi}, L., {et~al.} 2012,
  \href{http://dx.doi.org/10.1111/j.1365-2966.2012.21948.x}{\JournalTitle{\mnras},
  427, 127}

\bibitem[{{Calzetti} {et~al.}(2007){Calzetti}, {Kennicutt}, {Engelbracht},
  {Leitherer}, {Draine}, {Kewley}, {Moustakas}, {Sosey}, {Dale}, {Gordon},
  {Helou}, {Hollenbach}, {Armus}, {Bendo}, {Bot}, {Buckalew}, {Jarrett}, {Li},
  {Meyer}, {Murphy}, {Prescott}, {Regan}, {Rieke}, {Roussel}, {Sheth}, {Smith},
  {Thornley}, \& {Walter}}]{Calzetti07}
{Calzetti}, D., {Kennicutt}, R.~C., {Engelbracht}, C.~W., {et~al.} 2007,
  \href{http://dx.doi.org/10.1086/520082}{\JournalTitle{\apj}, 666, 870}

\bibitem[{{Carlson} {et~al.}(2012){Carlson}, {Sewi{\l}o}, {Meixner}, {Romita},
  \& {Lawton}}]{Carlson12}
{Carlson}, L.~R., {Sewi{\l}o}, M., {Meixner}, M., {Romita}, K.~A., \& {Lawton},
  B. 2012,
  \href{http://dx.doi.org/10.1051/0004-6361/201118627}{\JournalTitle{\aap},
  542, A66}

\bibitem[{{Castets} {et~al.}(1990){Castets}, {Duvert}, {Dutrey}, {Bally},
  {Langer}, \& {Wilson}}]{Castets90}
{Castets}, A., {Duvert}, G., {Dutrey}, A., {et~al.} 1990, \JournalTitle{\aap},
  234, 469

\bibitem[{{Chen} {et~al.}(2010){Chen}, {Indebetouw}, {Chu}, {Gruendl},
  {Testor}, {Heitsch}, {Seale}, {Meixner}, \& {Sewilo}}]{Chen10}
{Chen}, C.~H.~R., {Indebetouw}, R., {Chu}, Y.-H., {et~al.} 2010,
  \href{http://dx.doi.org/10.1088/0004-637X/721/2/1206}{\JournalTitle{\apj},
  721, 1206}

\bibitem[{{Cohen} {et~al.}(1988){Cohen}, {Dame}, {Garay}, {Montani}, {Rubio},
  \& {Thaddeus}}]{Cohen88}
{Cohen}, R.~S., {Dame}, T.~M., {Garay}, G., {et~al.} 1988,
  \href{http://dx.doi.org/10.1086/185243}{\JournalTitle{\apjl}, 331, L95}

\bibitem[{{Davies} {et~al.}(1976){Davies}, {Elliott}, \& {Meaburn}}]{Davies76}
{Davies}, R.~D., {Elliott}, K.~H., \& {Meaburn}, J. 1976,
  \JournalTitle{\memras}, 81, 89

\bibitem[{{Davis}(2021)}]{Davis21}
{Davis}. 2021, \JournalTitle{in prep}

\bibitem[{{Dobashi} {et~al.}(2008){Dobashi}, {Bernard}, {Hughes}, {Paradis},
  {Reach}, \& {Kawamura}}]{Dobashi08}
{Dobashi}, K., {Bernard}, J.~P., {Hughes}, A., {et~al.} 2008,
  \href{http://dx.doi.org/10.1051/0004-6361:20079151}{\JournalTitle{\aap}, 484,
  205}

\bibitem[{{Elmegreen}(2018)}]{Elmegreen18}
{Elmegreen}, B.~G. 2018,
  \href{http://dx.doi.org/10.3847/1538-4357/aaa770}{\JournalTitle{\apj}, 854,
  16}

\bibitem[{{Evans}(1999)}]{Evans99}
{Evans}, Neal~J., I. 1999,
  \href{http://dx.doi.org/10.1146/annurev.astro.37.1.311}{\JournalTitle{\araa},
  37, 311}

\bibitem[{{Fukui} {et~al.}(2008){Fukui}, {Kawamura}, {Minamidani}, {Mizuno},
  {Kanai}, {Mizuno}, {Onishi}, {Yonekura}, {Mizuno}, {Ogawa}, \&
  {Rubio}}]{NANTEN}
{Fukui}, Y., {Kawamura}, A., {Minamidani}, T., {et~al.} 2008,
  \href{http://dx.doi.org/10.1086/589833}{\JournalTitle{\apjs}, 178, 56}

\bibitem[{{Gao} \& {Solomon}(2004)}]{GaoSolomon04}
{Gao}, Y., \& {Solomon}, P.~M. 2004,
  \href{http://dx.doi.org/10.1086/382999}{\JournalTitle{\apj}, 606, 271}

\bibitem[{{Gruendl} \& {Chu}(2009)}]{GC09}
{Gruendl}, R.~A., \& {Chu}, Y.-H. 2009,
  \href{http://dx.doi.org/10.1088/0067-0049/184/1/172}{\JournalTitle{\apjs},
  184, 172}

\bibitem[{Harris {et~al.}(2020)Harris, Millman, van~der Walt, Gommers,
  Virtanen, Cournapeau, Wieser, Taylor, Berg, Smith, Kern, Picus, Hoyer, van
  Kerkwijk, Brett, Haldane, del R{'{\i}}o, Wiebe, Peterson,
  G{'{e}}rard-Marchant, Sheppard, Reddy, Weckesser, Abbasi, Gohlke, \&
  Oliphant}]{numpy}
Harris, C.~R., Millman, K.~J., van~der Walt, S.~J., {et~al.} 2020,
  \href{http://dx.doi.org/10.1038/s41586-020-2649-2}{\JournalTitle{Nature},
  585, 357}

\bibitem[{{Heikkil{\"a}} {et~al.}(1999){Heikkil{\"a}}, {Johansson}, \&
  {Olofsson}}]{Heikkila99}
{Heikkil{\"a}}, A., {Johansson}, L.~E.~B., \& {Olofsson}, H. 1999,
  \JournalTitle{\aap}, 344, 817

\bibitem[{{Henize}(1956)}]{Henize56}
{Henize}, K.~G. 1956,
  \href{http://dx.doi.org/10.1086/190025}{\JournalTitle{\apjs}, 2, 315}

\bibitem[{{Heyer} \& {Dame}(2015)}]{Heyer15}
{Heyer}, M., \& {Dame}, T.~M. 2015,
  \href{http://dx.doi.org/10.1146/annurev-astro-082214-122324}{\JournalTitle{\araa},
  53, 583}

\bibitem[{{Heyer} {et~al.}(2001){Heyer}, {Carpenter}, \& {Snell}}]{Heyer01}
{Heyer}, M.~H., {Carpenter}, J.~M., \& {Snell}, R.~L. 2001,
  \href{http://dx.doi.org/10.1086/320218}{\JournalTitle{\apj}, 551, 852}

\bibitem[{{Hughes} {et~al.}(2010){Hughes}, {Wong}, {Ott}, {Muller}, {Pineda},
  {Mizuno}, {Bernard}, {Paradis}, {Maddison}, {Reach}, {Staveley-Smith},
  {Kawamura}, {Meixner}, {Kim}, {Onishi}, {Mizuno}, \& {Fukui}}]{Hughes10}
{Hughes}, A., {Wong}, T., {Ott}, J., {et~al.} 2010,
  \href{http://dx.doi.org/10.1111/j.1365-2966.2010.16829.x}{\JournalTitle{\mnras},
  406, 2065}

\bibitem[{{Hunter}(2007)}]{matplotlib}
{Hunter}, J.~D. 2007,
  \href{http://dx.doi.org/10.1109/MCSE.2007.55}{\JournalTitle{Computing in
  Science Engineering}, 9, 90}

\bibitem[{{Indebetouw} {et~al.}(2020){Indebetouw}, {Wong}, {Chen}, {Kepley},
  {Lebouteiller}, {Madden}, \& {Oliveira}}]{Indebetouw20}
{Indebetouw}, R., {Wong}, T., {Chen}, C. H.~R., {et~al.} 2020,
  \href{http://dx.doi.org/10.3847/1538-4357/ab5db7}{\JournalTitle{\apj}, 888,
  56}

\bibitem[{{Indebetouw} {et~al.}(2008){Indebetouw}, {Whitney}, {Kawamura},
  {Onishi}, {Meixner}, {Meade}, {Babler}, {Hora}, {Gordon}, {Engelbracht},
  {Block}, \& {Misselt}}]{Indebetouw08}
{Indebetouw}, R., {Whitney}, B.~A., {Kawamura}, A., {et~al.} 2008,
  \href{http://dx.doi.org/10.1088/0004-6256/136/4/1442}{\JournalTitle{\aj},
  136, 1442}

\bibitem[{{Indebetouw} {et~al.}(2013){Indebetouw}, {Brogan}, {Chen}, {Leroy},
  {Johnson}, {Muller}, {Madden}, {Cormier}, {Galliano}, {Hughes}, {Hunter},
  {Kawamura}, {Kepley}, {Lebouteiller}, {Meixner}, {Oliveira}, {Onishi}, \&
  {Vasyunina}}]{Indebetouw13}
{Indebetouw}, R., {Brogan}, C., {Chen}, C. H.~R., {et~al.} 2013,
  \href{http://dx.doi.org/10.1088/0004-637X/774/1/73}{\JournalTitle{\apj}, 774,
  73}

\bibitem[{{Kennicutt}(1998)}]{Kennicutt98}
{Kennicutt}, Robert~C., J. 1998,
  \href{http://dx.doi.org/10.1086/305588}{\JournalTitle{\apj}, 498, 541}

\bibitem[{{Kennicutt} \& {Evans}(2012)}]{KennicuttEvans12}
{Kennicutt}, R.~C., \& {Evans}, N.~J. 2012,
  \href{http://dx.doi.org/10.1146/annurev-astro-081811-125610}{\JournalTitle{\araa},
  50, 531}

\bibitem[{{Khullar} {et~al.}(2019){Khullar}, {Krumholz}, {Federrath}, \&
  {Cunningham}}]{Khullar19}
{Khullar}, S., {Krumholz}, M.~R., {Federrath}, C., \& {Cunningham}, A.~J. 2019,
  \href{http://dx.doi.org/10.1093/mnras/stz1800}{\JournalTitle{\mnras}, 488,
  1407}

\bibitem[{{Koeppen} \& {Kegel}(1980)}]{KoeppenKegel80}
{Koeppen}, J., \& {Kegel}, W.~H. 1980, \JournalTitle{\aaps}, 42, 59

\bibitem[{{Lada} {et~al.}(2010){Lada}, {Lombardi}, \& {Alves}}]{Lada10}
{Lada}, C.~J., {Lombardi}, M., \& {Alves}, J.~F. 2010,
  \href{http://dx.doi.org/10.1088/0004-637X/724/1/687}{\JournalTitle{\apj},
  724, 687}

\bibitem[{{Li} {et~al.}(2021){Li}, {Wang}, {Gao}, {Liu}, {Zhang}, {Li}, {Gong},
  {Li}, \& {Shi}}]{Li21}
{Li}, F., {Wang}, J., {Gao}, F., {et~al.} 2021,
  \href{http://dx.doi.org/10.1093/mnras/stab745}{\JournalTitle{\mnras}},
  \href{http://arxiv.org/abs/2103.04126}{{\sffamily arXiv:2103.04126
  [astro-ph.GA]}}

\bibitem[{{Longmore} {et~al.}(2013){Longmore}, {Bally}, {Testi}, {Purcell},
  {Walsh}, {Bressert}, {Pestalozzi}, {Molinari}, {Ott}, {Cortese}, {Battersby},
  {Murray}, {Lee}, {Kruijssen}, {Schisano}, \& {Elia}}]{Longmore13}
{Longmore}, S.~N., {Bally}, J., {Testi}, L., {et~al.} 2013,
  \href{http://dx.doi.org/10.1093/mnras/sts376}{\JournalTitle{\mnras}, 429,
  987}

\bibitem[{{Maddalena} \& {Thaddeus}(1985)}]{Maddalena85}
{Maddalena}, R.~J., \& {Thaddeus}, P. 1985,
  \href{http://dx.doi.org/10.1086/163291}{\JournalTitle{\apj}, 294, 231}

\bibitem[{{Mangum} \& {Shirley}(2015)}]{MangumShirley15}
{Mangum}, J.~G., \& {Shirley}, Y.~L. 2015,
  \href{http://dx.doi.org/10.1086/680323}{\JournalTitle{\pasp}, 127, 266}

\bibitem[{{McMullin} {et~al.}(2007){McMullin}, {Waters}, {Schiebel}, {Young},
  \& {Golap}}]{McMullin07}
{McMullin}, J.~P., {Waters}, B., {Schiebel}, D., {Young}, W., \& {Golap}, K.
  2007, in Astronomical Society of the Pacific Conference Series, Vol. 376,
  Astronomical Data Analysis Software and Systems XVI, ed. R.~A. {Shaw},
  F.~{Hill}, \& D.~J. {Bell}, 127

\bibitem[{{Meixner} {et~al.}(2006){Meixner}, {Gordon}, {Indebetouw}, {Hora},
  {Whitney}, {Blum}, {Reach}, {Bernard}, {Meade}, {Babler}, {Engelbracht},
  {For}, {Misselt}, {Vijh}, {Leitherer}, {Cohen}, {Churchwell}, {Boulanger},
  {Frogel}, {Fukui}, {Gallagher}, {Gorjian}, {Harris}, {Kelly}, {Kawamura},
  {Kim}, {Latter}, {Madden}, {Markwick-Kemper}, {Mizuno}, {Mizuno}, {Mould},
  {Nota}, {Oey}, {Olsen}, {Onishi}, {Paladini}, {Panagia}, {Perez-Gonzalez},
  {Shibai}, {Sato}, {Smith}, {Staveley-Smith}, {Tielens}, {Ueta}, {van Dyk},
  {Volk}, {Werner}, \& {Zaritsky}}]{Meixner06}
{Meixner}, M., {Gordon}, K.~D., {Indebetouw}, R., {et~al.} 2006,
  \href{http://dx.doi.org/10.1086/508185}{\JournalTitle{\aj}, 132, 2268}

\bibitem[{{Meixner} {et~al.}(2013){Meixner}, {Panuzzo}, {Roman-Duval},
  {Engelbracht}, {Babler}, {Seale}, {Hony}, {Montiel}, {Sauvage}, {Gordon},
  {Misselt}, {Okumura}, {Chanial}, {Beck}, {Bernard}, {Bolatto}, {Bot},
  {Boyer}, {Carlson}, {Clayton}, {Chen}, {Cormier}, {Fukui}, {Galametz},
  {Galliano}, {Hora}, {Hughes}, {Indebetouw}, {Israel}, {Kawamura}, {Kemper},
  {Kim}, {Kwon}, {Lebouteiller}, {Li}, {Long}, {Madden}, {Matsuura}, {Muller},
  {Oliveira}, {Onishi}, {Otsuka}, {Paradis}, {Poglitsch}, {Reach},
  {Robitaille}, {Rubio}, {Sargent}, {Sewi{\l}o}, {Skibba}, {Smith},
  {Srinivasan}, {Tielens}, {van Loon}, \& {Whitney}}]{Meixner13}
{Meixner}, M., {Panuzzo}, P., {Roman-Duval}, J., {et~al.} 2013,
  \href{http://dx.doi.org/10.1088/0004-6256/146/3/62}{\JournalTitle{\aj}, 146,
  62}

\bibitem[{{Mizuno} {et~al.}(2001){Mizuno}, {Yamaguchi}, {Mizuno}, {Rubio},
  {Abe}, {Saito}, {Onishi}, {Yonekura}, {Yamaguchi}, {Ogawa}, \&
  {Fukui}}]{Mizuno01}
{Mizuno}, N., {Yamaguchi}, R., {Mizuno}, A., {et~al.} 2001,
  \href{http://dx.doi.org/10.1093/pasj/53.6.971}{\JournalTitle{\pasj}, 53, 971}

\bibitem[{{Nikoli{\'c}} {et~al.}(2007){Nikoli{\'c}}, {Garay}, {Rubio}, \&
  {Johansson}}]{Nikolic07}
{Nikoli{\'c}}, S., {Garay}, G., {Rubio}, M., \& {Johansson}, L.~E.~B. 2007,
  \href{http://dx.doi.org/10.1051/0004-6361:20067034}{\JournalTitle{\aap}, 471,
  561}

\bibitem[{{Nishimura} {et~al.}(2015){Nishimura}, {Tokuda}, {Kimura}, {Muraoka},
  {Maezawa}, {Ogawa}, {Dobashi}, {Shimoikura}, {Mizuno}, {Fukui}, \&
  {Onishi}}]{Nishimura15}
{Nishimura}, A., {Tokuda}, K., {Kimura}, K., {et~al.} 2015,
  \href{http://dx.doi.org/10.1088/0067-0049/216/1/18}{\JournalTitle{\apjs},
  216, 18}

\bibitem[{{Padoan} {et~al.}(2000){Padoan}, {Juvela}, {Bally}, \&
  {Nordlund}}]{Padoan00}
{Padoan}, P., {Juvela}, M., {Bally}, J., \& {Nordlund}, {\r{A}}. 2000,
  \href{http://dx.doi.org/10.1086/308229}{\JournalTitle{\apj}, 529, 259}

\bibitem[{{Pe{\~n}aloza} {et~al.}(2017){Pe{\~n}aloza}, {Clark}, {Glover},
  {Shetty}, \& {Klessen}}]{Penaloza17}
{Pe{\~n}aloza}, C.~H., {Clark}, P.~C., {Glover}, S. C.~O., {Shetty}, R., \&
  {Klessen}, R.~S. 2017,
  \href{http://dx.doi.org/10.1093/mnras/stw2892}{\JournalTitle{\mnras}, 465,
  2277}

\bibitem[{{Robitaille}(2017)}]{Robitaille17}
{Robitaille}, T.~P. 2017,
  \href{http://dx.doi.org/10.1051/0004-6361/201425486}{\JournalTitle{\aap},
  600, A11}

\bibitem[{{Robitaille} {et~al.}(2007){Robitaille}, {Whitney}, {Indebetouw}, \&
  {Wood}}]{Robitaille07}
{Robitaille}, T.~P., {Whitney}, B.~A., {Indebetouw}, R., \& {Wood}, K. 2007,
  \href{http://dx.doi.org/10.1086/512039}{\JournalTitle{\apjs}, 169, 328}

\bibitem[{{Sakamoto}(1994)}]{Sakamoto94}
{Sakamoto}, S. 1994,
  \href{http://dx.doi.org/10.1086/133486}{\JournalTitle{\pasp}, 106, 1112}

\bibitem[{{Schaefer}(2008)}]{Schaefer08}
{Schaefer}, B.~E. 2008,
  \href{http://dx.doi.org/10.1088/0004-6256/135/1/112}{\JournalTitle{\aj}, 135,
  112}

\bibitem[{{Seale} {et~al.}(2014){Seale}, {Meixner}, {Sewi{\l}o}, {Babler},
  {Engelbracht}, {Gordon}, {Hony}, {Misselt}, {Montiel}, {Okumura}, {Panuzzo},
  {Roman-Duval}, {Sauvage}, {Boyer}, {Chen}, {Indebetouw}, {Matsuura},
  {Oliveira}, {Srinivasan}, {van Loon}, {Whitney}, \& {Woods}}]{Seale14}
{Seale}, J.~P., {Meixner}, M., {Sewi{\l}o}, M., {et~al.} 2014,
  \href{http://dx.doi.org/10.1088/0004-6256/148/6/124}{\JournalTitle{\aj}, 148,
  124}

\bibitem[{{Shirley}(2015)}]{Shirley15}
{Shirley}, Y.~L. 2015,
  \href{http://dx.doi.org/10.1086/680342}{\JournalTitle{\pasp}, 127, 299}

\bibitem[{{Sidorin}(2017)}]{Sidorin17}
{Sidorin}, V. 2017, {Quickclump: Identify clumps within a 3D FITS datacube}

\bibitem[{{Skrutskie} {et~al.}(2006){Skrutskie}, {Cutri}, {Stiening},
  {Weinberg}, {Schneider}, {Carpenter}, {Beichman}, {Capps}, {Chester},
  {Elias}, {Huchra}, {Liebert}, {Lonsdale}, {Monet}, {Price}, {Seitzer},
  {Jarrett}, {Kirkpatrick}, {Gizis}, {Howard}, {Evans}, {Fowler}, {Fullmer},
  {Hurt}, {Light}, {Kopan}, {Marsh}, {McCallon}, {Tam}, {Van Dyk}, \&
  {Wheelock}}]{Skrutskie06}
{Skrutskie}, M.~F., {Cutri}, R.~M., {Stiening}, R., {et~al.} 2006,
  \href{http://dx.doi.org/10.1086/498708}{\JournalTitle{\aj}, 131, 1163}

\bibitem[{{Smith} \& {MCELS Team}(1998)}]{SmithMCELS98}
{Smith}, R.~C., \& {MCELS Team}. 1998,
  \href{http://dx.doi.org/10.1071/AS98163}{\JournalTitle{\pasa}, 15, 163}

\bibitem[{{Solomon} {et~al.}(1987){Solomon}, {Rivolo}, {Barrett}, \&
  {Yahil}}]{Solomon87}
{Solomon}, P.~M., {Rivolo}, A.~R., {Barrett}, J., \& {Yahil}, A. 1987,
  \href{http://dx.doi.org/10.1086/165493}{\JournalTitle{\apj}, 319, 730}

\bibitem[{{Trotta}(2008)}]{Trotta08}
{Trotta}, R. 2008,
  \href{http://dx.doi.org/10.1080/00107510802066753}{\JournalTitle{Contemporary
  Physics}, 49, 71}

\bibitem[{{Ulrich}(1976)}]{Ulrich76}
{Ulrich}, R.~K. 1976,
  \href{http://dx.doi.org/10.1086/154840}{\JournalTitle{\apj}, 210, 377}

\bibitem[{{van der Tak} {et~al.}(2007){van der Tak}, {Black}, {Sch{\"o}ier},
  {Jansen}, \& {van Dishoeck}}]{radex}
{van der Tak}, F.~F.~S., {Black}, J.~H., {Sch{\"o}ier}, F.~L., {Jansen}, D.~J.,
  \& {van Dishoeck}, E.~F. 2007,
  \href{http://dx.doi.org/10.1051/0004-6361:20066820}{\JournalTitle{\aap}, 468,
  627}

\bibitem[{Virtanen {et~al.}(2020)Virtanen, Gommers, Oliphant, Haberland, Reddy,
  Cournapeau, Burovski, Peterson, Weckesser, Bright, {van der Walt}, Brett,
  Wilson, Millman, Mayorov, Nelson, Jones, Kern, Larson, Carey, Polat, Feng,
  Moore, {VanderPlas}, Laxalde, Perktold, Cimrman, Henriksen, Quintero, Harris,
  Archibald, Ribeiro, Pedregosa, {van Mulbregt}, \& {SciPy 1.0
  Contributors}}]{scipy}
Virtanen, P., Gommers, R., Oliphant, T.~E., {et~al.} 2020,
  \href{http://dx.doi.org/10.1038/s41592-019-0686-2}{\JournalTitle{Nature
  Methods}, 17, 261}

\bibitem[{{Wang} {et~al.}(2011){Wang}, {Zhang}, \& {Shi}}]{Wang11}
{Wang}, J., {Zhang}, Z., \& {Shi}, Y. 2011,
  \href{http://dx.doi.org/10.1111/j.1745-3933.2011.01090.x}{\JournalTitle{\mnras},
  416, L21}

\bibitem[{{Whitney} {et~al.}(2008){Whitney}, {Sewilo}, {Indebetouw},
  {Robitaille}, {Meixner}, {Gordon}, {Meade}, {Babler}, {Harris}, {Hora},
  {Bracker}, {Povich}, {Churchwell}, {Engelbracht}, {For}, {Block}, {Misselt},
  {Vijh}, {Leitherer}, {Kawamura}, {Blum}, {Cohen}, {Fukui}, {Mizuno},
  {Mizuno}, {Srinivasan}, {Tielens}, {Volk}, {Bernard}, {Boulanger}, {Frogel},
  {Gallagher}, {Gorjian}, {Kelly}, {Latter}, {Madden}, {Kemper}, {Mould},
  {Nota}, {Oey}, {Olsen}, {Onishi}, {Paladini}, {Panagia}, {Perez-Gonzalez},
  {Reach}, {Shibai}, {Sato}, {Smith}, {Staveley-Smith}, {Ueta}, {Van Dyk},
  {Werner}, {Wolff}, \& {Zaritsky}}]{Whitney08}
{Whitney}, B.~A., {Sewilo}, M., {Indebetouw}, R., {et~al.} 2008,
  \href{http://dx.doi.org/10.1088/0004-6256/136/1/18}{\JournalTitle{\aj}, 136,
  18}

\bibitem[{{Williams} {et~al.}(1994){Williams}, {de Geus}, \&
  {Blitz}}]{clumpfind}
{Williams}, J.~P., {de Geus}, E.~J., \& {Blitz}, L. 1994,
  \href{http://dx.doi.org/10.1086/174279}{\JournalTitle{\apj}, 428, 693}

\bibitem[{{Wong} {et~al.}(2011){Wong}, {Hughes}, {Ott}, {Muller}, {Pineda},
  {Bernard}, {Chu}, {Fukui}, {Gruendl}, {Henkel}, {Kawamura}, {Klein},
  {Looney}, {Maddison}, {Mizuno}, {Paradis}, {Seale}, \& {Welty}}]{MAGMA}
{Wong}, T., {Hughes}, A., {Ott}, J., {et~al.} 2011,
  \href{http://dx.doi.org/10.1088/0067-0049/197/2/16}{\JournalTitle{\apjs},
  197, 16}

\bibitem[{{Wong} {et~al.}(2019){Wong}, {Hughes}, {Tokuda}, {Indebetouw},
  {Onishi}, {Band urski}, {Chen}, {Fukui}, {Glover}, {Klessen}, {Pineda},
  {Roman-Duval}, {Sewi{\l}o}, {Wojciechowski}, \& {Zahorecz}}]{Wong19}
{Wong}, T., {Hughes}, A., {Tokuda}, K., {et~al.} 2019,
  \href{http://dx.doi.org/10.3847/1538-4357/ab46ba}{\JournalTitle{\apj}, 885,
  50}

\bibitem[{{W{\"u}nsch} {et~al.}(2012){W{\"u}nsch}, {J{\'a}chym}, {Sidorin},
  {Ehlerov{\'a}}, {Palou{\v{s}}}, {Dale}, {Dawson}, \& {Fukui}}]{Wunsch12}
{W{\"u}nsch}, R., {J{\'a}chym}, P., {Sidorin}, V., {et~al.} 2012,
  \href{http://dx.doi.org/10.1051/0004-6361/201118061}{\JournalTitle{\aap},
  539, A116}

\bibitem[{{Yamaguchi} {et~al.}(2001){Yamaguchi}, {Mizuno}, {Mizuno}, {Rubio},
  {Abe}, {Saito}, {Moriguchi}, {Matsunaga}, {Onishi}, {Yonekura}, \&
  {Fukui}}]{Yamaguchi01}
{Yamaguchi}, R., {Mizuno}, N., {Mizuno}, A., {et~al.} 2001,
  \href{http://dx.doi.org/10.1093/pasj/53.6.985}{\JournalTitle{\pasj}, 53, 985}

\bibitem[{{Zhang} {et~al.}(2014){Zhang}, {Gao}, {Henkel}, {Zhao}, {Wang},
  {Menten}, \& {G{\"u}sten}}]{Zhang14}
{Zhang}, Z.-Y., {Gao}, Y., {Henkel}, C., {et~al.} 2014,
  \href{http://dx.doi.org/10.1088/2041-8205/784/2/L31}{\JournalTitle{\apjl},
  784, L31}

\end{thebibliography}

\appendix

\section{Beam Filling Factor} \label{append: bff}

A filling factor (defined as  $f = T_{B,45''}/T_{B,\text{true}}$) is required to get accurate physical conditions. We attempted two methods of dealing with this issue: fitting the filling factor as a fourth dimension in the fitting process, and fitting ratios of the intensities rather than their absolute values to avoid the need for the filling factor at all. However, both of these methods reduce the degrees of freedom in the fitting to zero, and we found they were unable to constrain the physical parameters or the beam filling factor reliably.

\begin{figure}[h]
    \centering
    \includegraphics[width=0.75\textwidth]{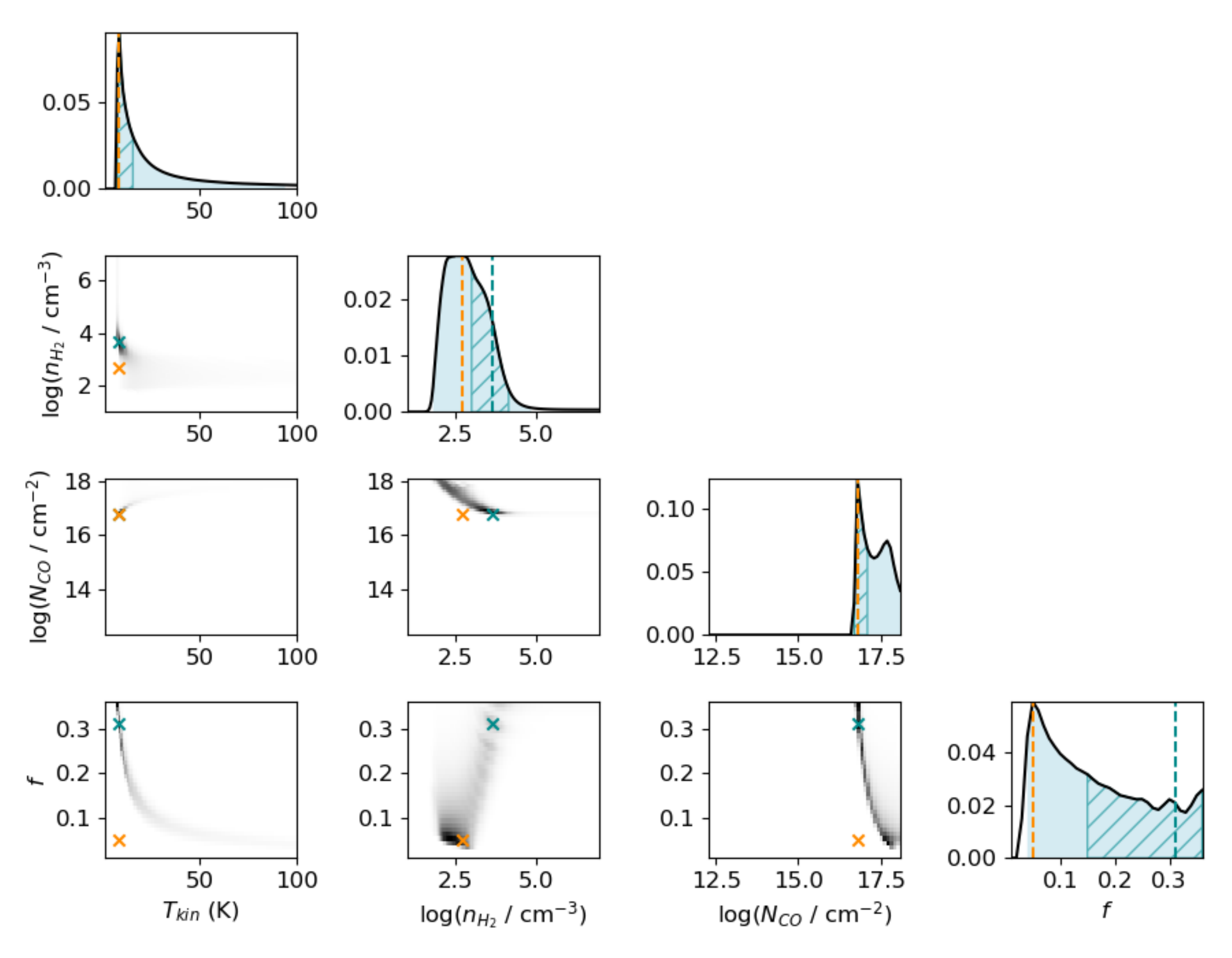}
    \caption{Example of a probability distribution from representative data (1.7 K, 0.2 K, 1.2 K, and 0.14 K for \twelveCO(1-0), \thirtCO(1-0), \twelveCO(2-1), and \thirtCO(2-1), respectively; same data, shading, and lines as in Figure\,\ref{fig:examplecorner}) with the beam filling factor, $f$, added a fourth dimension to fit, with $R_{13}$=100. This results in a poorly constrained filling factor and a multimodal 4-D probability distribution. The multimodality of the probability distribution results in a major inconsistency between the peak of the 4-D probability distribution and the peak of the collapsed probability profile for the filling factor.}
    \label{fig:example bff fitting}
\end{figure}

There are two common types of results when fitting the filling factor. An example of the first of these is shown in Figure\,\ref{fig:example bff fitting}, using the same representative data as in Figure\,\ref{fig:examplecorner} that has a high signal to noise ratio (ranging from 5 for \thirtCO(2-1) to 16 for \twelveCO(1-0)). In this plot, the filling factor is not constrained and the 4-dimensional probability distribution is multimodal. This results in a maximum likelihood point in 4-D parameter space (the blue line in Figure\,\ref{fig:example bff fitting}) that is inconsistent with the maximum of the collapsed probability profile for the filling factor (the orange line in Figure\,\ref{fig:example bff fitting}), making the results difficult to interpret. 

\begin{figure}
    \centering
    \includegraphics[width=0.75\textwidth]{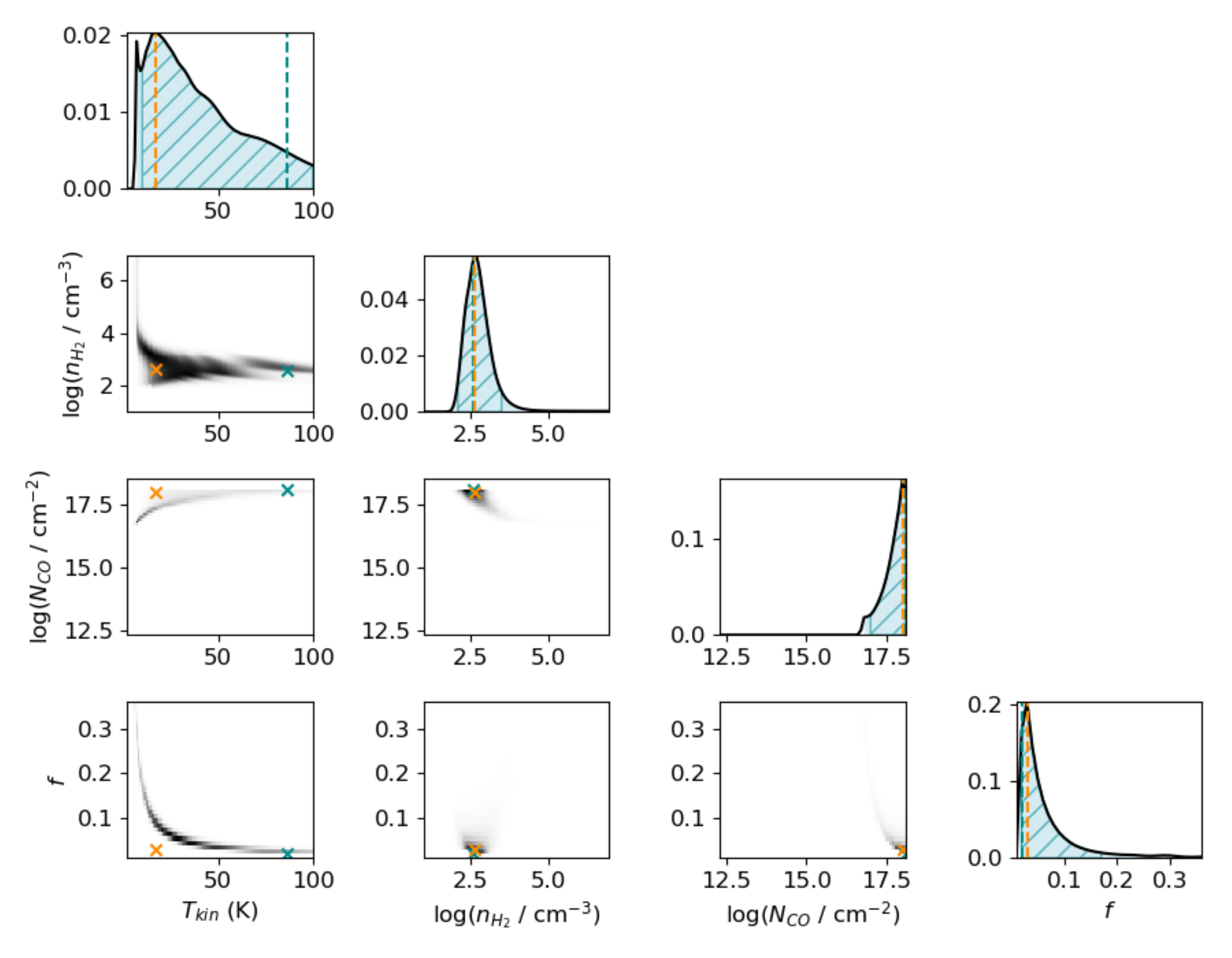}
    \caption{Example of a probability distribution from fainter data (0.96 K, 0.17 K, 0.91 K, and 0.10 K for \twelveCO(1-0), \thirtCO(1-0), \twelveCO(2-1), and \thirtCO(2-1), respectively) with the beam filling factor, $f$, added a fourth dimension to fit, and the same shading and lines as in Figure\,\ref{fig:examplecorner}. This results in a well-constrained fit for the filling factor at a low value, but an unconstrained, unphysically high \tkin and a combination of a high \NCO and low \nh which would require a much longer path length than is consistent with the size of the clump ($\gtrsim80$ pc compared to $\sim10$ pc). We consider this type of fit unreliable and unphysical.}
    \label{fig:faint bff fitting}
\end{figure}

The other common behavior when fitting the filling factor is shown in Figure\,\ref{fig:faint bff fitting} using the peak of one of the fainter clumps (numbered 21 in Figure\,\ref{fig:clump IDS}). In this case, the filling factor is well constrained on a low value, but the resulting fit for the other parameters are highly unlikely to occur physically. The temperature is poorly constrained but pushed to high temperatures, and \NCO is extremely high and \nh is low, which would require a very long path length along the line of sight ($\gtrsim80$ pc) that is inconsistent with the size of the clump being fitted ($\sim10$ pc). We concluded that this type of result was unphysical and unreliable, so could not be used to determine the filling factor. 

\begin{figure}
    \centering
    \includegraphics[width=0.65\textwidth]{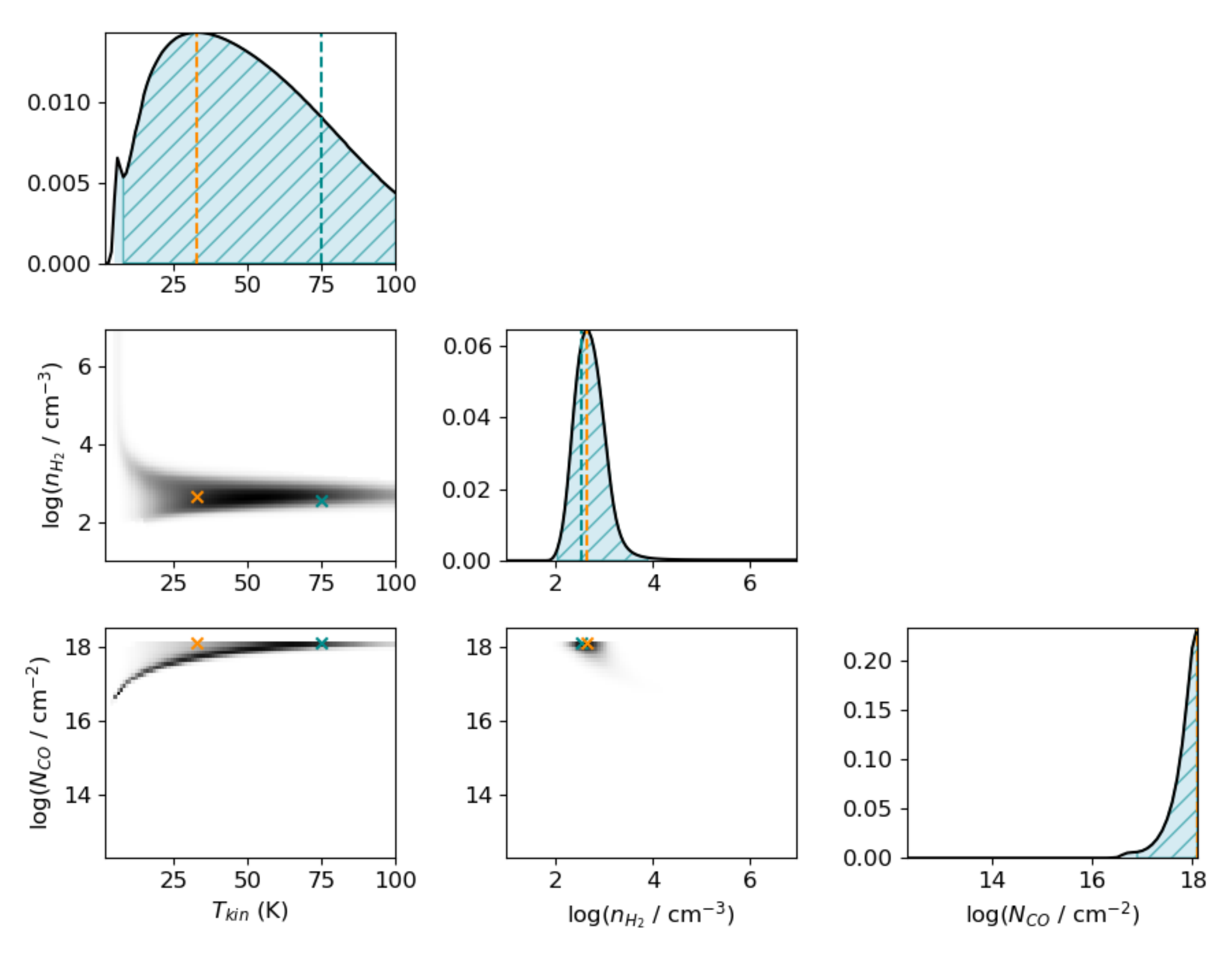}
    \caption{Example of a probability distribution from fitting the ratios of the line intensities, using \thirtCO(1-0) as the denominator. This plot uses the same intensities as Figure\,\ref{fig:faint bff fitting}, so the fitted ratios are 5.6, 5.3, and 0.59 for  \twelveCO(1-0)/\thirtCO(1-0), \twelveCO(2-1)/\thirtCO(1-0), and \thirtCO(2-1)/\thirtCO(1-0), respectively. The shading and lines are the same as in Figure\,\ref{fig:examplecorner}. This results in an unconstrained, unphysically high \tkin and a combination of a high \NCO and low \nh which would require a much longer path length than is consistent with the size of the clump ($\gtrsim80$ pc compared to $\sim10$ pc). We consider this type of fit unreliable and unphysical.}
    \label{fig:ratio fitting}
\end{figure}

Fitting ratios from the four observations removes the filling factor from the equation entirely, assuming that all four lines are tracing the same gas with the same filling factor. However, one line must be selected as the denominator, and so the number of data points to fit is reduced to three. The resulting fit has the same issues as fitting the filling factor: either the parameters are poorly constrained or the constrained values are unphysical. Figure\,\ref{fig:ratio fitting} shows an example of an unphysical fit from ratio fitting, where the fitted temperature is almost entirely unconstrained and the combination of a high \NCO and low \nh requires a long path length that is inconsistent with the size of the clump. 

Since fitting the filling factor or the ratios resulted in either an unreliable or unconstrained result, and because fitting the filling factor also significantly increases the computational requirements, we assumed a range of filling factors for the region, similar to our handling of $R_{13}$. The upper limits were determined by calculating the ratio of the high resolution (16\arcsec) ALMA \thirtCO(1-0) data to that same data convolved to 45\arcsec\ to get an observed upper limit.  

We ran \texttt{quickclump} on the high resolution ALMA \thirtCO(1-0), then took the ratio of the peak of the fifty brightest clumps and the peak of the corresponding low resolution clump as defined in \S\ref{subsec: clump defs}. This results in an upper limit for the filling factor of these high resolution clumps ($f < T_{B,45''}/T_{B,16''}$). Each low resolution clump had multiple corresponding high resolution clumps, so we looked at the minimum upper limit of those. All low resolution clumps had an upper limit above 20\%, so we adopt an upper limit on the filling factor of 20\% across the region. 

The lower limit of the range comes from the fits becoming unphysical below $f\sim$10\%. Taking lower beam filling factors results in unrealistically large temperatures with large errors, and large path lengths along the line of sight that are inconsistent with the projected size of the clumps. 

We make an exception to the 10\% filling factor lower limit for the clumps numbered 7, 8, and 9 in Figure\,\ref{fig:clump IDS}. Each of these clumps return unphysical values similar to those shown in Figure\,\ref{fig:faint bff fitting} when fitted with a 10\% filling factor. We show the resultant line-of-sight path length when adopting filling factors of 10\%, 15\%, and 20\% to demonstrate the unphysical nature of the 10\% results in that region. The path length is determined by dividing the H$_2$ column density ($N_{H_2}$) in units of cm$^{-2}$ by the H$_2$ volume density (\nh) in units of cm$^{-3}$, and converted to parsecs. These types of unphysical values occur primarily when the fitted filling factor is too low, and so for these three clumps we use a filling factor of 15\%, which is large enough that the results are no longer unphysical.

\begin{figure*}
    \centering
    \includegraphics[width=0.32\textwidth]{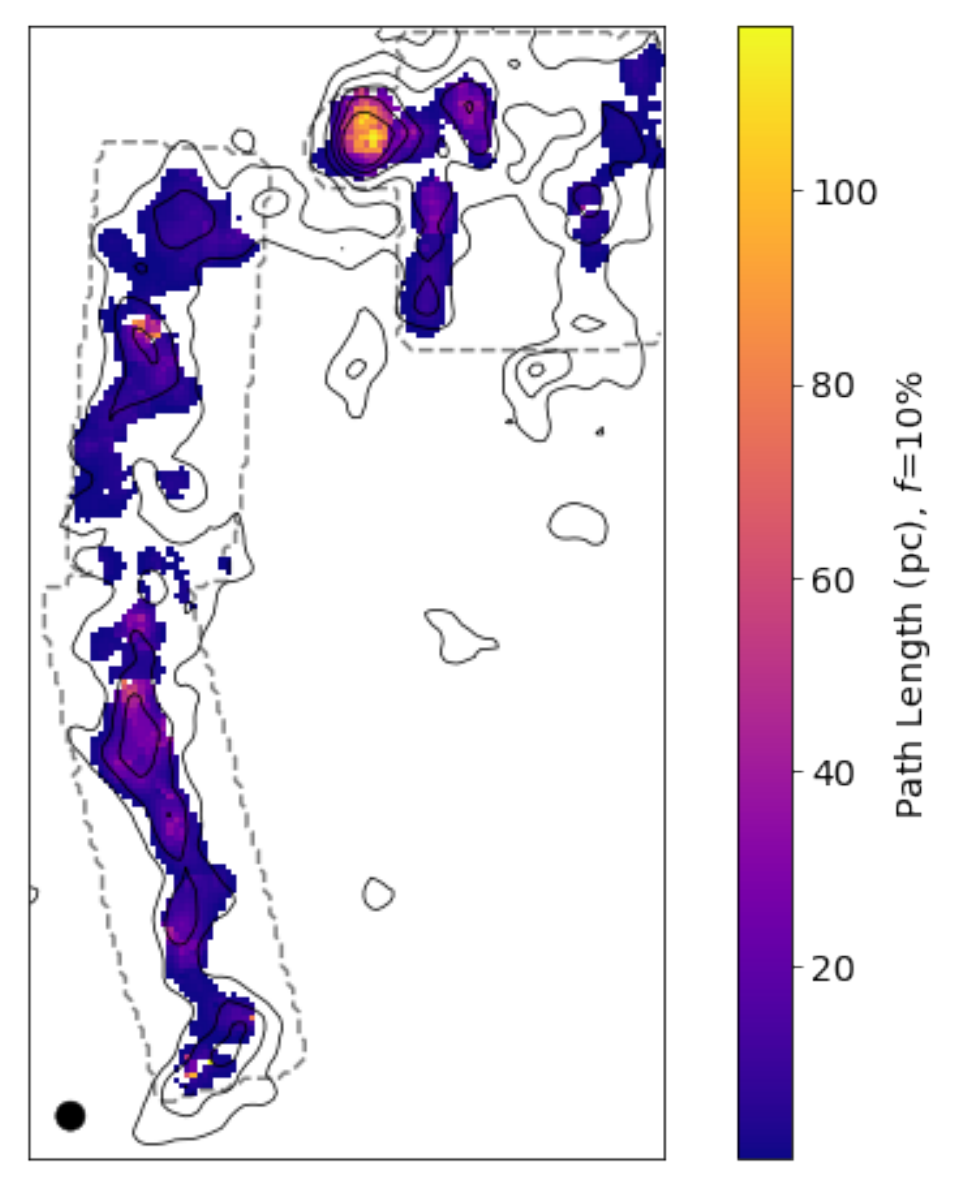}
    \includegraphics[width=0.32\textwidth]{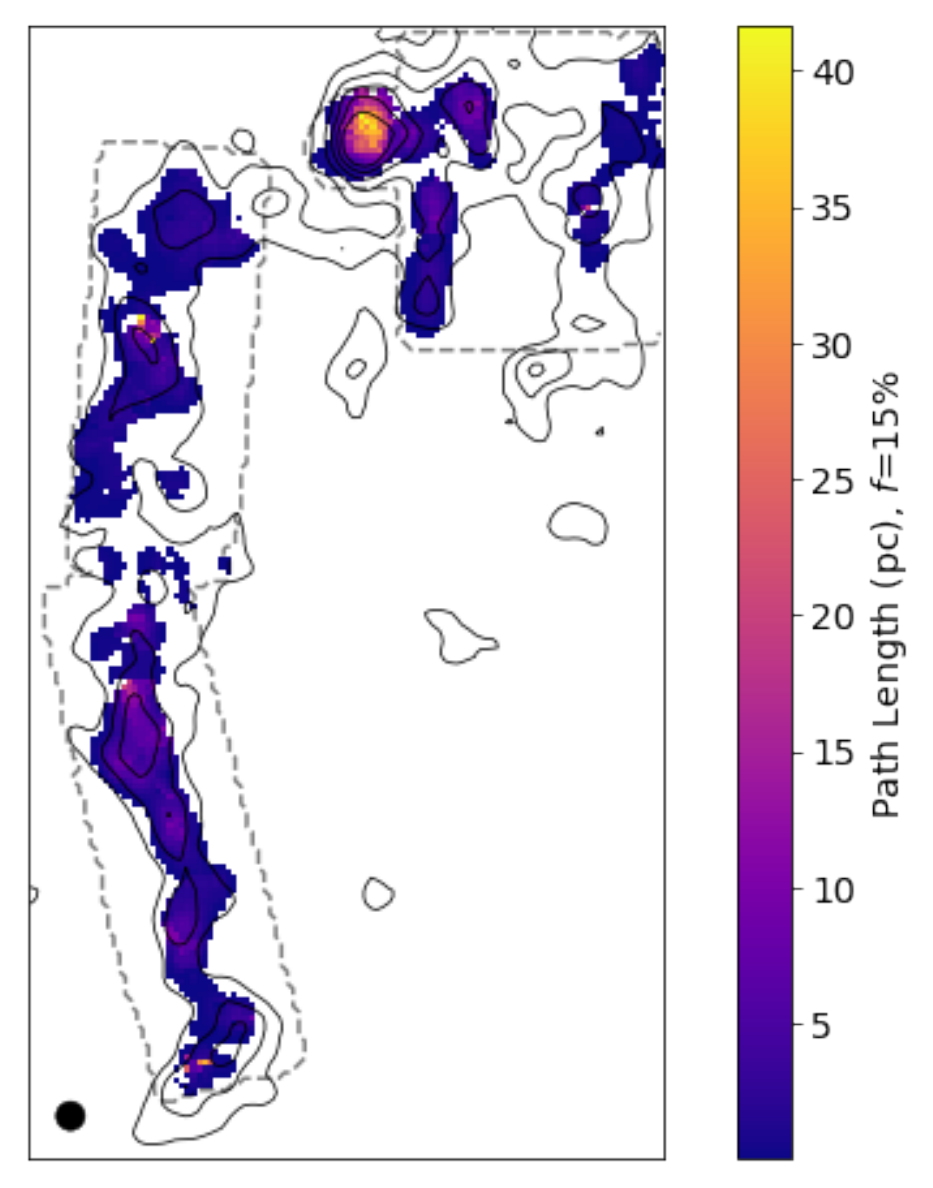}
    \includegraphics[width=0.32\textwidth]{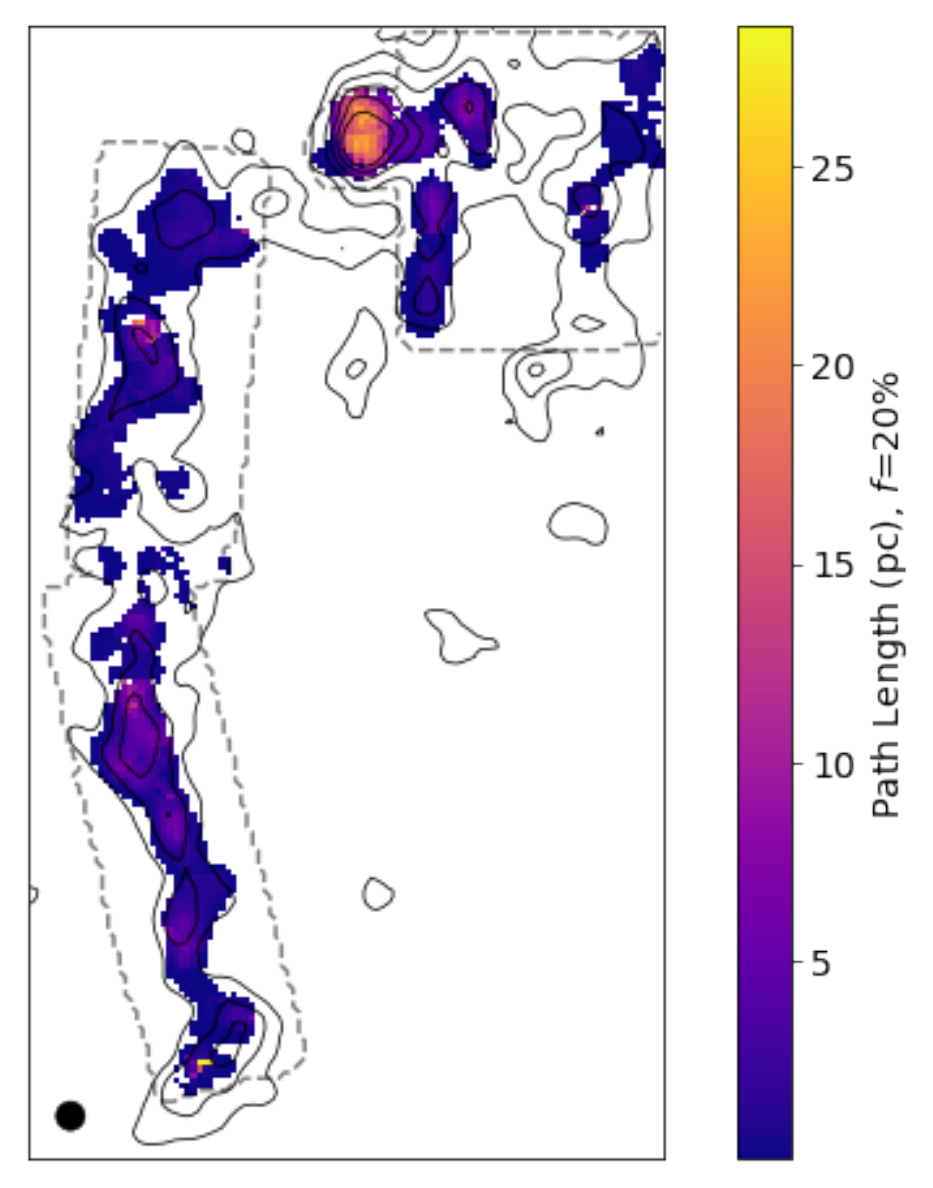}
    \caption{Line-of-sight path length of the fitting results with beam filling factors of 10\%, 15\%, and 20\%. The path length is determined by dividing the H$_2$ column density in units of cm$^{-2}$ by the H$_2$ volume density in units of cm$^{-3}$, and converted to parsecs. With a beam size of $\sim$11 pc, the path length becomes unrealistically large around clumps numbered 7, 8, and 9 (Figure\,\ref{fig:clump IDS}) when we assume $f=10$\%, so we instead use a filling factor lower limit of 15\% in that region. The contours are the \twelveCO(1-0) integrated intensity as shown in Figure\,\ref{fig:moments}, and the dashed line is the common observation footprint.}
    \label{fig:path lengths}
\end{figure*}

The assumed filling factor has the strongest effect on the fitted temperature. Due to the highest-probability line for \tkin and $f$ in the bottom left corner plots of Figures\,\ref{fig:example bff fitting} and \ref{fig:faint bff fitting}, at low filling factors the temperature quickly gets high with large errors, while at higher filling factors the variation in temperature levels off at lower values and varies much less. Because of this, taking a filling factor that is too low strongly affects the fitted temperature. 

The density is also affected by the filling factor, though not nearly as strongly as temperature. A higher filling factor results in a higher density most, but not all, of the time. When reporting values of \NCO, we multiply the filling factor back into the results (\NCO$\times f$) to get an accurate mass calculation. This results in a value of \NCO that is almost entirely unaffected by the assumed filling factor.

\section{Evaluating Fitting Performance} \label{append: simulated data}

We tested how well our multiline data with similar signal-to-noise as our Ridge measurements can be fit by this process and which of the five intervals described in \S\ref{subsec: fitting method} is best to accurately constrain the parameters (the five intervals are the three Bayesian intervals based on the 3-dimensional probability distribution and the two confidence intervals based on 1-dimensional probability profiles for each parameter). To do this evaluation, we simulated data for a range of physical parameters, covering the tested parameter space (\tkin between 5 and 55 K, \nh between $10^{2.5}$ cm$^{-3}$ and $10^{6}$ cm$^{-3}$, and \NCO between $10^{14.5}$ cm$^{-2}$ and $10^{18}$ cm$^{-2}$). For each combination of these three parameters, we used \radex to determine the expected emission from the four lines that we observe: \twelveCO(1-0), \thirtCO(1-0), \twelveCO(2-1), and \thirtCO(2-1). We then added random Gaussian noise based on the observed rms error for each line after convolving to a resolution of 45\arcsec\ in each cube to match the \twelveCO(1-0) resolution:  0.11 K, 0.017 K,0.1 K, and 0.035 K, respectively, for \twelveCO(1-0), \thirtCO(1-0), \twelveCO(2-1), and \thirtCO(2-1). We did not include any beam filling factors in this process. 

For each combination of physical parameters, we generated 100 instances of random Gaussian noise and then determined $\Vec{p}_\text{max}$, Bayesian intervals, and confidence intervals for each instance (see \S\ref{subsec: fitting method} for a description of these calculations). In all cases we used $R_{13}=100$. We then considered how often in these 100 instances the true model values were recovered within each of the five intervals. The recovery of each parameter depends on all three parameters (e.g., \NCO is better recovered at higher \tkin, as well as at higher \NCO), so to compare the five intervals' performances, we examine how each parameter is recovered as a function of each of the other two parameters as well as itself. 

\begin{figure}
    \centering
    \includegraphics[width=0.5\textwidth]{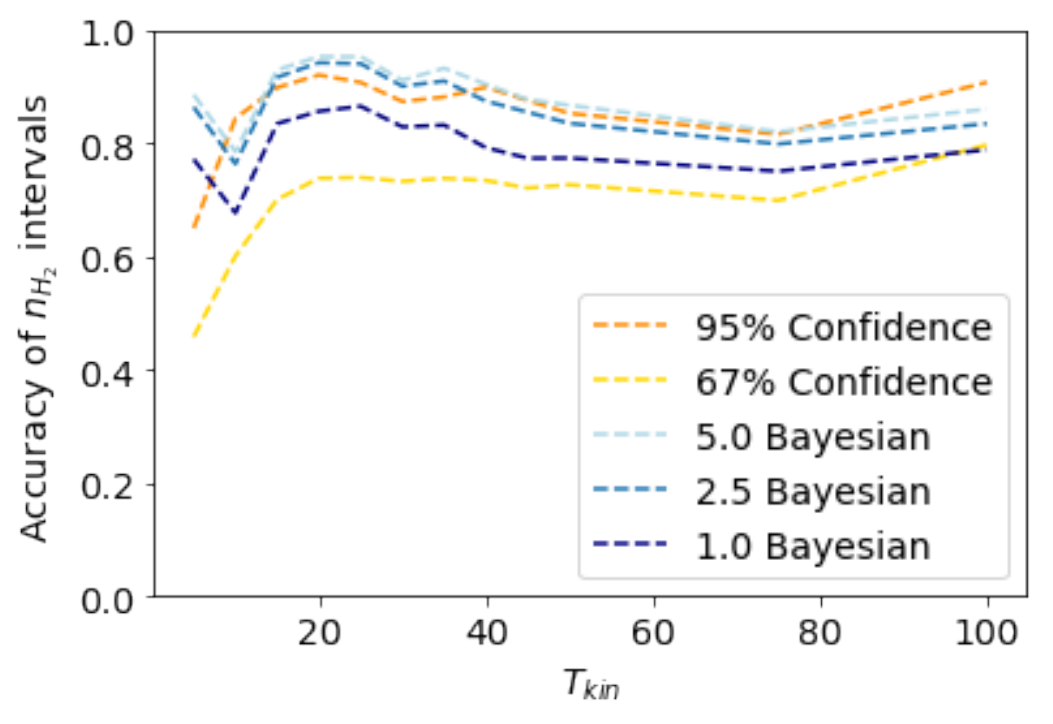}
    \caption{An example of how well we recover \nh as a function of \tkin. For each of the five intervals, we plot the fraction of runs in which the true model value was within the interval, averaged over all \nh and \NCO to get a function of \tkin. This plot shows that we do not robustly recover \nh when \tkin is low. Also, the 67\% confidence interval is much less accurate than the other four intervals.}
    \label{fig:accuracy1}
\end{figure}

An example of such a plot is shown in  Figure\,\ref{fig:accuracy1}, where we compare how often the correct value of \nh falls in each of the five intervals as a function of \tkin (and so collapsed over all values of \nh and \NCO). This shows that we do not robustly recover the true value of \nh when \tkin is low. Also, the 67\% confidence interval is a smaller range and contains the correct answer a much lower fraction of the time than the other wider confidence intervals, which is exactly what we would expect from a 1-sigma confidence interval.

\begin{figure}
    \centering
    \includegraphics[width=0.5\textwidth]{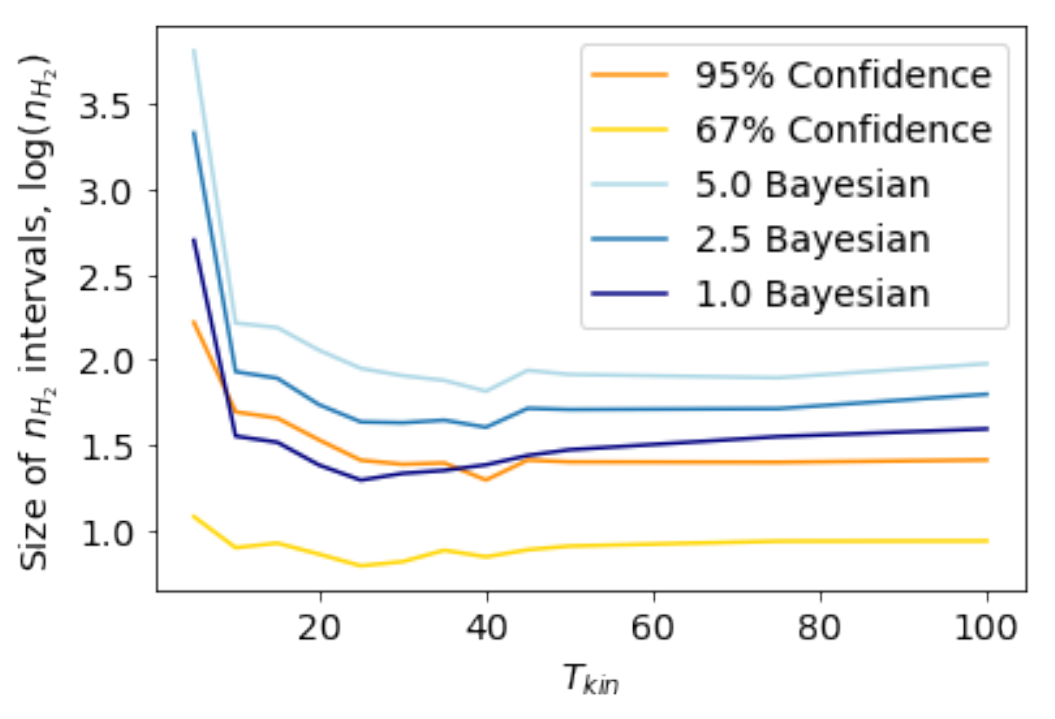}
    \caption{An example of how precise our fitting of \nh is as a function of \tkin. For each of the five intervals, we plot the total size of the \nh interval in dex, averaged over all \nh and \NCO to get a function of \tkin. The size shown here does not account for where within the interval the best fit value is, and it is frequently not symmetrical. At \tkin $<10$ K, \nh is less constrained, by approximately an order of magnitude in most cases. Also, the 67\% confidence interval constrains \nh most precisely (although we know from Fig\,\ref{fig:accuracy1} that it is much less accurate). Unsurprisingly, the most accurate intervals are also the least constrained.}
    \label{fig:precision1}
\end{figure}

How often the model value is in the interval tells us how accurate our fitting is, but we also want to know how precise our fitting can be. If the interval includes the entire parameter range, the true value will always be in it, but we have also done nothing to constrain it. So, we also consider the size of each interval for each parameter, as a function of each parameter. We show an example plot of this in Figure\,\ref{fig:precision1}, where the size of the five \nh intervals are again plotted as a function of \tkin. In this plot, \nh is better constrained when \tkin is high. As we would expect, the most constrained interval (the 67\% confidence interval) is the least accurate (as shown in Figure\,\ref{fig:accuracy1}), while the least constrained interval (the 5.0 Bayesian interval) is the most accurate.

\begin{figure}
    \centering
    \includegraphics[width=0.5\textwidth]{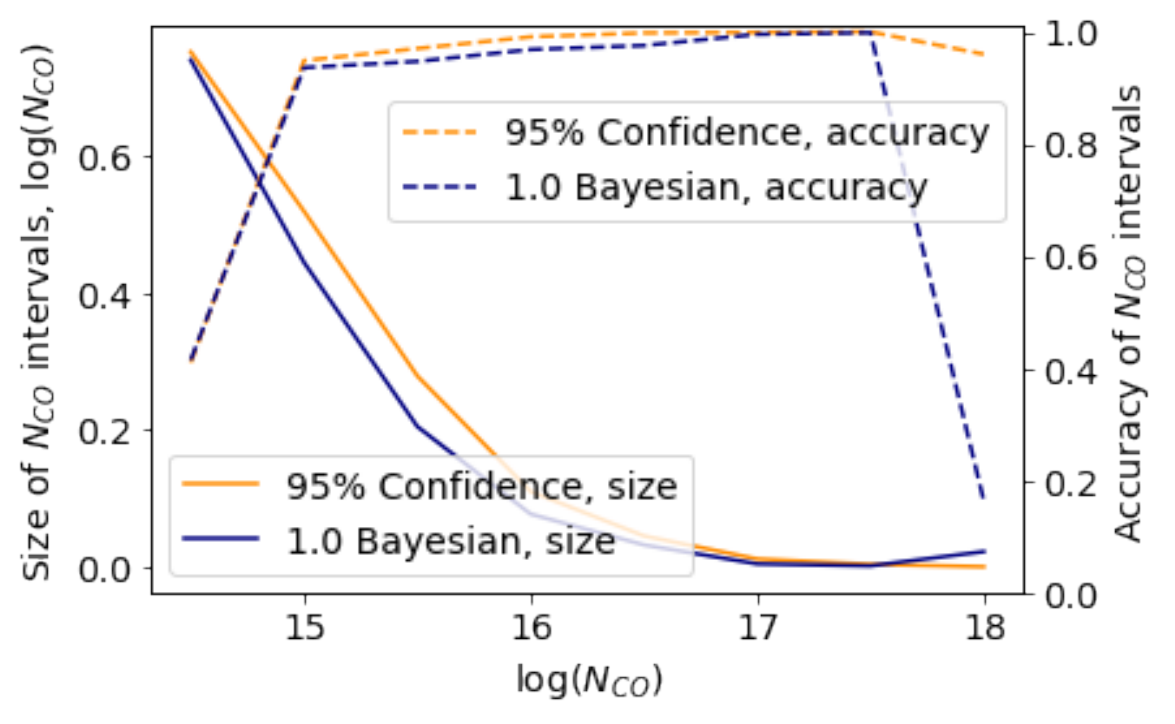}
    \caption{The accuracy and precision of the 95\% confidence interval and the 1.0 Bayesian interval for \NCO as a function of \NCO (averaged over \nh and \tkin). At low \NCO, neither interval is able to accurately predict the true \NCO value despite the size of the interval increasing. At high \NCO, the Bayesian interval drops sharply in accuracy, while the confidence interval does not. }
    \label{fig:NCO acc and prec}
\end{figure}

From an examination of plots similar to Figures\,\ref{fig:accuracy1} and\,\ref{fig:precision1} for all of the parameters, the combination of the 95\% confidence interval and the 1.0 Bayesian interval delivers the desired balance of accuracy and precision to fit the physical parameters to the data.
There are some regions of parameter space where one is preferred over the other and vice versa, and so a combination of the two is used to fit the actual data. This is demonstrated in Figure\,\ref{fig:NCO acc and prec}, where at high \NCO ($>10^{17.5}$ cm$^{-2}$), the 1.0 Bayesian interval returns a tight constraint on the wrong value, while the 95\% confidence interval tightly constrains the correct value. In other regions of parameter space, the 1.0 Bayesian interval is just as accurate as the 95\% confidence interval but more precise. We use a combination of the two intervals by selecting the more precise one to determine the correct value, with one exception: When the 95\% interval fits a high \NCO value ($>10^{17.5}$ cm$^{-2}$), we always use the 95\% interval since Figure\,\ref{fig:NCO acc and prec} demonstrates the Bayesian interval cannot be trusted in this range.

\begin{figure}
    \centering
    \includegraphics[width=0.5\textwidth]{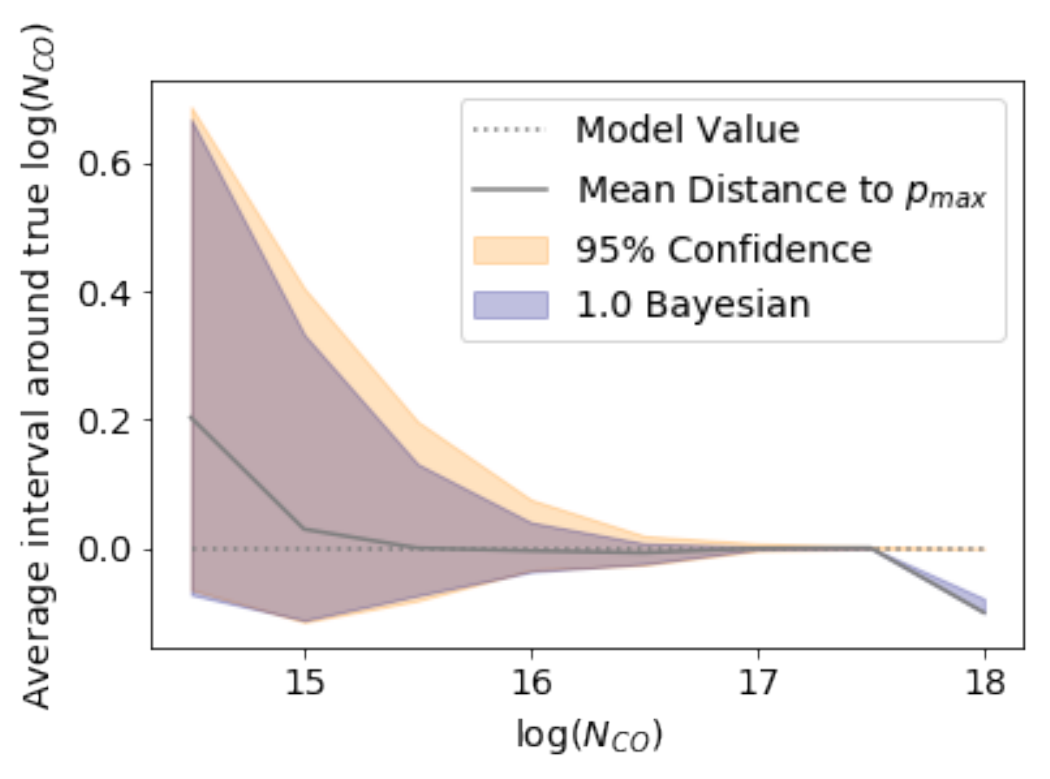}
    \caption{Offset of the 95\% confidence interval, the 1.0 Bayesian interval, and $\Vec{p}_\text{max}$ from the true value of \NCO, averaged over all \tkin and \nh, to get a function of \NCO. At low \NCO, both intervals as well as $\Vec{p}_\text{max}$ overestimate \NCO. At high \NCO, the confidence interval is accurate and well-constrained. The Bayesian interval however is precise and inaccurate - it tightly fits an under-estimate.}
    \label{fig:interval CI BI}
\end{figure}

We also consider systematic offsets (e.g., consistent over- or under-estimates) between the model values and the fitted intervals as measured by the mean distances of the model values to the edges of the intervals. We do this for the 95\% confidence interval, the 1.0 Bayesian interval, and $\Vec{p}_\text{max}$ for each parameter as a function of each parameter. We show an example of one such plot in Figure\,\ref{fig:interval CI BI}. At low \NCO, both intervals accurately include the true \NCO value, but most of the interval is an overestimate, while at high \NCO, the 1.0 Bayesian interval underestimates \NCO.

\subsection{Dependence on \twelveCO(1-0)} \label{append: 3 line fitting}

We tested the performance of the model fitting when we included only the highest resolution lines: \thirtCO(1-0), \twelveCO(2-1), and \thirtCO(2-1). We dropped the \twelveCO(1-0) since it had the worst resolution at 45\arcsec. We then simulated data once again using the \radex model from the same range of parameter space as above. This time we added random Gaussian noise based on the error after convolving each cube to 30\arcsec\ instead of 45\arcsec\ to match the new limiting resolution of \thirtCO(2-1). 

\begin{figure}
    \centering
    \includegraphics[width=0.5\textwidth]{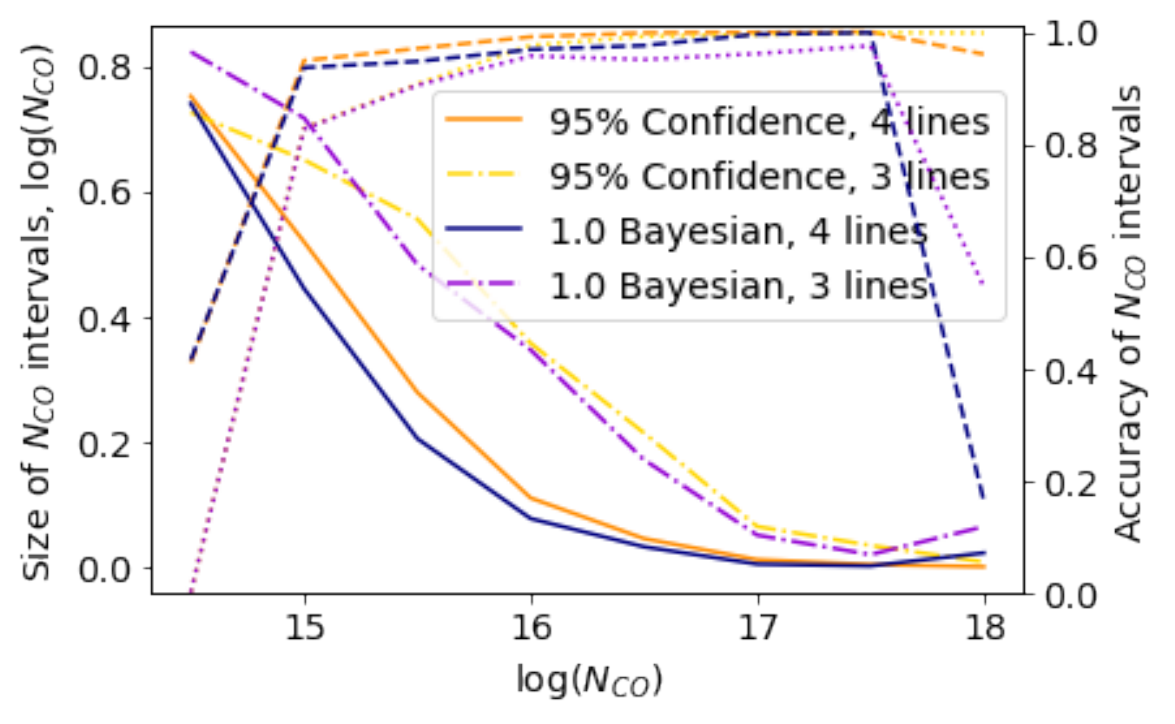}
    \caption{A comparison of how the 95\% confidence interval and 1.0 Bayesian interval fit \NCO as a function of \NCO when four lines are included (\twelveCO(1-0), \twelveCO(2-1), \thirtCO(1-0), and \thirtCO(2-1)), and when only three lines are included (no \twelveCO(1-0)). The solid lines show the sizes of the intervals, while the dashed lines show how often the true value is in the interval. Fitting with only three lines makes the biggest difference at moderate \NCO, where four lines can constrain the value much better than only three. At very high or very low \NCO, they perform similarly. }
    \label{fig:3v4 acc prec}
\end{figure}

\begin{figure}
    \centering
    \includegraphics[width=0.5\textwidth]{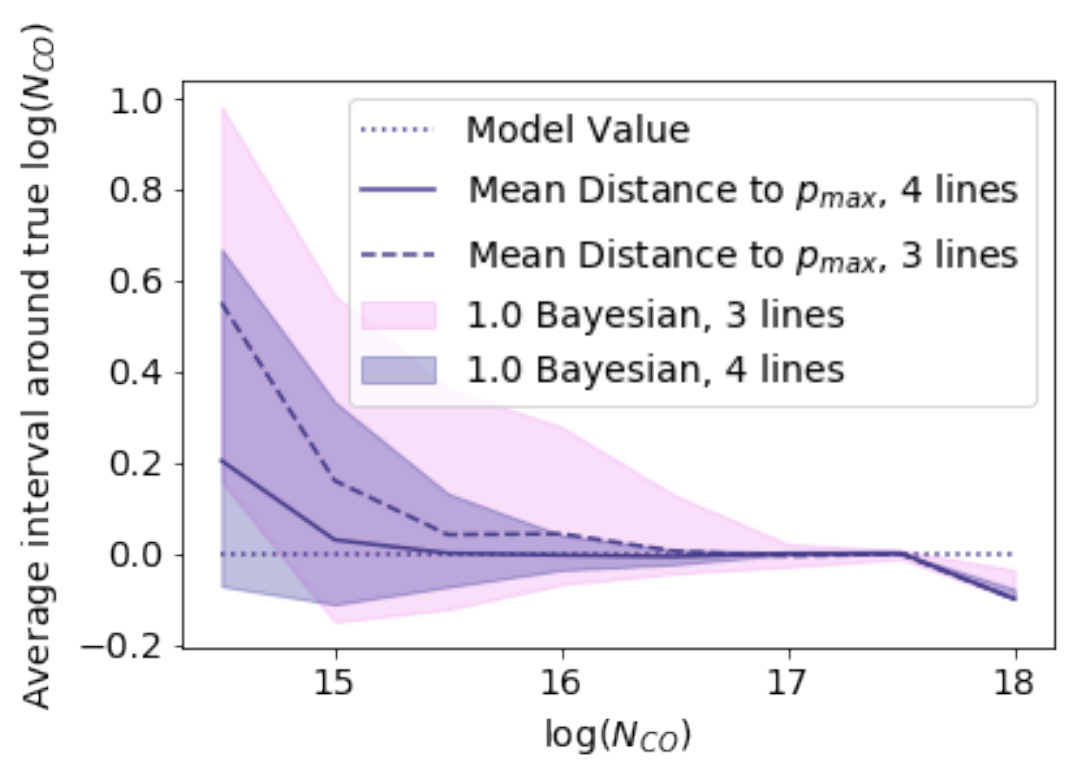}
    \caption{Offset of the 1.0 Bayesian interval and $\Vec{p}_\text{max}$ from the true value of \NCO as a function of \NCO when four lines are included (\twelveCO(1-0), \twelveCO(2-1), \thirtCO(1-0), and \thirtCO(2-1)), and when only three lines are included (no \twelveCO(1-0)). As in Figure\,\ref{fig:3v4 acc prec}, four lines more tightly constrain \NCO in general, especially at moderate values of \NCO. Note that at low \NCO, the average interval is  entirely above the true value if only three lines are fitted. }
    \label{fig:3v4 BI}
\end{figure}

We compared the fitting performance using the measures described above to determine how much sensitivity we lose by not including the information from \twelveCO(1-0). We show example plots from this comparison in Figures\,\ref{fig:3v4 acc prec} and \ref{fig:3v4 BI}. By dropping the \twelveCO(1-0), we are less sensitive to intermediate \NCO values: While the resulting fitted intervals still include the correct value almost all the time for \NCO $> 10^{15}$ cm$^{-2}$, they are only well-constrained for \NCO $> 10^{16}$ cm$^{-2}$. Since we expect that much of the Ridge may fall in this range of \NCO values, we decided that the improved sensitivity to the physical parameters is worth losing some resolution.

A comprehensive set of plots comparing the performance of fitting in the case of three lines instead of four for all parameters similar to Figures\,\ref{fig:accuracy1}--\ref{fig:3v4 BI} is available as supplementary material\footnote{\url{https://doi.org/10.5281/zenodo.4646288}}. 

\section{Radex Fitting Method} \label{append:comparing radex}

\begin{table}
    \caption{Comparison of \radex fitting methods}
    \begin{center}
    \begin{tabular}{c|c c c}
        \hline
        \hline
         & $M/M_\text{LF}$ & \tkin/$T_{\text{kin,LF}}$ & \nh/$n_{\text{H}_2,\text{LF}}$ \\
         \hline
         Full Fit      & 1.07$\pm$0.09 & 0.92$\pm$0.12 & 0.81$\pm$0.17 \\
         Clump Fixed   & 0.83$\pm$0.11 & 1.15$\pm$0.14 & 1.04$\pm$0.42 \\
    \end{tabular}
    \end{center}
    \textbf{Notes.} Comparison of different methods of determining mass (from \NCO), \tkin or $T_\text{ex}$, and \nh. These values are the average of the ratios for all the clumps that were fit with these methods. $M_\text{LF}$, $T_{\text{kin,LF}}$ are the values derived from the Line Fixed \radex fitting method. 
    \label{tab: radex fitting comparison}
\end{table}

In addition to the method described in \S\ref{subsec:getting maps}, we tested holding \tkin and \nh fixed for the whole clump (Clump Fixed method) and fully fitting \tkin and \nh for every velocity and every pixel (Full Fit method). We decided to hold \tkin and \nh fixed for only the line (Line Fixed method, described in \S\ref{subsec:getting maps}) because this had the best physical motivation and gave the most realistic results. The Full Fit method resulted in the line edges becoming unrealistically low in \tkin and high in \NCO and having poor fits at the line edges. The Full Fit method is also less consistent with the assumptions of escape probability codes including \radex, which assumes constant excitation across a finite-sized cloud. 

The Clump Fixed method had similar results to the Line Fixed method, but the \NCO fits were much less constrained on the edges of the clump. Also, a uniform temperature and density profile is unrealistic for a clump, and holding these values fixed for the whole clump made the fit overly dependent on the single peak pixel, which we wanted to avoid.

We compared the resulting masses, temperatures, and densities for each clump to the masses from the Full Fit and Clump Fixed methods, rather than the Line Fixed method. A table summarizing how our measurements would change with different fitting methods is given in Table\,\ref{tab: radex fitting comparison}, where we show the ratio of the derived mass, temperature, and density from alternate fitting methods and our preferred Line Fixed method,  averaged across all clumps included in this analysis and the standard deviation of these ratios.

The Full Fit method results in mass, temperature, and density estimates that seem largely consistent with the Line Fixed method. The difference between these two methods is in the fitting of the line wings, and since the line wings have a much lower column density than the line peak, the mass and the mass-weighted \tkin and \nh are not much higher or lower than when we use the Line Fixed method. When using the Full Fit method, the lower intensities in the line wings cause the fitting to prefer lower temperatures and densities, which is compensated for with higher \NCO values, causing a slightly higher mass than the Line Fixed method. 

The Clump Fixed method results in a lower mass and higher temperature than the Line Fixed method on average. The density is consistent with the Line Fixed method, but with larger variations. A lower mass is expected from this method, since the fitting compensates for a higher temperature by preferring a lower \NCO. On the edges of the clump where we would expect the temperature to be lower, it is instead fitted with the temperature from the center of the clump, causing lower \NCO estimates and a lower mass overall. The temperature is also expected to be higher, though the effect is not as large because \tkin is mass-weighted and the center of the clump where \NCO is highest is less affected by the constant temperature assumption.

\end{document}